\documentclass[journal]{rmaa}

\let\gtrsim\undefined
\let\lesssim\undefined
\usepackage{amssymb}
\usepackage{cancel}
\usepackage{epsf}
\usepackage{pstricks}
\usepackage{pst-plot}
\usepackage{pstricks-add}
\usepackage{epstopdf}
\usepackage{natbib}
\usepackage{pdflscape}	
\usepackage[figuresright]{rotating}   
\usepackage{longtable}
\usepackage{threeparttablex}
\usepackage{lscape} 
\usepackage{chngpage} 

\usepackage{paralist}

\usepackage{psfrag,color}

\usepackage[latin1]{inputenc}

\title{Tracing the assembly histories of galaxy clusters in the nearby universe}

\author{
  C.~A.~Caretta\altaffilmark{1},
  H.~Andernach\altaffilmark{1,2},
  M.~Chow-Mart\'inez\altaffilmark{1,3},
  R.~Coziol\altaffilmark{1},
  J.~De~Anda-Su\'arez\altaffilmark{4},
  C.~Hern\'andez-Aguayo\altaffilmark{5,6},
  J.~M.~Islas-Islas\altaffilmark{1,7},
  M.~M.~Mireles-Vidales\altaffilmark{1}, 
  M.~A.~Mu\~niz-Torres\altaffilmark{1},
  H.~Santoyo-Ruiz\altaffilmark{8},
  J.~J.~Trejo-Alonso\altaffilmark{1,9},
  Y.~Venkatapathy\altaffilmark{1,10},
  and J.~M.~Z\'u\~niga\altaffilmark{1}}

\altaffiltext{1}{Departamento de Astronom\'ia (DCNE-CGT), 
Universidad de Guanajuato, 
Callej\'on de Jalisco s/n, C.P. 36023, 
Guanajuato (Gto.), Mexico.} 

\altaffiltext{2}{Th\"uringer Landessternwarte, 
Sternwarte 5, D-07778, 
Tautenburg, Germany.}

\altaffiltext{3}{Instituto de Geolog\'ia y Geof\'isica,
Universidad Nacional Aut\'onoma de Nicaragua, 
Managua, Nicaragua.}

\altaffiltext{4}{
Tecnol\'ogico Nacional de M\'exico, ITS 
Pur\'isma del Rinc\'on,
Pur\'isma del Rinc\'on (Gto.), Mexico.}

\altaffiltext{5}{
Max-Planck-Institut f\"ur Astrophysik, Karl-Schwarzschild-Str. 1,
D-85748, 
Garching, Germany.}

\altaffiltext{6}{
Excellence Cluster ORIGINS, Boltzmannstrasse 2, D-85748, 
Garching, Germany.}

\altaffiltext{7}{
Universidad Tecnol\'ogica de Tulancingo, 
Tulancingo (Hgo.), Mexico.}

\altaffiltext{8}{
Descubre Museo Interactivo de Ciencia y Tecnolog\'ia, 
Aguascalientes (Ags.), Mexico.}

\altaffiltext{9}{Facultad de Ingenier\'ia, 
Universidad Aut\'onoma de Quer\'etaro, 
Quer\'etaro (Qro.), Mexico.}

\altaffiltext{10} {Instituto de Radioastronom\'ia y Astrof\'isica, Universidad Nacional Aut\'onoma de M\'exico, Morelia (Mich.), Mexico.}

\shortauthor{Caretta \textit{et al.}}
\shorttitle{Assembly histories of galaxy clusters}

\fulladdresses{
\item Corresponding author: C\'esar Augusto Caretta (c.augusto@ugto.mx)}

\listofauthors{C.~A.~Caretta, H.~Andernach, M.~Chow-Mart\'inez \textit{et al.}}
\indexauthor{Caretta, C.~A.}

\abstract{
We have compiled a sample of 67 nearby ($z <$ 0.15) clusters of galaxies, for which on average more than 150 spectroscopic members are available, and, by applying different methods to detect substructures in their galaxy distribution, we have studied their assembly history. Our analysis confirms that substructures are present in 70\% of our sample, having a significant dynamical impact in 57\% of them. 
A classification {of} the assembly state of the clusters based on the dynamical significance of their substructures is proposed. 
In 19\% of our clusters, the originally identified brightest cluster galaxy 
is not the {central gravitationally dominant galaxy (CDG)}, but turns out to be either the second-rank, or the dominant galaxy of a substructure {(a SDG, in our classification)}, or {even} a ``fossil'' galaxy in the periphery of the cluster. Moreover, no correlation was found in general between the projected offset of the {CDG} from the X-ray peak and {its} peculiar velocity.  
The comparison of the {CDGs} properties with the assembly states and dynamical state of the intracluster media, 
especially the core cooling status, suggests a complex assembly history, with clear evidence of co-evolution of the {CDG} and its host cluster in the innermost regions.
}
\resumen{
Hemos recopilado una muestra de 67 c\'umulos cercanos ($z <$ 0.15) de galaxias, para los cuales se dispone de un promedio de m\'as de 150 miembros espectrosc\'opicos, y, aplicando diferentes m\'etodos para detectar subestructuras en su distribuci\'on de galaxias, hemos estudiado su historia de ensamblaje.
Nues\-tro an\'alisis confirma la presencia de subestructuras en el 70\% de los c\'umulos de nuestra muestra, teniendo un impacto din\'amico significante en el 57\% de ellos.
Se propone una clasificaci\'on de los estados de ensamblaje de los c\'umulos basada en la significancia din\'amica de sus subestructuras.
En el 19\% de nuestros c\'umulos, la galaxia m\'as brillante 
identificada originalmente no es la galaxia gravitacionalmente dominante {central (CDG)}, pero suele ser 
una segunda galaxia dominante, 
o la galaxia dominante de una subestructura {(una SDG, en nuestra clasificaci\'on)}, 
o {hasta} una galaxia ``f\'osil'' en la periferia del c\'umulo. 
Adem\'as, no se encontr\'o correlaci\'on, en general, entre el desplazamiento proyectado de la {CDG} respecto al pico de emisi\'on en rayos-X y {su} velocidad peculiar. 
La comparaci\'on de las propiedades de las {CDGs} con los estados de ensamblaje y estados din\'amicos de los medios intracumulares 
especialmente el estado de enfriamiento del \textit{core}, sugiere una historia de ensamblaje compleja, con clara evidencia de co-evoluci\'on de la {CDG} y su c\'umulo hu\'esped en las regiones m\'as internas.
}

\addkeyword{galaxies: clusters: BCGs}
\addkeyword{galaxies: clusters: membership}
\addkeyword{galaxies: clusters: substructures}
\addkeyword{galaxies: groups: general}

\begin{document}
\maketitle

\section{Introduction}
\label{sec:Intro}

According to the hierarchical structure formation paradigm, gravity brings together smaller mass systems into larger, more massive ones: in a sequential process, galaxies assemble in groups, and groups merge to form clusters, which, at the present epoch, have started to congregate over the largest scale appearing as superclusters. At each mass scale, the environment where the object (galaxy, group or cluster) forms influences how it grows and evolves, through complex physical processes that still need to be investigated further and clarified. This makes retracing the assembly histories of such objects/systems a difficult {but paramount} task. 

The effects of the environment on galaxy formation and evolution have been extensively studied \citep[\textit{e.g.},][]{Dre80,Cal93,Pog06}, originally in terms of the cluster-field dichotomy.  
However, the discussion has recently taken a new turn {after realizing that groups and clusters are part of} 
the cosmic web, namely, the large-scale structure (LSS). Within this new paradigm, the global environment of galaxies has a foamy texture, a structure full of voids \citep[\textit{e.g.},][]{Tem09,Var12,Ein14,Dup19} that are encircled by a web of filaments \citep[\textit{e.g.},][]{Por08,Pou17,Iris20}, where the bulk of the intergalactic gas is found \citep[\textit{e.g.},][]{Fra11,Pla14,Rei21}, and connected through nodes, where the density of matter is highest. It is along the backbone of the filaments that groups of galaxies form, before migrating and merging within rich clusters of galaxies in the nodes.

Information about the fundamental properties of these structures and their member galaxies has also improved significantly thanks to many surveys: large-scale redshift surveys \citep[\textit{e.g.},][]{She96,daC98,Fal99,Col05,Jon09,Bal10,Huc12,Alb17}, optical CCD-based photometric surveys \citep[\textit{e.g.},][]{Ham01a,Skr06,Aih11,Sha15,Dey19,Cha19,Abb21}, and interferometric radio surveys \citep[\textit{e.g.},][]{Bec94,Con98,Brn01,Lac20}. 
These studies allowed various physical processes for the formation and evolution of galaxies to be identified, with efficiencies varying with the density of the environments of structures on different scales. Fundamentally, this has shown that understanding how galaxies form and evolve requires understanding how the structure and characteristics of their environments affect their intrinsic properties: their mass, morphology, star formation history and even BH formation and AGN activity.


However, this also implies being able to distinguish environmental effects from those related to secular 
evolution (the question of whether this is due to ``nature or nurture'').
A panoply of galaxy characteristics 
are used to achieve this goal, like their colors, their shapes, their orientations and spins, or equivalent parameters extracted from comparing their spectra with synthetic stellar population models. When studying groups and clusters, distinguishing between nature and nurture also necessitates recognizing the dynamical states of these systems, as reflected by the different distributions of galaxies, intergalactic gas (intracluster medium, ICM, or intragroup medium, IGM) and dark matter (the so-called halo-occupation problem). 
Reconciling all these different aspects is theoretically demanding and observationally expensive, which complicates the task of building a comprehensive model of their formation and evolution.

Usually, studies related to the structure and evolution of galaxy groups and clusters {suffer from} one or more of the following {limitations}:
i) only {projected positional} data are used for substructure analyses \citep[\textit{e.g.},][]{Lop06,Ram07,WeH13}; 
ii) {the use of photometric redshifts \citep[\textit{e.g.},][]{WeH15,Bon18} --clearly the estimation of redshifts using only photometry has improved a lot during recent years, but photometric redshifts still lack the accuracy to determine cluster membership and dynamical state in the way that is possible with spectroscopic redshifts--}; 
iii) only a small number of member galaxies with spectroscopic redshifts are available ({frequently} affecting high-redshift cluster studies); 
iv) many spectroscopic redshifts are available but only for a small number of clusters \citep[\textit{e.g.},][]{Tyl14,Son17,Liu18}; 
v) cluster samples {that} are biased in richness and mass, or focused on special aspects, like regularly shaped clusters, dominated by cD galaxies, showing strong X-ray emission or an ICM with strong Sunyaev-Zel'dovich (SZ) signal \citep[\textit{e.g.},][]{OeH01,Rum18,Lop18}. 
To palliate these limitations, we aim to build a database  {collecting information related to}
the environments of different structures in the nearby Universe (from groups to superclusters), that is as complete and homogeneous as possible. In this paper, we concentrate more specifically on defining a sample of galaxy clusters that have a large range of richness, to establish their dynamical and evolutionary states in order to trace their assembly histories. 

For that we need to better investigate the 
{importance of} substructures and their dynamically dominant galaxies {for the cluster evolution as a whole}.
{We make a clear distinction here between the photometric ranking of member galaxies of a galaxy system (cluster or group), {which has lead to}  
the terms BCG (Brightest Cluster Galaxy) and BGG (Brightest Group Galaxy), and a ranking that takes into account their dynamical relevance and evolution. Because today 
we have enough information to study the assembly and evolution of galaxy clusters, this distinction becomes necessary. 
Thus, we define, for each cluster or group, a CDG (Central Dominant Galaxy), and one or more 
SDGs (Substructure/Sub-cluster/Satellite Dominant Galaxies) for each of the cluster substructures when they are present. 
The CDG and SDGs of a cluster are usually the brightest galaxies of this cluster, and we will retain the term BCGs to refer colectivelly to them. 
In other words, BCGs and BGGs are photometrically defined prior to a dynamical analysis, while CDGs and SDGs are a reclassification of the BCGs and BGGs according to their host sub-systems and dynamical importance.} 

Moreover, assuming that {CDGs} with a cD (or D) type morphology form by cannibalizing galaxies falling toward the center of the potential wells of the clusters \citep[\textit{e.g.},][]{Coz09,Zha15}, one would naturally expect their masses to show some specific relation with the masses of their parent structures, $M_{Cl}$--$M_{CDG}$ \citep[\textit{e.g.},][]{Sto10,Lav16}. 
In particular, we would expect {CDGs} in dynamically relaxed clusters to lie at the bottom of the potential wells of their systems. 
However, observations show {that, for most of the clusters}, the positions of many cDs {are} offset from
the peak in X-ray emission, the {latter} assumed to settle more rapidly to the bottom of the potential well, or having high peculiar velocities within the cluster compared to the center of the radial velocity distribution \citep[][]{Coz09,Mar14,Lau14}. {This points to most of the clusters being unrelaxed or maybe to the presence of some undetected projection effects.}

Another difficulty lies in the cannibalism mechanism itself. How can mergers happen efficiently in a systems where the velocity dispersion of galaxies increases as they fall into deep potential wells \citep[\textit{e.g.},][]{Mer85,1985Tonry,2004Mihos}? 
Alternatively, an important part of the formation of galaxies now in clusters could have happened in smaller-mass systems, like groups, where the velocity dispersion (and thus the amount of ICM) is smaller, the groups then merging to form or enrich more massive clusters. This phenomenon is known in literature as pre-processing \citep[\textit{e.g.},][]{Cal93,Car08,Don21}.

Within the cosmic web paradigm, one needs to ponder how the constant feeding of clusters by the merging and accretion of groups forming in filaments tempered these expectations. For instance, assuming mergers take place regularly, substructures in the distribution of galaxies would be expected to be common at low redshifts. This would naturally explain the {CDG}--X-ray offsets, since the ICM having a higher impact parameter than galaxies would follow a different path towards virialization, reaching equilibrium more rapidly. 
 
Common mergers of groups within a cluster would also be expected to disrupt the cool core (CC) of this system, making the {CDG} wobble around the distorted potential well, explaining its peculiar velocity \citep[\textit{e.g.},][]{Har17}. This could also have an important impact on the formation of cD galaxies. In the evolution scenario proposed by \citet{Lav16}, for example, it is proposed that a {CDG} transforms into a cD by cannibalism only when, after a cluster-scale merger event, the most massive galaxies of the merging groups, displaced from their initial potential, migrate towards the potential center of the newly formed cluster; this temporary imbalance increases dynamical friction and thus favors cannibalism. Consequently, one would expect the magnitude gaps to increase between the {CDG} and its luminous neighbors, but not necessarily between the second and third-rank galaxies, due to their high velocity dispersion. 

All these considerations suggest that, assuming that groups in filaments continuously merge to form clusters, several primeval group {CDGs} might appear among the BCGs of a cluster. 
Moreover, due to the different time-scales for the relaxation of such complex systems, we might also expect the galaxy distributions and characteristics to reflect some specific aspects of their merger processes. 
By compiling and studying a well characterized sample of galaxy clusters, therefore, it should be possible to distinguish different states of the merger process, and better document their assembly histories. 

The sample we present in this article is an effort in this direction. It is composed of 67 optically selected Abell galaxy clusters that are nearby ($z <$ 0.15), and for which a large number (above 100) of spectroscopically confirmed potential members are available. 
This sample includes a fair distribution of all Bautz-Morgan (BM) type clusters and various levels of ICM X-ray properties (from luminous to under-luminous, AXU). The article is organized in the following way. 
In \S~\ref{sec:Dat}, we present the data used in our study: we introduce the cluster and galaxy samples and describe the information retrieved from the photometric, astrometric and spectroscopic observations. 
In \S~\ref{sec:Met}, we describe the methods we used to characterize the galaxy systems and their structures: center and membership determination, characterization of dynamical parameters (like cluster redshift, velocity dispersion, richness, mass, gravitational binding, and {CDG} offsets and gaps), and optical substructuring analyses.
Our results about the dynamical properties and level of substructuring, for the outer, inner and core regions of the systems, are discussed in \S~\ref{sec:Res}. 
This is followed by a brief summary and conclusions in \S~\ref{sec:Con}. For our analysis, we assume a standard $\Lambda$CDM cosmology, with $\Omega_\Lambda = 0.7$, $\Omega_{\mathrm{M}} = 0.3$ and H$_0 =$ 70 $h_{70}$ km s$^{-1}$ Mpc$^{-1}$.

\section{Data}
\label{sec:Dat}
\subsection{Cluster sample} 
\label{sec:Clust}

To build our sample of galaxy clusters, we started with the compilation maintained by {one of us} (\citealt{And05}, see also \citealt{Cho14}), where we included clusters for which at least $N_z = 100$ spectroscopic-redshifts from the literature were available. These are nearby, optically selected Abell-ACO \citep{ACO89} galaxy clusters, with richness varying from poor to rich, and 
located within the redshift range 0.005 $< z <$ 0.150. 
In each cluster, a galaxy is identified as potential member when its apparent position puts it inside a projected Abell radius, R$_{\mathrm{A}}$ = 2.14 $h_{70}^{-1}$ Mpc, and its radial velocity has a value within $\pm\ 2\,500$ km s$^{-1}$ of a preliminary {estimate} for the central velocity of the system. 
However, its final acceptance as member will depend on a more thorough analysis, which is explained in \S~\ref{sec:Met}.

Our cluster sample is presented in Table~\ref{T_Sample}, together with some relevant data taken from the literature. 
The first seven columns reproduce the original ACO data for the clusters: the cluster ID (col.~1), its equatorial coordinates, RA and Dec, in J2000 (col.~2 and col.~3), its richness ($\cal{R}$ in col.~4), distance class ($\cal{D}$ in col.~5), BM-type \citep[BM in col.~6;][]{BM_70} (we have converted the original scale I, II, III to 1, 3, 5 and intermediate types 2 and 4), and Rood-Sastry type when available \citep[RS in col.~7;][]{RS_71, SR_82}. The X-ray characteristics for each cluster are given in columns~8 to 14:  alternative X-ray name when existent (col.~8), equatorial coordinates of the X-ray emission peak (centroid position; J2000 RA and Dec in col.~9 and col.~10), the X-ray luminosity inside $r_{500}$ (col.~11; this is the radius at which the mean interior overdensity is 500 times the critical density, $\rho_c$, at the respective redshift), $r_{500}$ itself \citep[col.~12; mostly from][]{Pif11}, and the X-ray temperature as measured by \citet{Mig20} in col.~13, or by others as indicated in col.~14. Note that there are new temperatures for four clusters, based on XMM-Newton and Chandra, presented for the first time in this table (see Appendix \ref{sec:ap-A}). Finally, we list the membership of a cluster in a supercluster (col.~15), based on the \emph{Master SuperCluster Catalog} \citep[MSCC;][]{Cho14}, followed in col.~16 by an alternate or common name, when available, or the name of the pair when it is the case, and multiplicity, $m$, of the supercluster in col.~17, the multiplicity being the number of Abell clusters forming the supercluster.

This sample is well balanced in terms of BM types, covering all the possible different dynamical states: containing 17, 17, 9, 9 and 15 clusters, respectively with BM types 1 to 5. 
It also follows roughly the distribution of richness of ACO clusters, clearly favoring low richness systems, in accordance with the power-law mass distribution function for clusters: 20 are classified as $\cal{R}$ = 0 (poorest; 30-49 galaxies), 24 as $\cal{R}$ = 1 (50-79), 20 as $\cal{R}$ = 2 (80-129) but only 2 as $\cal{R}$ = 3 (130-199) and 1 as $\cal{R} \ge$ 4 (richest; {more than 200 galaxies}). 
However, due to the spectroscopic selection criterion, the distribution of Abell distance classes ($\cal{D}$ varying from 0 to 7) is not equally represented, favoring nearby clusters.

\clearpage
\onecolumn
\begin{center}
\begin{tiny}
 \begin{changemargin}{-1cm}{-1cm}
  \begin{minipage}{165mm} 
 \setlength{\tabcolsep}{1.5mm}
   \begin{longtable}{lrrcccc c lrrrrrr c rlr} 
 \caption{Cluster sample.}
 \label{T_Sample} \\
\toprule
\multicolumn{7}{c}{ACO data} & & 
\multicolumn{7}{c}{X-ray data} & & 
\multicolumn{3}{c}{LSS data} \\
\cline{1-7} \cline{9-15} \cline{17-19}
  \multicolumn{1}{c}{ACO} &
  \multicolumn{1}{c}{RA$_{\mathrm{ACO}}$} &
  \multicolumn{1}{c}{Dec$_{\mathrm{ACO}}$} &
  \multicolumn{1}{c}{$\cal{R}$} &
  \multicolumn{1}{c}{$\cal{D}$} &
  \multicolumn{1}{c}{BM\footnote{\textbf{BM} types I, I-II, II, II-III and III, coded as 1, 2, 3, 4 and 5.}} &
  \multicolumn{1}{c}{RS} & &
  \multicolumn{1}{c}{Alt\_Name} &
  \multicolumn{1}{c}{RA$_{\mathrm{X}}$} &
  \multicolumn{1}{c}{Dec$_{\mathrm{X}}$} &
  \multicolumn{1}{c}{L$_{500}$} &
  \multicolumn{1}{c}{$r_{500}$} &
  \multicolumn{1}{c}{$k$T$_{\mathrm{X}}$} &
  \multicolumn{1}{c}{Ref.\footnote{[1] \citet{Ste84}, 
 [2] \citet{Oba98}, 
 [3] \citet{Fin01}, 
 [4] \citet{Ike02}, 
 [5] \citet{Cru02}, 
 [6] \citet{Fuk04}, 
 [7] \citet{Vik09}, 
 [8] \citet{Sat10}, 
 [9] \citet{Pla11}, 
[10] \citet{Mig20}, 
[11] This work}} & & 
  \multicolumn{1}{c}{MSCC\footnote{[iso] Isolated, [out] not in MSCC ($z >$ 0.15); S-clusters not in MSCC, but with [$^{*}$] percolated in SSCC \citep[\emph{Southern SuperCluster Catalog;}][]{Cho14} with the respective MSCC supercluster, and clusters with [$^{\#}$] percolated only in SSCC (SSCC number used here).}} &
  \multicolumn{1}{c}{SC\_{Name}} &
  \multicolumn{1}{c}{$m$} \\
 \multicolumn{1}{c}{} & 
 \multicolumn{1}{c}{[deg]$_{\mathrm{J2000}}$} & 
 \multicolumn{1}{c}{[deg]$_{\mathrm{J2000}}$} & 
 \multicolumn{1}{c}{} & 
 \multicolumn{1}{c}{} &  
 \multicolumn{1}{c}{} & 
 \multicolumn{1}{c}{} & & 
 \multicolumn{1}{c}{} & 
 \multicolumn{1}{c}{[deg]$_{\mathrm{J2000}}$} & 
 \multicolumn{1}{c}{[deg]$_{\mathrm{J2000}}$} &
 \multicolumn{1}{c}{[$10^{44}$ erg/s]} & 
 \multicolumn{1}{c}{[Mpc]} & 
 \multicolumn{1}{c}{[keV]} & 
 \multicolumn{1}{c}{} & &
 \multicolumn{1}{c}{} & 
 \multicolumn{1}{c}{} &
 \multicolumn{1}{c}{} \\
 \multicolumn{1}{c}{(1)} & 
 \multicolumn{1}{c}{(2)} & 
 \multicolumn{1}{c}{(3)} & 
 \multicolumn{1}{c}{(4)} & 
 \multicolumn{1}{c}{(5)} &  
 \multicolumn{1}{c}{(6)} & 
 \multicolumn{1}{c}{(7)} & & 
 \multicolumn{1}{c}{(8)} & 
 \multicolumn{1}{c}{(9)} & 
 \multicolumn{1}{c}{(10)} &
 \multicolumn{1}{c}{(11)} & 
 \multicolumn{1}{c}{(12)} & 
 \multicolumn{1}{c}{(13)} & 
 \multicolumn{1}{c}{(14)} & & 
 \multicolumn{1}{c}{(15)} &
 \multicolumn{1}{c}{(16)} & 
 \multicolumn{1}{c}{(17)} \\
\endfirsthead
\endhead
\endfoot
\endlastfoot 
\midrule
A2798 &   9.3916 & $-$28.5417 & 1 & 5 & 2 & -- & & J0037.4-2831 &   9.3625 & $-$28.5311 & 0.5455 & 0.7476 & 3.39 &  5 & &  33  & Scl(C) & 24\\
A2801 &   9.6404 & $-$29.0752 & 1 & 6 & 1 & -- & & \nodata      &   9.6346 & $-$29.0789 & \nodata& \nodata& 3.20 &  2 & &  33  & Scl(C) & 24\\
A2804 &   9.9149 & $-$28.9088 & 1 & 5 & 2 & -- & & \nodata      &   9.9113 & $-$28.8892 & \nodata& \nodata& 1.00 &  8 & &  33  & Scl(C) & 24\\
A0085 &  10.4075 &  $-$9.3425 & 1 & 4 & 1 & cD & & J0041.8-0918 &  10.4587 &  $-$9.3019 & 5.1001 & 1.2103 & 7.23 & 10 & &  39  & PisCet-N & 11\\
A2811 &  10.5386 & $-$28.5426 & 1 & 5 & 2 & -- & & J0042.1-2832 &  10.5363 & $-$28.5358 & 2.7341 & 1.0355 & 5.89 & 10 & &  33  & Scl(C) & 24\\
A0118 &  13.9329 & $-$26.4127 & 1 & 5 & 5 & I  & & \nodata      &  \nodata &    \nodata &\nodata &\nodata &\nodata&\nodata& &  33  & Scl(NE) & 24\\
A0119 &  14.0890 &  $-$1.2629 & 1 & 3 & 4 & C  & & J0056.3-0112 &  14.0762 &  $-$1.2167 & 1.4372 & 0.9413 & 5.82 & 10 & &  45  & - & 4\\
A0122 &  14.3571 & $-$26.2799 & 1 & 5 & 2 & B  & & J0057.4-2616 &  14.3529 & $-$26.2806 & 0.8612 & 0.8165 & 3.70 & 11 & &  33  & Scl(NE) & 24\\
A0133 &  15.6610 & $-$21.7982 & 0 & 4 & 1 & cD & & J0102.7-2152 &  15.6754 & $-$21.8736 & 1.4602 & 0.9379 & 4.25 & 10 & &  27  & PisCet-C & 9\\
A2870 &  16.9299 & $-$46.9165 & 0 & 3 & 1 & -- & & \nodata      &  \nodata &    \nodata &\nodata &\nodata & 1.07 & 11 & &  41  & Phe            & 8 \\
A2877 &  17.4554 & $-$45.9006 & 0 & 2 & 1 & C  & & J0110.0-4555 &  17.5017 & $-$45.9228 & 0.1815 & 0.6249 & 3.28 & 10 & &  41  & Phe & 8\\
A3027 &  37.6300 & $-$33.0953 & 0 & 4 & 5 & -- & & J0230.7-3305 &  37.6812 & $-$33.0986 & 0.4186 & 0.7200 & 3.12 &  5 & & iso  & - & 1\\
A0400 &  44.4107 &     6.0333 & 1 & 1 & 4 & I  & & J0257.6+0600 &  44.4121 &     6.0061 & 0.2211 & 0.6505 & 2.25 & 10 & & iso  & Southern GW & 1\\
A0399 &  44.4851 &    13.0164 & 1 & 3 & 2 & cD & & J0257.8+1302 &  44.4575 &    13.0492 & 3.5929 & 1.1169 & 6.69 & 10 & & 108  & +A0401 & 2\\
A0401 &  44.7373 &    13.5823 & 2 & 3 & 1 & cD & & J0258.9+1334 &  44.7396 &    13.5794 & 6.0886 & 1.2421 & 7.06 & 10 & & 108  & +A0399 & 2\\
A3094 &  47.8608 & $-$26.9289 & 2 & 4 & 2 & -- & & J0311.4-2653 &  47.8542 & $-$26.8997 & 0.3343 & 0.6907 & 3.15 & 11 & & 114  & - & 3\\
A3095 &  48.1094 & $-$27.1464 & 0 & 4 & 2 & -- & & \nodata      &  \nodata &    \nodata &\nodata &\nodata &\nodata&\nodata& & 114  & - & 3\\
A3104 &  48.5788 & $-$45.4150 & 0 & 4 & 1 & -- & & J0314.3-4525 &  48.5825 & $-$45.4242 & 1.0275 & 0.8662 & 3.56 & 10 & & 115  & HorRet-B & 9\\
S0334 &  49.0794 & $-$45.1168 & 0 & 4 & 3 & -- & & \nodata      &  \nodata &    \nodata &\nodata &\nodata &\nodata&\nodata& & 115* & HorRet-B & 9\\
S0336 &  49.3815 & $-$44.7012 & 0 & 4 & 3 & -- & & \nodata      &  \nodata &    \nodata &\nodata &\nodata &\nodata&\nodata& & 115* & HorRet-B & 9\\
A3112 &  49.4845 & $-$44.2349 & 2 & 4 & 1 & cD & & J0317.9-4414 &  49.4937 & $-$44.2389 & 3.8159 & 1.1288 & 5.49 & 10 & & 115  & HorRet-B & 9\\
A0426 &  49.6517 &    41.5151 & 2 & 0 & 4 & L  & & J0319.7+4130 &  49.9467 &    41.5131 & 6.2174 & 1.2856 & 6.42 &  4 & &  96  & PerPis & 3\\
S0373 &  54.6289 & $-$35.4545 & 0 & 0 & 1 & C  & & J0338.4-3526 &  54.6163 & $-$35.4483 & 0.0197 & 0.4017 & 1.56 &  4 & & iso  & Southern SC & 1\\
A3158 &  55.7526 & $-$53.6426 & 2 & 4 & 2 & -- & & J0342.8-5338 &  55.7246 & $-$53.6353 & 2.7649 & 1.0667 & 5.42 & 10 & & 117  & HorRet-A & 26\\
A0496 &  68.4045 & $-$13.2462 & 1 & 3 & 1 & cD & & J0433.6-1315 &  68.4100 & $-$13.2592 & 1.8530 & 0.9974 & 4.64 & 10 & & iso  & - & 1\\
A0539 &  79.1463 &     6.4540 & 1 & 2 & 5 & F  & & J0516.6+0626 &  79.1554 &     6.4378 & 0.5377 & 0.7773 & 3.04 &  4 & & iso  & - & 1\\
A3391 &  96.5644 & $-$53.6812 & 0 & 4 & 1 & -- & & J0626.3-5341 &  96.5950 & $-$53.6956 & 1.1601 & 0.8978 & 5.89 & 10 & & 160  & - & 3\\
A3395 &  96.8796 & $-$54.3994 & 1 & 4 & 2 & F  & & J0627.2-5428 &  96.9000 & $-$54.4463 & 1.3755 & 0.9298 & 5.10 &  7 & & 160  & - & 3\\
A0576 & 110.3506 &    55.7389 & 1 & 2 & 5 & I  & & J0721.3+5547 & 110.3425 &    55.7864 & 0.7571 & 0.8291 & 4.27 & 10 & & iso  & - & 1\\
A0634 & 123.6404 &    58.0479 & 0 & 3 & 5 & F  & & \nodata      &  \nodata &    \nodata &\nodata &\nodata &\nodata&\nodata& & iso  & - & 1\\
A0754 & 137.2086 &  $-$9.6366 & 2 & 3 & 2 & cD & & J0909.1-0939 & 137.1978 &  $-$9.6412 & 3.8497 & 1.1439 & 8.93 &  9 & & 198  & +A0780 & 2\\
A1060 & 159.2137 & $-$27.5265 & 1 & 0 & 5 & C  & & J1036.6-2731 & 159.1742 & $-$27.5244 & 0.3114 & 0.7015 & 2.79 & 10 & & 365  & HyaCen & 10\\
A1367 & 176.1231 &    19.8390 & 2 & 1 & 4 & F  & & J1144.6+1945 & 176.1521 &    19.7589 & 1.1046 & 0.9032 & 3.81 & 10 & & 295  & ComLeo & 5\\
A3526 & 192.2157 & $-$41.3058 & 0 & 0 & 2 & F  & & J1248.7-4118 & 192.1996 & $-$41.3078 & 0.6937 & 0.8260 & 3.40 & 10 & & 365  & HyaCen & 9\\
A3530 & 193.9037 & $-$30.3540 & 0 & 4 & 2 & -- & & J1255.5-3019 & 193.8937 & $-$30.3306 & 0.6805 & 0.8043 & 3.62 & 10 & & 389  & Shapley(W) & 24\\
A1644 & 194.3115 & $-$17.3535 & 1 & 4 & 3 & cD & & J1257.1-1724 & 194.2904 & $-$17.4003 & 1.8975 & 0.9944 & 5.25 & 10 & & 370  & +A1631 & 2\\
A3532 & 194.3299 & $-$30.3702 & 0 & 4 & 4 & C  & & J1257.2-3022 & 194.3204 & $-$30.3769 & 1.3233 & 0.9201 & 4.63 & 10 & & 389  & Shapley(W) & 24\\
A1650 & 194.6926 &  $-$1.7530 & 2 & 5 & 2 & cD & & J1258.6-0145 & 194.6712 &  $-$1.7569 & 3.4706 & 1.1015 & 5.72 & 10 & & 376  & SGW & 6\\
A1651 & 194.8456 &  $-$4.1862 & 1 & 4 & 2 & cD & & J1259.3-0411 & 194.8396 &  $-$4.1947 & 3.8536 & 1.1252 & 7.47 & 10 & & 376  & SGW & 6\\
A1656 & 194.9530 &    27.9807 & 2 & 1 & 3 & B  & & J1259.7+2756 & 194.9296 &    27.9386 & 3.4556 & 1.1378 & 7.41 & 10 & & 295  & ComLeo & 5\\
A3556 & 201.0260 & $-$31.6605 & 0 & 4 & 1 & -- & & \nodata      & 200.9350 & $-$31.8380 &\nodata &\nodata & 3.08 &  6 & & 389  & Shapley(C) & 24\\
A1736 & 201.7173 & $-$27.1093 & 0 & 2 & 5 & I  & & J1326.9-2710 & 201.7250 & $-$27.1833 & 1.6675 & 0.9694 & 3.34 & 10 & & 389  & Shapley(N) & 24\\
A3558 & 201.9782 & $-$31.4922 & 4 & 3 & 1 & -- & & J1327.9-3130 & 201.9896 & $-$31.5025 & 3.1385 & 1.1010 & 5.83 & 10 & & 389  & Shapley(C) & 24\\
A3562 & 203.3825 & $-$31.6729 & 2 & 3 & 1 & -- & & J1333.6-3139 & 203.4012 & $-$31.6611 & 1.3458 & 0.9265 & 5.10 & 10 & & 389  & Shapley(C) & 24\\
A1795 & 207.2522 &    26.5852 & 2 & 4 & 1 & cD & & J1348.8+2635 & 207.2208 &    26.5956 & 5.4781 & 1.2236 & 6.42 & 10 & & 414  & Boo & 24\\
A2029 & 227.7447 &     5.7617 & 2 & 4 & 1 & cD & & J1510.9+0543 & 227.7292 &     5.7200 & 8.7267 & 1.3344 & 8.45 & 10 & & 457  & - & 6\\
A2040 & 228.1884 &     7.4300 & 1 & 4 & 5 & C  & & \nodata      & 228.2113 &     7.4317 &\nodata &\nodata & 2.41 &  1 & & 454  & - & 6\\
A2052 & 229.1896 &     7.0003 & 0 & 3 & 2 & cD & & J1516.7+0701 & 229.1833 &     7.0186 & 1.4421 & 0.9465 & 2.88 & 10 & & 458  & Her-S & 4\\
A2065 & 230.6776 &    27.7226 & 2 & 3 & 5 & C  & & J1522.4+2742 & 230.6104 &    27.7094 & 2.6279 & 1.0480 & 6.59 & 10 & & 463  & CrB & 14\\
A2063 & 230.7578 &     8.6394 & 1 & 3 & 3 & cD & & J1523.0+0836 & 230.7725 &     8.6025 & 1.1388 & 0.9020 & 3.34 & 10 & & 458  & Her-S & 4\\
A2142 & 239.5672 &    27.2246 & 2 & 4 & 3 & B  & & J1558.3+2713 & 239.5858 &  27.2269 & 10.6761 & 1.3803 & 11.63 & 10 & & 472  & +A2148 & 2\\
A2147 & 240.5716 &    15.8954 & 1 & 1 & 5 & F  & & J1602.3+1601 & 240.5779 &    16.0200 & 1.3584 & 0.9351 & 4.26 & 10 & & 474  & Her-C & 5\\
A2151 & 241.3125 &    17.7485 & 2 & 1 & 5 & F  & & J1604.5+1743 & 241.2863 &    17.7300 & 0.5088 & 0.7652 & 2.10 & 10 & & 474  & Her-C & 5\\
A2152 & 241.3435 &    16.4486 & 1 & 1 & 5 & F  & & J1605.5+1626 & 241.3842 &    16.4419 & 0.1283 & 0.5783 & 2.41 &  6 & & 474  & Her-C & 1\\
A2197 & 247.0436 &    40.9072 & 1 & 1 & 5 & L  & & J1627.6+4055 & 246.9175 &    40.9197 & 0.0674 & 0.5093 & 2.21 &  3 & & 485  & Her-N & 4\\
A2199 & 247.1540 &    39.5243 & 2 & 1 & 1 & cD & & J1628.6+3932 & 247.1583 &    39.5486 & 1.9007 & 1.0040 & 4.04 & 10 & & 485  & Her-N & 4\\
A2204 & 248.1903 &     5.5785 & 3 & 5 & 3 & C  & & J1632.7+0534 & 248.1937 &   5.5706 & 13.6256 & 1.3998 & 10.24 & 10 & & out  & - & -\\
A2244 & 255.6834 &    34.0468 & 2 & 5 & 2 & cD & & J1702.7+3403 & 255.6787 &    34.0619 & 4.0452 & 1.1295 & 5.99 & 10 & & 492  & - & 3\\
A2256 & 255.9313 &    78.7174 & 2 & 3 & 4 & B  & & J1703.8+7838 & 255.9533 &    78.6444 & 3.5435 & 1.1224 & 8.23 & 10 & & 495  & - & 3\\
A2255 & 258.1293 &    64.0926 & 2 & 3 & 4 & C  & & J1712.7+6403 & 258.1967 &    64.0614 & 2.9491 & 1.0678 & 7.01 & 10 & & iso  & NEP SC & 1\\
A3716 & 312.8866 & $-$52.7121 & 1 & 3 & 4 & F  & & \nodata      & 312.9873 & $-$52.6301 &\nodata &\nodata & 2.19 & 11 & & 309\# & - & 3\\
S0906 & 313.1034 & $-$51.9613 & 0 & 4 & 3 & -- & & \nodata      &  \nodata &    \nodata &\nodata &\nodata &\nodata&\nodata& & 309\# & - & 3\\
A4012 & 352.9398 & $-$33.8239 & 0 & 6 & 4 & -- & & \nodata      &  \nodata &    \nodata &\nodata &\nodata &\nodata&\nodata& & 584  & - & 3\\
A2634 & 354.5766 &    27.0270 & 1 & 1 & 3 & cD & & J2338.4+2700 & 354.6071 &    27.0125 & 0.4414 & 0.7458 & 3.71 & 10 & & 592  & +A2666 & 2\\
A4038 & 356.9246 & $-$28.1387 & 2 & 2 & 5 & B  & & J2347.7-2808 & 356.9300 & $-$28.1414 & 1.0295 & 0.8863 & 2.84 & 10 & & 595  & +A4049 & 2\\
A4049 & 357.8971 & $-$28.3718 & 0 & 3 & 5 & -- & & \nodata      &  \nodata &    \nodata &\nodata &\nodata &\nodata&\nodata& & 595  & +A4038 & 2 \\
A2670 & 358.5571 & $-$10.4190 & 3 & 4 & 2 & cD & & J2354.2-1024 & 358.5560 & $-$10.4130 & 1.3365 & 0.9113 & 4.45 & 10 & & iso  & - & 1\\
\bottomrule
 \end{longtable}
  \end{minipage}
 \end{changemargin}
\end{tiny}
\end{center}
\clearpage
\twocolumn

\noindent Although we cannot claim completeness, this sample can be considered a fair representation of optically selected Abell clusters at low redshifts.

Most of these clusters (59 or 88\%) are detected in X-rays. Fifty-three are included in the compilation of X-ray clusters by \citet{Pif11}.
The other six X-ray clusters in our sample, namely A0118, A2040, A2801, A2804, A3556 and A3716, were detected by previous surveys \citep[respectively by][the last reference applying to the last two clusters]{Kow84,Ste84,Oba98,Sat10,Ebe96}.
A3716 was also identified as a SZ source by the Planck satellite \citep{Pla16}, with which catalog we have 43 clusters (64\%) in common. 
The range in temperature, $k$T$_{\mathrm{X}}$, is also quite large, varying from 1 to 12 keV, which is typical of low-mass to relatively massive clusters. 
Only 8 clusters in our sample, namely A0634, A2870, A3095, A4012, A4049, S0334, S0336 and S0906, have not yet been detected {in X-rays}. 
These might be considered as ``Abell X-ray Underluminous'' Cluster candidates \citep[AXUs, for short, \textit{e.g.},][]{Tre14}.


\subsection{Spectroscopic data for member galaxies} 
\label{sec:Spect}

From the information gathered in the compilation {described} by \citet[][]{And05}, we retrieved, for each galaxy, the celestial coordinates and line-of-sight (LOS) heliocentric radial velocity with their uncertainties. 
Due to the diversity of the sources, these data are not homogeneous. 
To assess this problem, we treated the different quality of redshift data by adopting distinct approaches: 
1) eliminating data with large ($\ge$ 400 km s$^{-1}$) estimated uncertainties, 
2) eliminating obvious outliers as described further below, 
3) using the average of the radial velocities for every single galaxy, 
4) taking advantage of the statistics to minimize the stochastic errors (for example, calculating mean velocities and velocity dispersion for the clusters).

For the celestial coordinates, we adopted the strategy of inspecting every close pair of entries {with separations larger than 3\arcsec} (they rarely exceed 30\arcsec), {tentatively associated to the same galaxy,} directly on a DSS2 image (Digitized Sky Survey, STScI) using the \textsc{Aladin} interface \citep[][]{aladin}.
Multiple redshift entries for the same galaxy are not uncommon
and, once the multiple velocities for the same galaxy are judged {consistent}, we proceed to averaging the different measurements, taking great care in excluding outlier values (above 3 sigma, when there are at least three independent velocity measurements).
After applying this process to each cluster, we obtained a list of potential member galaxies all with a single position, average LOS velocity and respective {uncertainties (typically $\pm\ 0.5$" and} $\pm\ 60$ km~s$^{-1}$, respectively, per galaxy).

\subsection{Astrometric and photometric data {of galaxies}}
\label{sec:Phot}

{Although the galaxy coordinates, calculated as in the last paragraph, are precise enough, they are not homogeneous, combining more accurate positions with poorer ones.
To calculate mean pairwise separations (see below), for example, we need to improve these positions.}
To this end, we cross-correlated the position of each galaxy in our lists with the positions in two astrometric and photometric catalogues covering the whole sky, SuperCOSMOS \citep{Ham01a} and Two Micron All Sky Survey \citep[2MASS;][]{Skr06}.\footnote{ The most recent CCD-based surveys, like the Sloan Digital Sky Survey \citep[SDSS;][]{Ahn12}, the Mayall z-band Legacy Survey (MzLS) + Dark Energy Camera Legacy Survey (DECaLS) \citep[][]{Dey19}, and the Dark Energy Survey \citep[DES;][]{Abb21}, were not used because they only cover small regions of the sky; at most about $1/3$, and $3/4$ in the case of the Panoramic Survey Telescope and Rapid Response System \citep[PanSTARRS;][]{Cha19}.} 
SuperCOSMOS is a relatively deep optical survey, 
reaching b$_{J} \sim$ 19.5 with acceptable levels for completeness, $>$ 95\%, and contamination, $<$ 5\% \citep{Ham01b}. 
Although it has good astrometry, with an uncertainty of order $\pm\ 0.25\arcsec$ \citep[][]{Ham01c}, the typical uncertainty on the magnitudes is relatively large, $\pm\ 0.3$ mag \citep[][]{Ham01b}, because it was obtained from the digitization of photographic sky survey plates. 
2MASX (the catalogue of 2MASS eXtended sources), on the other hand, is a digital NIR survey reaching K$_s \sim$ 13.5, with nominal levels of completeness and contamination respectively of $>$ 90\% and $<$ 2\%. 
Although the uncertainty in astrometry is about the same as in SuperCOSMOS (0.3\arcsec, for extended sources), the quality of the magnitudes is better, with typical uncertainty  $\pm\ 0.03$ \citep[][]{Jar00a}. 
Also, since the NIR is less affected by dust extinction \citep[\textit{e.g.},][]{Fit99}, the \emph{K-correction} is minimal \citep[\textit{e.g.},][]{Fuk95,Chi10} compared to the optical. 
Particular care is devoted to the photometry of the brightest galaxies in 2MASS because the data come mostly from a special catalogue, the Large Galaxy Atlas \citep{Jar03}, which is dedicated to galaxies that are more extended than 1\arcmin.

The only drawback of 2MASS is the depth of the survey: at the magnitude limit K$_s \sim 13.5$ the mean redshift of galaxies is $z \sim$ 0.08 \citep{Jar04}. This implies that only a fraction (about 50\%) of the galaxies, the brightest in our sample, have an entry in 2MASX. 
Another disadvantage is the limited capacity of 2MASS to separate galaxies that are very close in projection, which is the case of faint ``dumbbell'' {CDGs/SDGs} in our sample (although, this does not happen for galaxies in the Large Galaxy Atlas). 
SuperCOSMOS performs better in ``deblending'' galaxies, but then fails in accuracy  determining their magnitudes once they are separated, their brightness being usually underestimated. 
Therefore, although we matched our data with the astrometric and photometric data in both catalogues, for the sake of homogeneity we used only 2MASS in the present paper.
Absolute magnitudes for the BCGs were calculated after correcting for Galactic extinction, using the re-calibration done by \citet{SeF11} of the dust maps of \citet{Sch98}, and applying a \emph{K-correction} as determined by \citet{Chi10}.  

\section{Methods to characterize the galaxy systems and their structures}
\label{sec:Met}
\subsection{{Central and Substructure Dominant} Galaxies and the definition of the cluster center} 
\label{sec:FRGs}
                                     
In any cluster, group or substructure of galaxies 
we classify as {SDG/CDG} the galaxy that, being among the BCGs, occupies the most central position around the gravitational potential well of its {(sub-)}system. 
From a practical observational point of view, a {CDG/SDG} must be coincident or very close to the {local} surface density peak in the sky distribution of the member galaxies.
This also implies that its position {is expected to} be located near the X-ray emission peak (c.f. \S~\ref{sec:Core}) {when this emission is detected}.

{In} each substructured cluster, we identify (based on criteria to be explained below) one substructure as the ``main'' (or gravitationally dominant) substructure and adopt its {SDG} as the {CDG} for the whole system. 
Also, as a general rule, we adopt the position of the {CDG/SDG} as the location of the dynamical center of the {cluster/substructure}.

In the literature, the physical characteristics of the BCG are usually used as a trade mark of its system. However, identifying which galaxy is the BCG obviously depends on which band-pass is used. 
For example, a very bright spiral galaxy, especially in a starburst phase, in the outskirts of a cluster, could easily be brighter in B or V than a giant and red elliptical near its center (\textit{e.g.}, NGC\,1365 in Fornax/S0373). 
This is why we define our BCGs to be more luminous in K$_s$, which is a better proxy for the stellar mass of galaxies, and thus consistent with the idea that a {CDG/SDG} should also be the most massive galaxy of its {cluster/substructure}. 

Adopting these definitions, in 81\% of our cluster sample we found the {CDG} to be coincident with the original BCG, according to prescriptions given by, for example, Abell \citep[or others like][]{BM_70,RS_71,SR_82}. 
However, in the remaining 19\% of the clusters the BCG is not the {CDG}, for {one of the} following three reasons: 
i) the BCG is really {a SDG}, like in the case of Fornax (NGC\,1316), forming 13\% of our cluster sample (we classified them as ``Fornax-like'' clusters); 
ii) the BCG is located close to the center of the cluster (as confirmed by X-ray emission or based on other dynamical analyses) but is only slightly brighter than the real {CDG}, as in Coma (BCG: NGC\,4889, {CDG}: NGC\,4874), forming 4\% of our sample (they are classified as ``Coma-like'' clusters);
iii) the BCG is a giant elliptical galaxy located at the periphery of the cluster, forming 2\% of our sample (we classify these galaxies as ``fossil group candidates'', {without, at this point, gathering more data to confirm this candidacy}).

\subsection{Cluster membership} 
\label{sec:Caust}                                         

When clusters are separated by less than 6 $h_{70}^{-1}$ Mpc on the sky, the areas subtended by their R$_{\mathrm{A}}$ overlap. 
{There are} 38 clusters in our sample for which this is the case. 
To separate the members (or populations) of these clusters, we first merged the lists of their potential member galaxies, eliminating the duplicated entries as assigned by different sources in the literature to the different systems. Then we proceed in separating the galaxies that are members of each cluster, by applying a method similar to the one described in \S~\ref{sec:Substr} for identifying substructures.

\begin{table*}[!t]
 \small                             
 \centering
 \tablecols{8}
 \setlength{\tabcolsep}{1mm}
 \setlength{\tabnotewidth}{1\columnwidth}
   \caption{Superposed clusters in our sample.}
   \label{T_Pairs}                                
 \begin{minipage}{180mm}                    
  \begin{tabular}{p{5cm} >{\centering}p{2.5cm}<{\centering} rrrcrr}
\toprule
\multicolumn{5}{c}{Projected information} & & 
\multicolumn{2}{c}{Gravitational binding} \\
\cline{1-5} \cline{7-8} 
Clusters & \multicolumn{1}{c}{Sep.$^{\dagger}$ [$h_{70}^{-1}$ Mpc]} & 
$N_z$ & $N_g$ & $N_c$ & &  
\multicolumn{1}{c}{Bound} & \multicolumn{1}{c}{Unbound} \\                             
\midrule                                                                         
A0118--A0122                     &      2.95       &   190 & 119 & 111 & & A0118-A0122  &                 \\
A0399--A0401                     &      3.03       &   245 & 217 & 184 & & A0399-A0401  &                 \\
A1736A--A1736B$^{\#}$            &      0.83       &   464 & 219 & 215 & &              & A1736A, A1736B  \\
A2052--A2063A                    &      5.53       &   959 & 378 & 369 & & A2052-A2063A &                 \\
A2147--A2151--A2152A             & 4.94,3.69       &  2096 & 936 & 880 & & A2147-A2151  & A2152A          \\
A2197--A2199                     &      3.05       &  1684 & 815 & 774 & & A2197-A2199  &                 \\
A2798B--A2801--A2804--A2811B     & 4.36,2.40,5.32  &   424 & 381 & 342 & & A2798B-A2801-A2804 &  A2811B   \\
A2870--A2877                     &      1.92       &   428 & 237 & 174 & & A2877-A2870* &                 \\
A3094A-A3095                     &      1.52       &   253 & 170 & 154 & &              & A3094A, A3095$^{\natural}$ \\
A3104--S0334--S0336--A3112B      & 2.96,2.23,3.85  &   563 & 268 & 221 & & S0336$^{\natural}$-A3112B-S0334$^{\natural}$-A3104 &  \\
A3391--A3395                     &      3.03       &   761 & 343 & 318 & &              & A3391, A3395    \\
A3526A--A3526B$^{\#}$            &      0.29       &  1041 & 336 & 330 & &              & A3526A, A3526B  \\
A3530--A3532                     &      1.69       &   411 & 238 & 213 & & A3530-A3532  &                 \\
A3556--A3558--A3562              & 3.32,4.83       &  2057 & 863 & 800 & & A3556-A3558-A3562$^{\S}$ &            \\
A3716--S0906                     &      1.67       &   409 & 219 & 194 & &              & A3716, S0906$^{\natural}$ \\
A4038A--A4049                    &      2.09       &   816 & 247 & 237 & & A4038A-A4049*   &                 \\
\bottomrule
   \end{tabular}
 \end{minipage}
\noindent [$^{\dagger}$]:   Projected separation between nearest clumps. \\
\noindent [$^{\#}$]:        Clusters slightly separated in projection and separable in redshift. \\
\noindent [$*$]:            Clusters considered to be substructures. \\
\noindent [$^{\natural}$]:  Clusters that are possible satellite groups. \\
\noindent [$^{\S}$]:        A fourth non-Abell cluster was clearly identified within this system (AM~1328$-$313).
\end{table*}

To double-check our results, we also applied to the superposed clusters a two-body Newtonian criterion for gravitationally bound systems \citep[see][and \S~\ref{sec:Bind}]{Bee82}. 
The results (bound \textit{vs.} unbound) obtained by this process are in good agreement with previous results in the literature \citep[][]{GeT84,KK02,PeB13,Yua05}. 
It is worth noting that 4 of the 11 bound complexes in Table~\ref{T_Pairs} are consistent with supercluster ``cores'' (Z\'u\~niga \textit{et al.}, \emph{in preparation}); those are: Her-C (MSCC\,474), Scl (MSCC\,033), HorRet-B (MSCC\,115) and Shapley (MSCC\,389). 
Among the other bound systems, three are typical pairs,  A0399-A0401, A2052-A2063A and A3530-A3532, and four, A0122-A0118, A2199-A2197, A2877-A2870 and A4038A-A4049, are examples of a massive cluster (main cluster, first of the pair) linked by a filament made of various groups (secondary system; c.f. \S~\ref{sec:Out}). 

\begin{figure}
  \centering
  \includegraphics[width=0.5\textwidth]{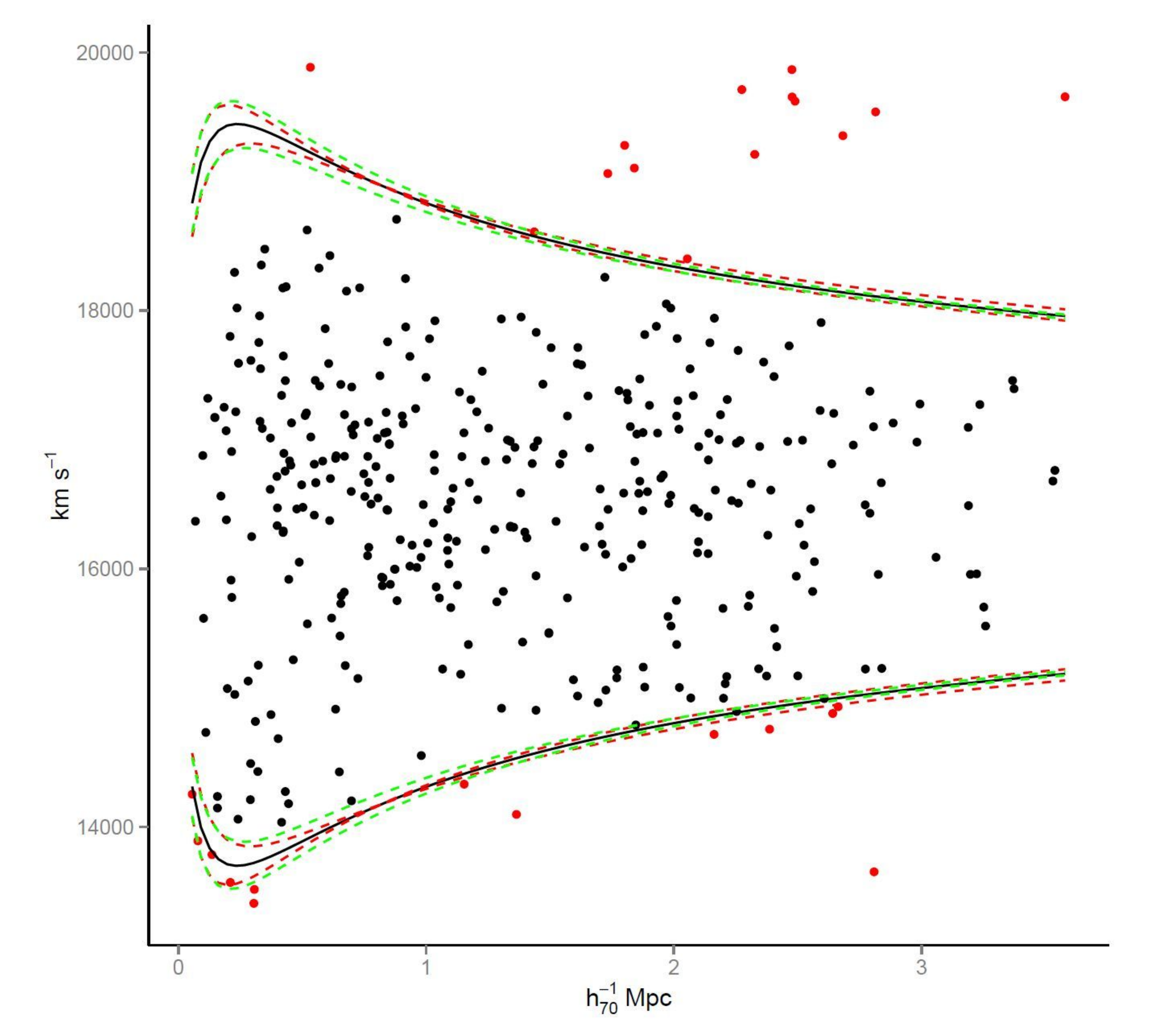}
  \caption{Example of applying the caustics method to the cluster A0085; the caustics are the black solid lines, while the red dashed lines indicate the rms of the fit and the green dashed lines the bootstrap uncertainty (1\,000 simulations). {Black points (that is, the ones within the caustics) are taken to be the bound members, while red points are discarded as cluster members.}}
  \label{F_Caus}
\end{figure}

After homogenizing the galaxy coordinates by applying the match with photometric data, and after defining the different centers and correcting for superposed clusters, we proceeded {to check}  
the membership of the galaxies in their respective clusters using a more robust approach. 
The principle is simple: considering that, in a gravitationally-bound system, the galaxies must have velocities that do not exceed the escape velocity, their distribution in a projected phase-space (PPS) diagram, formed by the LOS velocity as a function of  projected cluster-centric distance, \textit{e.g.}, Fig.~\ref{F_Caus}, must be enclosed within a trumpet-shaped curve, usually called ``caustic'', as defined by the escape velocity \citep[\textit{e.g.},][]{ReG89,Lop22}. Details about the method for defining and fitting caustics are presented in \citet{Cho19}.
In Fig.~\ref{F_Caus}, all the galaxies falling outside the caustic for the cluster A0085 are {discarded} leaving only those that are considered as gravitationally bound. 

Applying the caustics analysis, we found that on average 10\% of the candidate member galaxies in each cluster, {at least in relatively isolated systems}, must be discarded. 
{For the overlapping clusters above, these galaxies are usually bound to another cluster of the complex. Obviously, no single
galaxy is assigned to more than one cluster}.

\subsection{Determination of cluster dynamical parameters} 
\label{sec:Param}

To establish the dynamical properties of each system, we first measure two robust kinematical parameters, known as the biweight central value, $C_{\mathrm{BI}}$, and scale, $S_{\mathrm{BI}}$ \citep{Bee90}. 
Using these parameters, we calculate preliminary values for the systemic radial velocities, $v_{\rm c}$, and velocity dispersions, $\sigma_{\rm c}$, considering the $N_{c}$ 
members. 
Then we proceed by defining, for each cluster, a projected aperture on the sky consistent with the virial radius. 

This implies first estimating $r_{200}$, the radius inside which the mean density of galaxies exceeds $200 \times \rho_c$ at the redshift of the cluster. Following the prescription by \citet{Carlb97}:

\begin{equation}
r_{200} = \frac{\sqrt{3} \; \sigma_{\mathrm{c}}}{10 \, H(z)}
\end{equation}

\noindent where $H(z) = H_0 \sqrt{\Omega_r (1+z)^4 + \Omega_m (1+z)^3 + \Omega_k (1+z)^2 + \Omega_{\Lambda}}$, assuming $\Omega_r$ (radiation) and $\Omega_k$ (curvature) are 
about zero. 
Since the virial radius depends on the redshift and cosmology \citep[\textit{e.g.},][]{BeN98}, a local value of $1.3 \times r_{200}$, corresponding to about $r_{100}$, is usually adopted \citep[\textit{e.g.},][]{KeK18}. Once the circular aperture corresponding to the virial radius is determined, we counted the number of galaxies inside it, $N_a$, and recalculated the systemic radial velocity, $v_{\mathrm{cl}}$, and its velocity dispersion, $\sigma_{\mathrm{cl}}$, using once again $C_{\mathrm{BI}}$ and $S_{\mathrm{BI}}$.  

These $N_a$ galaxies are also used to calculate the projected radius, $R_{\mathrm{p}}$, equal to twice the harmonic mean projected separation, using the relation: 

\begin{equation}
R_{\mathrm{p}} = {N_a(N_a-1)} \left( \sum_{i<j}^{N_a} \frac{1}{|r_{ij}|} \right)^{-1}
\end{equation}

\noindent where $r_{ij}$ is the pairwise projected separation between galaxies. Estimation of the virial mass, $M_{\mathrm{vir}}$, 
follows the relation:

\begin{equation}
M_{\mathrm{vir}} = \frac{\alpha \pi}{2 G} \; \sigma_{\mathrm{cl}}^2 \, R_{\mathrm{p}}
\end{equation}

\noindent where the factor $\alpha$ quantifies the isotropy level of the system ($\alpha= 3$ when complete isotropy is assumed), applied to $\sigma_{\mathrm{cl}}$, while the factor $\pi/2$ is applied to deproject  $R_{\mathrm{p}}$ \citep{LeM60}. Finally, the  virial radius is obtained using the relation:

\begin{equation}
R_{\mathrm{vir}}^3 = \frac{3}{4 \pi} \frac{M_{\mathrm{vir}}}{\rho_{\mathrm{vir}}} = 
\frac{\alpha \; \sigma_{\mathrm{cl}}^2 \; R_{\mathrm{p}}} {18 \pi \, H^2(z)}
\end{equation}


\noindent where the virial density is defined as $\rho_{\mathrm{vir}} = 18 \pi^2 (3 H^2(z) / 8 \pi G)$.

\subsection{Substructure analysis} 
\label{sec:Substr}

Several tests, in different spatial dimensions, have been proposed for the detection of substructures in galaxy clusters based on optical data: in 1D, as applied to the redshifts \citep[\textit{e.g.},][]{BeB93, Hou09}, in 2D, as applied to galaxy projected celestial coordinates \citep[\textit{e.g.},][]{GeB82, FeW87, Wes88, KeB97, FeK06}, and in 3D, as applied to both redshifts and coordinates \citep[\textit{e.g.},][]{DeS88, SeG96, Pis96, Ein10, Yu15}. 
However, not all these tests have the same efficiency. 
Through a study of 31 different tests, \citet{Pin96} found the 3D test developed by \citet[][DS herafter]{DeS88} to be the most sensitive, concluding that, in general, the higher the dimension of a test, the more powerful it is in distinguishing substructures.
This was confirmed subsequently by \citet{Ein12}, who also showed that 3D tests are more robust than 2D tests, and 2D tests more robust than 1D tests. However, both groups recommended the application of more than one test.

Tests for detecting substructures using X-ray data have also been proposed \citep[\textit{e.g.},][]{Moh93, BeT95, And12}. 
However, because the tests are based on X-ray surface brightness, the detection of substructures is usually limited to the densest, most concentrated ($R < r_{500}$) regions of the clusters \citep[\textit{e.g.},][]{Pif08}, which might complicate comparisons with substructures detected in the optical. 
In the study made by \citet{Lop18}, for example, only $\sim$ 60\% of the substructures detected in optical were also detected in X-ray, both inside $r_{500}$. 
They also found the fraction of substructures to increase with the aperture used, as well as with the mass of the cluster and its redshift \citep[up to $z \sim$ 1; see also][]{Jel05}. 
Although these trends are consistent with various levels of relaxation (for example, clusters being less relaxed in the past than now), establishing a firm connection as well as a time scale to reconstruct the assembly histories of clusters is not straightforward and needs independent confirmation. 

In principle, this is what a study of the {CDG} dynamical characteristics can contribute.
More specifically, the projected position offset of a {CDG} relative to the X-ray peak and its peculiar velocity relative to the centroid of the distribution of galaxy members are two parameters expected to be correlated with the level of relaxation of the clusters \citep[\textit{e.g.},][]{Zha11, Lav16, Lop18}: the smaller the offset and peculiar velocity, the higher the level of relaxation. This assumes migration through dynamical friction of the {CDG} towards the center of the cluster is less rapid than that of the hot gas. 
Comparing these two parameters with the level of substructuring in clusters --the higher the number of substructures the lower the level of relaxation-- should consequently complement our view about their assembly histories. 

{Thus}, to reconstruct the assembly histories of the clusters in our sample, we will develop our study of substructures applying different tests, with different dimensions, in the optical, 
comparing with X-ray substructuring information and radio data from the literature whenever available, and each time comparing the results (as an independent test) with the dynamical characteristics of the {CDGs} in their respective clusters. 

\paragraph*{Radial velocity distributions (1D test):}

We start by directly inspecting the LOS velocity distributions of galaxies within the clusters, comparing them with a Gaussian distribution. 
The physical motivation of this test is the following: as the system tends toward relaxation, the absolute values of \textsl{skewness} and excess \textsl{kurtosis} (with respect to the value of 3 for a Gaussian distribution) also tend to decrease. 
This is a straightforward test that is easy to quantify.

Adopting a level below 0.3 as an upper limit for relaxation, only 25\% of the clusters in our sample present LOS velocity histograms consistent with a Gaussian. 
This indicates that as much as 75\% of the clusters in our sample show a possible signal of being substructured, as disclosed, more specifically, by two or more noticeable peaks in the LOS velocity distribution or platykurtic \textsl{kurtosis} values. 

\paragraph*{Projected distribution of galaxies (2D-test):}

The second test consists in tracing the isodensity contour map of the projected galaxy distribution in each cluster, where any galaxy density peak may be considered as a significant substructure \citep[][]{GeB82} since only spectroscopically confirmed members were considered. 
The (probability) density maps were obtained using a bivariate adaptive kernel, fitted by the function:

\begin{equation}
\label{eq:kernel2}
G(x,y)=\frac{1}{2\pi\sigma_x\sigma_y\sqrt{1-\rho^2}}\exp{\left(-\frac{z}{2(1-\rho^2)}\right)}
\end{equation}

\noindent where $z$ {(do not confuse with redshift)} is equal to: 

\begin{equation}
\label{eq:A3}
z\equiv\frac{(x-\mu_x)^2}{\sigma_x^2}-\frac{2\rho(y-\mu_y)(y-\mu_y)}{\sigma_x\sigma_y}+\frac{(y-\mu_y)^2}{\sigma_y^2}
\end{equation}

\noindent and where, when the parameters $x,y$ are not strongly correlated, one can assume $\rho=0$.

\begin{figure}[!t]
  \centering
  \includegraphics[width=0.5\textwidth, trim=0cm 4cm 0cm 4cm]{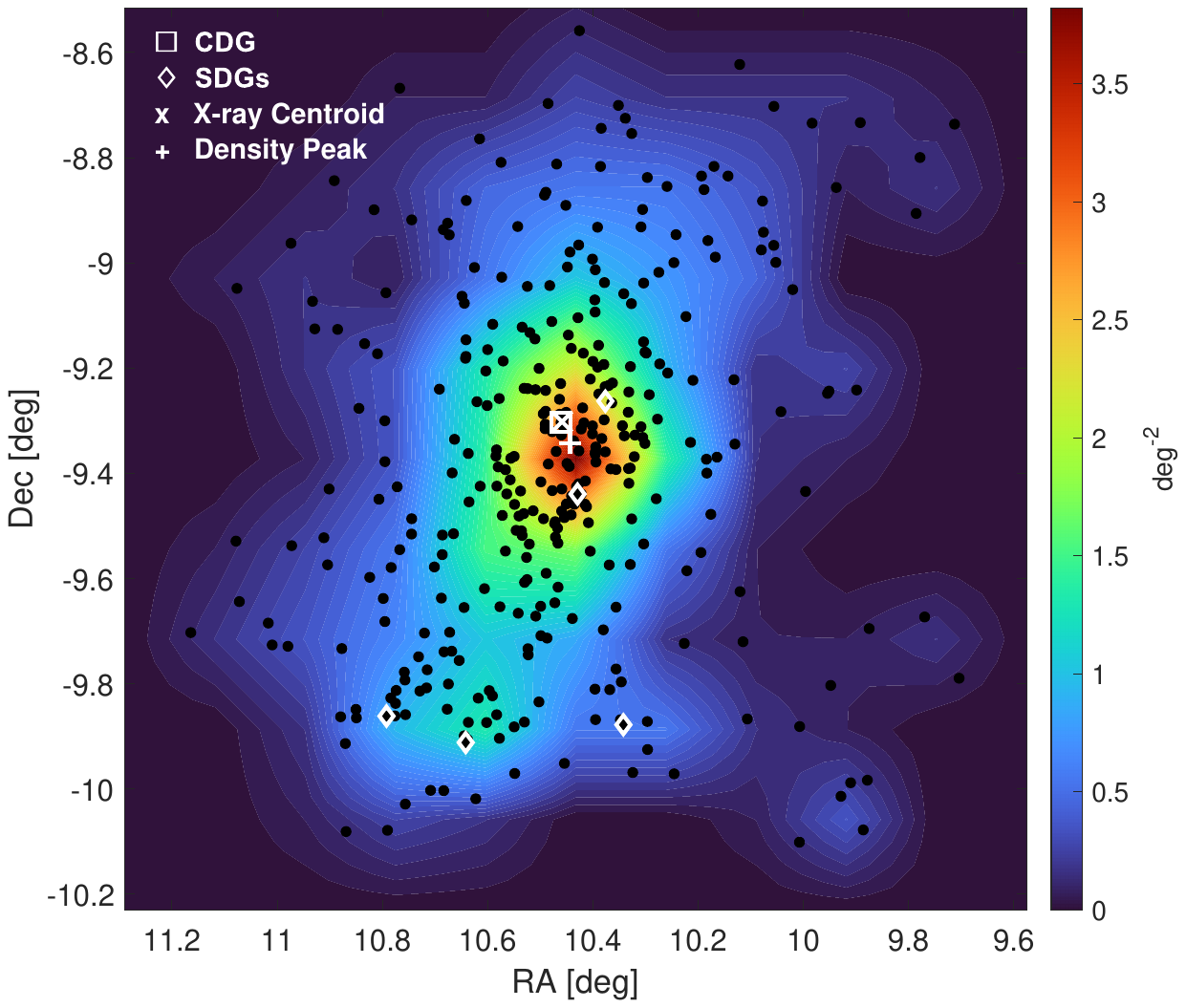}
  \caption{Example of an isodensity map for the 
  cluster A0085. The {isodensity levels} are coded by colors, in units of the  probability density to find spectroscopic member galaxies/deg$^{2}$ (mean probability density equals to 1). The position of the peak in density is indicated, as well as the position of the {CDG, SDGs} and X-ray centroid.}
  \label{F_Isod}
\end{figure}

In Fig.~\ref{F_Isod} the 
{surface density map of the member galaxy distribution} 
of the cluster A0085 is shown as example. In this figure, the isodensity contours are codified in colors, with a cross indicating the density peak. 
Also shown are the positions of the {CDG} (open square) and peak in X-ray ($\times$ symbol). 
Note that although the {CDG} is slightly off-centered from the distribution of the galaxies, its position is almost the same as the peak in X-ray. 
Despite the substructures, this looks like a relatively well evolved cluster.

As it is difficult to resume the information on substructure in a table we offer the isodentity maps of all of our clusters on an 
accompanying web site (\url{www.astro.ugto.mx/recursos/HP_SCls/Top70.html}).

\paragraph*{X-ray surface brightness maps (2D test):}

X-ray surface brightness maps are constructed and used as a supplementary information for identifying the substructures. Apart from applying an algorithm to detect independently substructures in these data, we checked every substructure detected in the optical for its counterpart in X-rays. 
This made possible, for example, to find cases of multimodal main structures that would not be identifiable from the optical data alone.

Using the \textsc{Aladin} interface, we traced the X-ray surface brightness maps for all our clusters, overlaid in red contours on top of the respective DSS2 R-band\footnote{\ \url{https://archive.stsci.edu/cgi-bin/dss\_form}} image. The X-ray data come from ROSAT\footnote{\ \url{http://heasarc.gsfc.gov/docs/rosat/rosat3.html}} soft band (surface brightness in the 0.1--2.4 keV). 
{It is worth to note that the all-sky sensitivity of ROSAT is limited to about 10$^{-13}$ erg s$^{-1}$ cm$^{-2}$ \citep[\textit{e.g.},][]{Vik98,Bur07}. 
This is enough for detecting $k$T$_{\mathrm{X}} \ge$ 1 keV clusters, but not enough for identifying substructures in the cooler ones.}
All these maps can also be examined in the webpage accompanying this article. 

\begin{figure}[!b]
  \centering
  \includegraphics[width=0.5\textwidth]{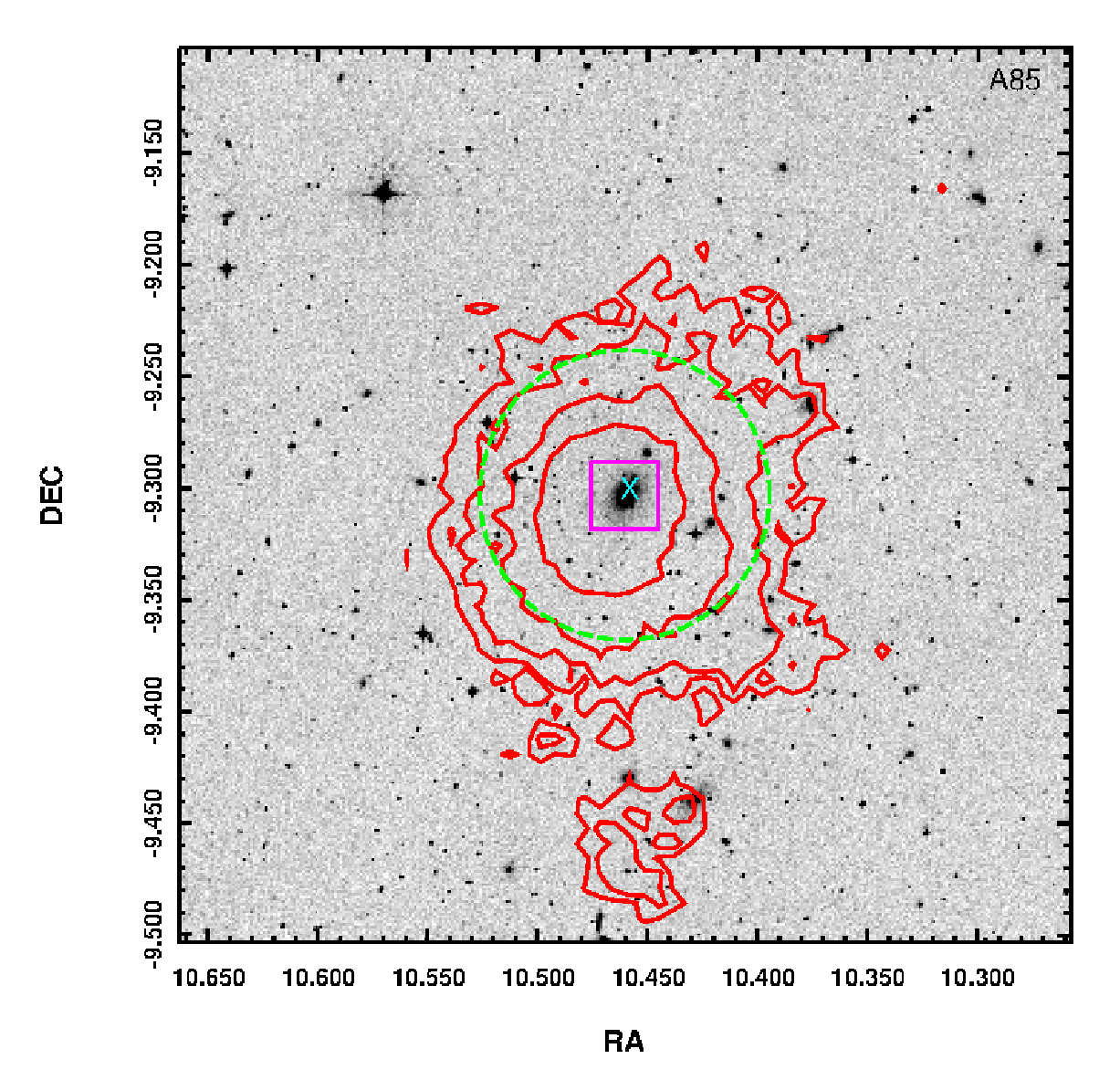}
  \caption{Example of an X-ray map (red contours over the DSS image), for the cluster A0085. The position of the X-ray peak is indicated (cyan `$\times$' symbol), as well as the position of the {CDG} (magenta square). The green circle marks a 0.5 $h_{70}^{-1}$ Mpc radius around the {CDG}.}
  \label{F_Xray}
\end{figure}

The example shown in Fig.~\ref{F_Xray} is once again for A0085. 
A smoothing parameter of 4 in \textsl{ds9} was used. 
The lowest contour in X-ray was traced at the 3$\sigma$ level, followed by contours at levels of 6, 12, 24 and 48 $\sigma$.
The cyan $\times$ symbol is the X-ray peak emission, the magenta square locates the {CDG} and the green circle corresponds to a 0.5 $h_{70}^{-1}$ Mpc radius around it. 
Usually, an optical image of 
size 30\arcmin$\times$30\arcmin\ in the plane of the sky is sufficient to show the distribution of the X-ray emission.
Comparing with Fig.~\ref{F_Isod}, one can see that the detected gas in A0085 covers a smaller area (volume) than the distribution of the galaxies in the cluster and that the {CDG} is only slightly displaced from the X-ray peak.

\paragraph*{Dressler \& Shectman test (3D-test):}

The DS-test \citep{DeS88} is performed in two steps. The first step consists in calculating the $\delta_i$ parameter for each member galaxy:

\begin{equation}
\delta_i^2 = \frac{N_{nn} + 1}{\sigma_{\mathrm{c}}^2} [(\bar{v}_{local} - v_{\mathrm{c}})^2 + 
(\sigma_{local} - \sigma_{\mathrm{c}})^2]
\end{equation}

\noindent where $v_{\mathrm{c}}$ and $\sigma_{\mathrm{c}}$ are the cluster global parameters, while $\bar{v}_{local}$ and $\sigma_{local}$ are local parameters, calculated for the $N_{nn} = 10$\ nearest neighbors of each member galaxy \citep[see][for a discussion of these local parameters and the number of nearest neighbors]{Bra09}.

The second step consists in calculating, for each cluster, the parameter $\Delta = \sum \delta_i$ and comparing its value with a set of 1000 Monte Carlo simulations, obtaining the probability $p$ that a value $\Delta > \Delta_{observed}$ would have been obtained by chance. {We, thus, calculate} P$_{\mathrm{sub}} = 100 * (1-p)$, which is the probability that the cluster is substructured. 
Based on this test, a cluster with P$_{\mathrm{sub}} >$ 90\% can be considered to be significantly substructured.

Because $\Delta$ tends to equal the number of galaxy members when the cluster is close to relaxation \citep[\textit{e.g.},][]{Pin96}, we used the ratio $\Delta/N_c$ as an iterative parameter for the test.
Note that, different from the traditional way substructures are identified by this test, we do not consider only specific concentrations of galaxies with high $\delta$ values in the projected distribution.\footnote{ By considering only high $\delta$ values, one may lose bimodal clusters, where the cluster is formed by two equally massive substructures with small $\delta$ \citep[see a discussion on the limitations of the DS method in][]{Islas15}.} 
Specifically, when $\Delta/N_c >$ 1.2, we analyse both the 3D distribution of galaxies in RA, Dec and $v_{local}$ and the respective 2D PPS diagram to identify, in the former, the volume separation surfaces between the substructures \citep[\textit{e.g.},][]{Lop22}. 
In this pseudo-3D volume, substructure members {are more smoothly distributed, defining a} more isolated \textit{locus} (local concentration) than in a RA-Dec-$z$-volume, while in the PPS they show the typical caustics-shape distribution. 
Therefore, after separating the substructures from the remaining main structure, we recalculate $\Delta/N_c$ to see whether it is below 1.2, and if not, iterate again to isolate new substructures.
 
Note that in applying this test there are cases for which it is not correct to assume there is only one main structure. 
This happens when there are two or more substructures that are comparable in mass, as well as being much more massive than all the other substructures in the cluster. 
These are examples of ``bimodal'' or ``multi-modal'' clusters.

{The parameters calculated from the dynamical and substructure analyses} are reported in Table~\ref{T_Clusters}. 
In col.~1, we give the updated cluster ID. Note that the entries in Table~\ref{T_Clusters} slightly differ from those in Table~\ref{T_Sample}. More specifically, the clusters A2870 and A4049 were determined (see \S~\ref{sec:Caust}) to be part of other massive clusters, respectively A2877 and A4038A. On the other hand, two clusters had their well-known LOS components considered separately, A1736 becoming A1736A and A1736B, and A3526 becoming A3526A and A3526B. Consequently, although the number of entries in both tables are the same, 67 clusters, their identities are somewhat different. In Table~\ref{T_Clusters}, the equatorial coordinates of the corresponding {CDGs} (and by convention, cluster centers) are given in cols.~2 and 3, followed in col.~4 by their LOS velocities. 
Column~5 reports the number of candidate members after splitting up the intersecting neighbors, $N_{g}$, and col.~6, the number of galaxies, $N_{c}$, considered to be bound (included within the caustics) in each cluster.
Other dynamical parameters are reported in col.~7 ($v_{\rm c}$), col.~8 ($\sigma_{\rm c}$), {both for the $N_{c}$ members}, col.~9 ($r_{200}$), col.~10 ($N_a$), col.~11 ($v_{\mathrm{cl}}$), col.~12 ($\sigma_{\mathrm{cl}}$), {the last three for the members inside the circular aperture}, col.~13 ($R_{\rm p}$), col.~14 ($R_{\mathrm{vir}}$) and col.~15 ($M_{\mathrm{vir}}$). 
{Parameters associated to the substructure analyses are shown in columns~16 to 18:} the \textsl{skewness}, the excess \textsl{kurtosis}, and P$_{\mathrm{sub}} = 100 * (1-p)$, respectively.

Note that some clusters in our sample do not have their ICM emission centered on the \textbf{main} structure of their clusters, but on a substructure (A0754, A1736B and A2151). These are marked with a '$^{\cancel{x}}$' in the first column of the Table~\ref{T_Clusters}.

\subsection{Gravitational binding} 
\label{sec:Bind}

For the 16 superposed clusters appearing in Table~\ref{T_Pairs}, as well as for the substructures reported in Table~\ref{T_ApB}, we complemented our analysis by applying a test for gravitational binding. 
Since the evolutionary state of a system like a galaxy cluster is also related to its geometry in redshift space, any density enhancement present in real space will also appear as a density enhancement in  redshift space: systems representing small overdensities, where the Hubble flow has not yet been significantly perturbed, appear essentially undistorted, while those that are clearly collapsing, the Hubble flow being slowed down, appear flattened along the LOS. 
On the other hand, systems that are close to virial equilibrium appear as particularly elongated condensations in redshift space, a phenomenon known in the literature as \emph{Fingers-of-God}. 
Because of this effect, it is therefore possible to 
assess what is the global dynamical state of a galaxy system at the scale of a cluster by evaluating its distortions in redshift space.

As explained by \citet{SeT77}, this level of distortion can be determined, for a pair of objects (galaxies, groups or clusters), by determining the separation between the members of the pair and the angle $\chi$ between the separation vector and the plane of the sky. Such angle is measured as follows: let $\theta_{12}$ be the angular separation between the center of the two galaxies, $z_1$ and $z_2$ (with $z_2 \ge z_1$) being their respective redshifts, then the physical distance ($d_{12}$) and projected separation ($\ell_{12}$) in the plane of the sky are given, respectively, by:

\begin{equation}
d_{12}=\frac{c}{H_0}\left[ z_1^2+z_2^2 -2z_1 z_2\cos{\theta_{12}}\right]^{1/2}
\end{equation}

\noindent and

\begin{equation}
\ell_{12}=\frac{c}{H_0}\left( z_1+z_2\right)\tan{\frac{\theta_{12}}{2}}
\end{equation}

\noindent the angle $\chi$ between the separation vector being equal to:

\begin{equation}
\chi=\arctan\left[\frac{1}{2}\left(\frac{z_2}{z_1}-1\right)\cot{\frac{\theta_{12}}{2}}\right]
\end{equation}
\noindent where $0 \le \chi \le \pi/2$.

For a homogeneous spherical system following the expansion flow, $\langle\chi\rangle$ approaches the isotropic value of 32.7$^{\circ}$; $\langle\chi\rangle$ tends to lower values for a collapsing (flattened) system and larger values for a virialized (elongated) one. 
Note, however, that for a non-spherical system, the geometrical elongation/flattening could dominate $\langle\chi\rangle$, masking their real dynamical state.

\clearpage
\onecolumn
\begin{sidewaystable}
 \tiny
 \centering
 \tablecols{22}
 \setlength{\tabcolsep}{1mm}
 \setlength{\tabnotewidth}{0.9\columnwidth}
   \caption{Basic data on clusters of the sample.}
   \label{T_Clusters}                                
    \begin{tabular}{lrrrrrrrrrrrrrrrrrccrr}
\toprule
  \multicolumn{1}{c}{ID\tabnotemark{a}} &
  \multicolumn{1}{c}{RA$_{\mathrm{CDG}}$} &
  \multicolumn{1}{c}{Dec$_{\mathrm{CDG}}$} &
  \multicolumn{1}{c}{$v_{\mathrm{CDG}}$} &
  \multicolumn{1}{c}{$N_{g}$} &
  \multicolumn{1}{c}{$N_{c}$} &
  \multicolumn{1}{c}{$v_{c}$}&
  \multicolumn{1}{c}{$\sigma_{c}$}&
  \multicolumn{1}{c}{$r_{200}$} &
  \multicolumn{1}{c}{$N_{a}$} &
  \multicolumn{1}{c}{$v_{\mathrm{cl}}$} &
  \multicolumn{1}{c}{$\sigma_{\mathrm{cl}}$} &
  \multicolumn{1}{c}{$R_{\mathrm{p}}$} &
  \multicolumn{1}{c}{$R_{\mathrm{vir}}$} &
  \multicolumn{1}{c}{$M_{\mathrm{vir}}$} &
  \multicolumn{1}{c}{skew} &
  \multicolumn{1}{c}{kurt} &
  \multicolumn{1}{c}{P$_{\mathrm{sub}}$} &
  \multicolumn{1}{c}{$N_{\mathrm{sub}}$} &
  \multicolumn{1}{c}{$\cal{A}$} &
  \multicolumn{1}{c}{$r_{\mathrm{ox}}$} &
  \multicolumn{1}{c}{$v_{\mathrm{pec}}$} \\
  \multicolumn{1}{c}{} &
  \multicolumn{1}{c}{[deg]$_{\mathrm{J2000}}$} &
  \multicolumn{1}{c}{[deg]$_{\mathrm{J2000}}$} &
  \multicolumn{1}{c}{[km/s]} &
  \multicolumn{1}{c}{} &
  \multicolumn{1}{c}{} &
  \multicolumn{1}{c}{[km/s]} &
  \multicolumn{1}{c}{[km/s]} &
  \multicolumn{1}{c}{[Mpc]\tabnotemark{b}} &		
  \multicolumn{1}{c}{} &
  \multicolumn{1}{c}{[km/s]} &
  \multicolumn{1}{c}{[km/s]} &
  \multicolumn{1}{c}{[Mpc]\tabnotemark{b}} &		
  \multicolumn{1}{c}{[Mpc]\tabnotemark{b}} &		
  \multicolumn{1}{c}{[$M_{\odot}$]\tabnotemark{b}} &	
  \multicolumn{1}{c}{} &
  \multicolumn{1}{c}{} &
  \multicolumn{1}{c}{[\%]} &
  \multicolumn{1}{c}{\textsl{m,hs,ls}} &
  \multicolumn{1}{c}{} &
  \multicolumn{1}{c}{[kpc]\tabnotemark{b}} &		
  \multicolumn{1}{c}{[km/s]} \\
\multicolumn{1}{c}{(1)} & 
 \multicolumn{1}{c}{(2)} & 
 \multicolumn{1}{c}{(3)} & 
 \multicolumn{1}{c}{(4)} & 
 \multicolumn{1}{c}{(5)} & 
 \multicolumn{1}{c}{(6)} & 
 \multicolumn{1}{c}{(7)} & 
 \multicolumn{1}{c}{(8)} & 
 \multicolumn{1}{c}{(9)} & 
 \multicolumn{1}{c}{(10)} &
 \multicolumn{1}{c}{(11)} & 
 \multicolumn{1}{c}{(12)} & 
 \multicolumn{1}{c}{(13)} & 
 \multicolumn{1}{c}{(14)} & 
 \multicolumn{1}{c}{(15)} & 
 \multicolumn{1}{c}{(16)} & 
 \multicolumn{1}{c}{(17)} & 
 \multicolumn{1}{c}{(18)} & 
 \multicolumn{1}{c}{(19)} & 
 \multicolumn{1}{c}{(20)} & 
 \multicolumn{1}{c}{(21)} & 
 \multicolumn{1}{c}{(22)} \\
\midrule
  A2798B   &   9.37734 & -28.52947 & 33338 &   81 &  78 & 33604 & 697  & 1.353 &  60 & 33544 &  757 & 1.148 & 1.748 &  6.01 & -0.313 & -0.502 &  87.4 & 1,0,0 & U &  96.5 &  -185.3  \\
  A2801    &   9.62876 & -29.08160 & 33660 &   50 &  45 & 33773 & 652  & 1.259 &  35 & 33640 &  699 & 1.553 & 1.833 &  6.94 & -0.397 & -0.024 &  15.5 & 1,0,0 & U &  42.4 &    18.0  \\
  A2804    &   9.90753 & -28.90620 & 33546 &   88 &  80 & 33378 & 663  & 1.292 &  48 & 33669 &  516 & 1.277 & 1.403 &  3.11 & -0.346 & -0.982 &  97.6 & 2,0,0 & M & 126.3 &  -110.6  \\
  A0085A   &  10.46051 &  -9.30304 & 16613 &  368 & 352 & 16561 & 1011 & 2.027 & 321 & 16570 & 1045 & 2.030 & 2.668 & 20.20 & -0.444 & -0.118 &  99.9 & 1,2,3 & S &   8.2 &    32.2  \\
  A2811B   &  10.53717 & -28.53577 & 32466 &  146 & 139 & 32354 & 831  & 1.625 & 103 & 32329 &  947 & 1.701 & 2.316 & 13.90 &  0.162 & -0.042 &  94.3 & 1,0,2 & S &   5.5 &   123.7  \\
  A0118    &  13.74348 & -26.37515 & 34068 &  119 &  80 & 34384 & 681  & 1.341 &  59 & 34287 &  689 & 1.206 & 1.667 &  5.23 &  0.349 & -0.635 &  90.4 & 1,2,0 & S & \nodata & -212.7  \\
  A0119    &  14.06709 &  -1.25549 & 13323 &  339 & 333 & 13299 & 810  & 1.633 & 294 & 13301 &  853 & 1.430 & 2.082 &  9.51 & -0.175 &  0.246 &  99.9 & 1,2,1 & S & 125.2 &    21.1  \\
  A0122    &  14.34534 & -26.28134 & 33804 &  111 &  31 & 34076 & 659  & 1.265 &  28 & 34062 &  677 & 1.190 & 1.641 &  4.98 &  0.337 & -0.266 &  46.8 & 1,0,0 & U &  50.6 &  -231.7  \\
  A0133A   &  15.67405 & -21.88215 & 17048 &  137 & 132 & 16830 & 713  & 1.425 &  86 & 16838 &  778 & 1.320 & 1.899 &  7.30 &  0.077 & -0.393 &  77.7 & 1,2,0 & S &  33.9 &   198.8  \\
  A2877-70 &  17.48166 & -45.93122 &  7213 &  237 & 174 & 7034  & 652  & 1.326 & 112 &  7143 &  679 & 0.999 & 1.596 &  4.20 &  0.273 & -0.398 & 100.0 & 1,2,0 & S &  27.8 &    68.4  \\
  A3027A   &  37.70600 & -33.10375 & 23541 &  167 & 102 & 23429 & 668  & 1.335 &  82 & 23494 &  713 & 1.618 & 1.904 &  7.52 & -0.278 & -0.814 &  97.2 & 1,1,1 & S & 113.9 &    43.6  \\
  A0400    &  44.42316 &   6.02700 &  6789 &  115 &  61 & 6959  & 323  & 0.682 &  51 &  6947 &  343 & 0.635 & 0.870 &  0.68 &  0.100 & -1.007 &  76.0 & 1,1,0 & S &  39.8 &  -154.4  \\
  A0399    &  44.47119 &  13.03080 & 21483 &  101 &  69 & 21138 & 957  & 1.894 &  69 & 21146 &  950 & 1.414 & 2.209 & 11.70 & -0.225 & -0.460 &  13.3 & 1,0,0 & U & 110.1 &   314.8  \\
  A0401    &  44.74091 &  13.58287 & 22297 &  116 & 115 & 22053 & 1028 & 2.036 & 114 & 22061 & 1026 & 1.574 & 2.407 & 15.10 &  0.263 & -0.352 &  40.8 & 1,0,0 & U &  18.6 &   219.8  \\
  A3094A   &  47.85423 & -26.93122 & 20552 &  126 & 114 & 20489 & 548  & 1.090 &  84 & 20539 &  637 & 1.305 & 1.648 &  4.83 &  0.628 &  0.243 &  88.3 & 1,0,0 & U & 148.3 &    12.2  \\
  A3095    &  48.11077 & -27.14016 & 19314 &   44 &  40 & 19485 & 306  & 0.606 &  21 & 19557 &  327 & 0.664 & 0.845 &  0.65 & -0.018 & -0.932 &  10.5 & 1,0,0 & L & \nodata & -228.1  \\
  A3104    &  48.59055 & -45.42024 & 21785 &   62 &  53 & 21777 & 413  & 0.823 &  28 & 21681 &  498 & 0.782 & 1.178 &  1.77 &  0.039 & -0.106 &  39.5 & 1,2,0 & S &  34.4 &    97.0  \\
  S0334    &  49.08556 & -45.12110 & 22401 &   22 &  27 & 22373 & 518  & 1.030 &  26 & 22363 &  534 & 0.811 & 1.249 &  2.11 &  0.449 &  1.037 &  27.4 & 1,0,0 & L & \nodata &  35.4  \\
  A3112B   &  49.49025 & -44.23821 & 22764 &  120 &  97 & 22631 & 672  & 1.337 &  74 & 22669 &  705 & 1.861 & 1.982 &  8.46 &  0.271 & -0.572 &  99.9 & 1,1,0 & S &  13.2 &    88.3  \\
  S0336    &  49.45997 & -44.80069 & 22849 &   54 &  44 & 23223 & 506  & 1.006 &  32 & 23186 &  538 & 1.140 & 1.405 &  3.02 &  0.178 & -0.606 &   3.2 & 1,0,0 & L & \nodata & -312.8  \\
  A0426A   &  49.95042 &  41.51167 &  5231 &  360 & 314 & 5254  & 1030 & 2.104 & 314 &  5262 & 1029 & 1.395 & 2.359 & 13.50 &  0.030 & -0.535 &  97.1 & 1,0,2 & P &   4.0 &   -30.5  \\
  S0373    &  54.62118 & -35.45074 &  1452 &  272 & 229 & 1452  & 334  & 0.688 &  98 &  1461 &  390 & 0.308 & 0.749 & 0.430 & -0.025 & -0.428 & 100.0 & 1,2,1 & S &   1.7 &    -9.0  \\
  A3158    &  55.72063 & -53.63130 & 17327 &  258 & 249 & 17764 & 1064 & 2.138 & 249 & 17735 & 1066 & 1.341 & 2.353 & 13.90 &  0.260 & -0.478 &  49.6 & 1,2,2 & S &  19.1 &  -385.2  \\
  A0496    &  68.40767 & -13.26196 &  9841 &  358 & 351 & 9892  & 688  & 1.395 & 279 &  9925 &  712 & 1.364 & 1.822 &  6.31 &  0.030 & -0.535 &  99.9 & 1,1,0 & S &   8.5 &   -81.3  \\
  A0539    &  79.15555 &   6.44092 &  8257 &  143 & 132 & 8679  & 571  & 1.160 &  92 &  8631 &  698 & 0.882 & 1.557 &  3.92 & -0.249 &  0.825 & 100.0 & 1,2,0 & S &   6.5 &  -363.5  \\
  A3391    &  96.58521 & -53.69330 & 16361 &  119 & 100 & 16831 & 760  & 1.519 &  75 & 16776 &  817 & 1.270 & 1.936 &  7.74 &  0.159 & -0.482 &   9.8 & 1,0,0 & U &  24.4 &  -393.0  \\
  A3395    &  96.90105 & -54.44936 & 14571 &  224 & 218 & 14875 & 722  & 1.450 & 199 & 14878 &  746 & 1.264 & 1.823 &  6.42 & -0.158 & -0.494 & 100.0 & 2,2,1 & M &  10.9 &  -292.5  \\
  A0576    & 110.37600 &  55.76158 & 11435 &  238 & 220 & 11351 & 810  & 1.638 & 191 & 11350 &  866 & 1.633 & 2.202 & 11.20 &  0.031 & -0.114 & 100.0 & 1,2,0 & S &  84.3 &    81.9  \\
  A0634    & 123.93686 &  58.32109 &  8029 &  140 & 132 & 8006  & 318  & 0.646 &  70 &  8037 &  395 & 0.792 & 1.029 &  1.13 &  0.057 & -0.061 &  73.4 & 1,0,0 & L & \nodata &  -7.8  \\
  A0754$^{\cancel{x}}$ & 137.13495 &  -9.62974 & 16451 & 468 & 386 & 16246 & 757  & 1.520 & 333 & 16258 &  820 & 1.484 & 2.045 &  9.10 & -0.024 &  0.373 & 100.0 & 2,2,2 & M & 239.1 &   183.1  \\
  A1060    & 159.17796 & -27.52858 &  3808 &  382 & 380 & 3709  & 652  & 1.335 & 343 &  3694 &  678 & 0.951 & 1.574 &  3.99 &  0.122 & -0.532 & 100.0 & 1,0,5 & P &   4.9 &   112.6  \\
  A1367    & 176.00905 &  19.94982 &  6260 &  339 & 286 & 6451  & 547  & 1.115 & 226 &  6444 &  597 & 1.157 & 1.539 &  3.76 & -0.026 & -0.629 & 100.0 & 2,0,2 & M & 366.0 &  -180.1  \\
  A3526A   & 192.20392 & -41.31166 &  2948 &  235 & 235 & 3088  & 491  & 1.005 & 126 &  2993 &  564 & 0.836 & 1.335 &  2.43 & -0.482 & -0.191 & 100.0 & 1,0,3 & P &   3.8 &   -44.6  \\
  A3526B   & 192.51645 & -41.38207 &  4593 &  101 &  95 & 4580  & 276  & 0.552 &  45 &  4636 &  317 & 0.480 & 0.754 &  0.44 &  0.300 & -0.359 &  99.9 & 1,1,0 & S & \nodata & -42.3  \\
  A3530    & 193.90001 & -30.34749 & 16187 &  126 & 110 & 16036 & 615  & 1.231 &  94 & 16064 &  631 & 1.272 & 1.632 &  4.63 &  0.305 & -0.567 &  99.5 & 1,1,0 & S &  66.5 &   116.7  \\
  A1644    & 194.29825 & -17.40957 & 14225 &  307 & 301 & 14085 & 987  & 1.989 & 288 & 14095 & 1008 & 1.507 & 2.365 & 14.00 & -0.049 &  0.201 &  86.6 & 1,0,2 & P &  39.6 &   124.2  \\
  A3532    & 194.34134 & -30.36348 & 16303 &  112 & 103 & 16753 & 423  & 0.849 &  58 & 16709 &  443 & 0.929 & 1.160 &  1.66 & -0.130 & -0.798 &  93.5 & 1,2,0 & S &  87.9 &  -384.6  \\
  A1650    & 194.67290 &  -1.76139 & 25328 &  220 & 192 & 25216 & 673  & 1.330 & 146 & 25249 &  723 & 1.581 & 1.903 &  7.55 &  0.188 &  0.016 &   2.1 & 1,0,0 & U &  27.3 &    72.9  \\
  A1651    & 194.84383 &  -4.19612 & 25622 &  221 & 191 & 25454 & 833  & 1.651 & 158 & 25453 &  876 & 1.782 & 2.250 & 12.50 &  0.132 & -0.569 &  59.2 & 1,0,2 & P &  25.5 &   155.8  \\
  A1656    & 194.89879 &  27.95939 &  7157 &  969 & 969 & 6976  & 993  & 2.025 & 919 &  6997 &  995 & 1.734 & 2.474 & 15.70 & -0.100 & -0.018 & 100.0 & 1,1,8 & S &  57.9 &   156.4  \\
  A3556    & 201.02789 & -31.66996 & 14406 &  102 & 102 & 14435 & 505  & 1.016 &  90 & 14436 &  520 & 1.048 & 1.347 &  2.59 &  0.354 & -0.606 &  99.0 & 2,0,0 & M & 630.5 &   -28.6  \\
  A1736A   & 201.68378 & -27.43940 & 10506 &   74 &  74 & 10363 & 417  & 0.840 &  36 & 10499 &  386 & 0.955 & 1.075 &  1.30 &  0.102 & -0.997 & 100.0 & 1,3,0 & S & \nodata &   6.8  \\
  A1736B$^{\cancel{x}}$ & 201.86685 & -27.32468 & 13574 &  145 & 141 & 13665 & 839 & 1.689 & 126 & 13678 &  844 & 1.355 & 2.029 &  8.82 &  0.022 & -0.500 &  99.0 & 1,1,1 & S & 610.8 &   -99.5  \\
  A3558    & 201.98701 & -31.49547 & 14073 &  548 & 548 & 14455 & 950  & 1.912 & 469 & 14476 &  955 & 1.893 & 2.460 & 15.80 & -0.230 & -0.478 & 100.0 & 1,0,1 & P &  25.0 &  -384.4  \\
  A3562    & 203.39474 & -31.67227 & 14693 &  231 &  98 & 14541 & 564  & 1.138 &  82 & 14561 &  594 & 1.221 & 1.549 &  3.94 &  0.037 & -0.634 &  43.6 & 1,0,0 & U &  42.6 &   125.9  \\
  A1795    & 207.21880 &  26.59301 & 18968 &  179 & 164 & 18893 & 764  & 1.525 & 154 & 18889 &  780 & 1.278 & 1.876 &  7.09 & -0.049 &  0.210 &  98.3 & 1,0,0 & U &  13.8 &    74.3  \\
  A2029    & 227.73376 &   5.74491 & 23353 &  202 & 155 & 23052 & 934  & 1.850 & 155 & 23051 &  931 & 0.989 & 1.931 &  7.82 &  0.096 & -0.791 &  92.8 & 1,0,0 & U & 132.8 &   280.4  \\
  A2040B   & 228.19781 &   7.43426 & 13713 &  153 & 150 & 13472 & 567  & 1.141 & 104 & 13527 &  627 & 1.327 & 1.653 &  4.77 &  0.312 &  0.026 &  97.8 & 1,1,0 & S &  43.3 &   178.0  \\
  A2052    & 229.18536 &   7.02167 & 10332 &  178 & 176 & 10295 & 581  & 1.179 & 120 & 10416 &  648 & 1.115 & 1.600 &  4.28 &  0.356 &  0.003 & 100.0 & 1,1,0 & S &   9.1 &   -81.2  \\
  A2065    & 230.62053 &  27.71228 & 21828 &  204 & 169 & 21889 & 1043 & 2.071 & 168 & 21878 & 1043 & 1.712 & 2.504 & 17.00 &  0.129 & -0.720 &  99.0 & 1,0,0 & U &  47.1 &   -46.6  \\
  A2063A   & 230.77209 &   8.60918 & 10377 &  200 & 193 & 10312 & 667  & 1.350 & 142 & 10345 &  762 & 1.165 & 1.809 &  6.18 & -0.069 & -0.132 &  99.4 & 1,0,1 & P &  16.5 &    30.9  \\
  A2142    & 239.58345 &  27.23335 & 27254 &  232 & 191 & 26975 & 820  & 1.618 & 157 & 27036 &  828 & 1.767 & 2.157 & 11.10 & -0.087 & -0.276 &  66.7 & 1,0,1 & P &  41.0 &   200.0  \\
  A2147    & 240.57086 &  15.97451 & 10595 &  481 & 453 & 10929 & 918  & 1.858 & 397 & 10889 &  935 & 1.966 & 2.466 & 15.70 &  0.215 & -0.286 & 100.0 & 2,2,0 & M & 119.9 &  -283.7  \\
  A2151$^{\cancel{x}}$ & 241.28754 &  17.72997 &  9378 & 331 & 311 & 10906 & 743  & 1.502 & 276 & 10898 &  768 & 1.551 & 1.999 &  8.35 & -0.064 & -0.573 & 100.0 & 3,2,2 & M &   3.1 &  -1466.6 \\
  A2152    & 241.37175 &  16.43579 & 13268 &  124 & 116 & 13283 & 398  & 0.799 &  64 & 13266 &  406 & 1.038 & 1.139 &  1.56 &  0.934 &  0.724 &  98.6 & 2,0,0 & M &  42.1 &     1.9  \\
  A2197    & 246.92114 &  40.92690 &  9511 &  294 & 276 & 9108  & 546  & 1.109 & 185 &  9114 &  573 & 1.402 & 1.593 &  4.21 &  0.258 & -0.797 & 100.0 & 3,0,1 & M &  16.9 &   385.3  \\
  A2199    & 247.15948 &  39.55138 &  9296 &  521 & 498 & 9089  & 753  & 1.529 & 459 &  9090 &  779 & 1.302 & 1.907 &  7.21 & -0.055 & -0.058 & 100.0 & 1,0,3 & P &   6.4 &   199.9  \\
  A2204A   & 248.19540 &   5.57583 & 45528 &  111 &  96 & 45274 & 760  & 1.456 &  38 & 45497 & 1101 & 1.838 & 2.588 & 20.30 &  0.669 &  0.092 &  95.5 & 1,2,0 & S &  52.1 &    26.9  \\
  A2244    & 255.67697 &  34.06010 & 29543 &  106 & 102 & 29811 & 1154 & 2.282 & 102 & 29778 & 1161 & 1.491 & 2.546 & 18.30 &  0.166 & -0.851 &  63.8 & 1,0,0 & U &  15.2 &  -213.8  \\
  A2256    & 256.11352 &  78.64056 & 17778 &  295 & 280 & 17567 & 1222 & 2.449 & 280 & 17567 & 1222 & 1.514 & 2.683 & 20.60 &  0.009 & -0.624 &  99.9 & 1,2,2 & S & 129.2 &   199.3  \\
  A2255    & 258.11981 &  64.06070 & 22068 &  189 & 181 & 24100 & 992  & 1.973 & 179 & 24126 & 1000 & 1.802 & 2.470 & 16.40 & -0.316 & -0.518 & 100.0 & 1,2,0 & S & 183.7 &  -1904.7 \\
  A3716    & 312.98715 & -52.62983 & 14112 &  157 & 140 & 13517 & 746  & 1.501 & 123 & 13508 &  783 & 1.247 & 1.878 &  6.99 &  0.168 & -0.915 &  88.8 & 2,0,1 & M &   0.9 &   578.0  \\
  S0906    & 313.18576 & -52.02746 & 13947 &   62 &  54 & 14481 & 340  & 0.680 &  26 & 14446 &  440 & 0.824 & 1.113 &  1.46 &  0.002 &  0.014 &  57.4 & 1,0,0 & L & \nodata & -476.1  \\
  A4012A   & 352.96231 & -34.05553 & 16241 &   93 &  85 & 16219 & 473  & 0.947 &  39 & 16249 &  575 & 1.043 & 1.436 &  3.15 & -0.126 &  0.431 &  73.1 & 1,0,0 & L & \nodata &  -7.6  \\
  A2634    & 354.62244 &  27.03130 &  9117 &  192 & 185 & 9262  & 695  & 1.409 & 166 &  9268 &  717 & 1.251 & 1.780 &  5.87 & -0.274 & -0.177 &  94.9 & 1,1,0 & S &  51.7 &  -146.5  \\
  A4038A-49 & 356.93768 & -28.14070 &  8672 &  247 & 237 & 8799  & 683  & 1.385 & 180 &  8873 &  753 & 1.150 & 1.789 &  5.95 &  0.252 & -0.017 & 100.0 & 1,2,0 & S &  14.4 &  -195.2  \\
  A2670    & 358.55713 & -10.41900 & 23157 &  308 & 288 & 22791 & 909  & 1.805 & 251 & 22799 &  970 & 1.153 & 2.089 &  9.91 &  0.066 & -0.258 & 100.0 & 1,0,0 & U &  31.7 &   332.7  \\
\bottomrule
\tabnotetext{a}{A capital letter after the ACO name indicates the  
line-of-sight component of the cluster \citep[see][]{Cho14}.
A $^{\cancel{x}}$ symbol indicates the cluster center is not at the X-peak substructure.}

\tabnotetext{b}{Longitud and mass parameters are also in units of $h_{70}$.}
   \end{tabular}
\end{sidewaystable}
\clearpage
\twocolumn

Assuming a symmetric geometry, on the other hand, the same angle $\chi$ can be used to test the Newtonian criterion for gravitational binding of two systems \citep{Bee82}. 
This allows one to determine whether the pairs are either relaxed, collapsing or expanding, or not bound but simply close in space. Within the context of a two-body-problem, the orbits of the two galaxies or systems, with masses $M_1$ and $M_2$, are assumed to be linear, with no rotations or discontinuities around the center of mass. 
The projected separation between their centers would then be $R_p = R\cos{\chi} \, (= \ell_{12})$ and their relative velocity projected along LOS, $V_r = V\sin{\chi}$, where $R$ is the physical distance ($= d_{12}$) between them and $V$ is their relative velocity. 
Considering that the energy criterion for gravitational binding is, $\frac{1}{2}v_{esc}^2\le\frac{GM}{R}$, where $v_{esc}$ is the escape velocity, we assume $V = v_{esc}$ and estimate the total mass to be $M = M_1 + M_2$, yielding the condition for the pair to be bound:

\begin{equation}
V_r^2 R_p \le 2 G M \sin^2{\chi} \cos{\chi}
\end{equation} 

Having evaluated these parameters for the 16 cases of superposed clusters in our sample, this is how we determined, as reported in Table~\ref{T_Pairs}, that 5 pairs and 2 clusters are only apparent superpositions.

\subsection{{CDG} related parameters} 
\label{sec:Offs}


\paragraph*{CDG--X-ray offset:}
The offsets for 52 of the 59 clusters detected in X-rays in our sample were calculated based on the coordinates of the peak emission in X-ray compiled by \citet{Pif11}. However, for A2151 we did not use this source because the peak reported by these authors, although the brightest in this cluster, corresponds to the emission of a subcluster. Instead, we used the coordinates of the main structure as reported by \citet[][]{Tiw21}. For 5 of the 6 remaining clusters, the coordinates for the X-ray peaks came from three different studies \citep[][]{Ebe96, Led03, Sat10}. This leaves one cluster, namely A0118, for which information is missing. 
The offsets, $r_{\mathrm{ox}}$, reported in col.~21 of Table~\ref{T_Clusters}, correspond to the angular separations transformed into the physical separations in kpc at the redshift of each cluster.

To compare these offsets with those reported in the literature, we also calculated the relative offsets, using the relation:   

\begin{equation}
\Delta r_{\mathrm{ox}} = r_{\mathrm{ox}}/r_{500}
\end{equation}

\noindent Note that since \citet{Pif11} is our only source for $r_{500}$, we only calculated $\Delta r_{\mathrm{ox}}$ for the 52 clusters in common with these authors (to be reported further in Table~\ref{T_Assembl}). 

Consistent with a typical cooling time of 4~Gyr \citep[see Figure~2 in][]{Zha11}, a cluster with an offset $r_{\mathrm{ox}} < 30\ h_{70}^{-1}$ kpc, equivalent to $\Delta r_{\mathrm{ox}} \lesssim 0.03$, can be considered to be relaxed. Compared with the literature, this relaxation threshold is between two previously proposed values: \citet{Lav16} used $\Delta r_{\mathrm{ox}}= 0.05$ and \citet{Lop18} used $\Delta r_{\mathrm{ox}}= 0.01$.

\paragraph*{Peculiar velocity:}
We calculated the peculiar velocity of the {CDGs} using the formula \citep[\textit{e.g.},][]{Coz09}:

\begin{equation}
v_{\mathrm{pec}} = \frac{v_{\mathrm{CDG}} - v_{\mathrm{cl}}}{1 + z_{\mathrm{cl}}}
\end{equation}

We also calculated the respective relative peculiar velocity using the definition \citep[\textit{e.g.},][]{Lau14}: 

\begin{equation}
\Delta v_{\mathrm{pec}} = |v_{\mathrm{pec}}|/\sigma_{\mathrm{cl}}
\end{equation}

We report the values obtained for $v_{\mathrm{pec}}$ in col.~22 of Table~\ref{T_Clusters}, while $\Delta v_{\mathrm{pec}}$ will be reported in Table~\ref{T_Assembl}. According to this parameter, we consider a system to be relaxed when $|v_{\mathrm{pec}}| < 175$ km s$^{-1}$, which is equivalent to $\Delta v_{\mathrm{pec}} \lesssim 0.21$.

\paragraph*{CDG luminosities:}
As described in \S~\ref{sec:Phot}, we used the K$_s$ absolute magnitudes of the {CDGs}, M$_\mathrm{Ks}$, as a proxy for their stellar masses. 
Comparison of these masses with the masses (or number of galaxies) of the substructure hosting the {CDGs} can yield important information about the assembly histories of the clusters. 
In particular, one could expect the most massive {CDGs} to be located in the most massive substructures, and these massive substructures to form the main subclusters of their respective clusters.
The absolute magnitudes {of the CDGs} are reported further in Table~\ref{T_Assembl}.

\paragraph*{Magnitude gaps:}
Another important parameter relating the assembly history of the {CDG} to its cluster is its magnitude gap: assuming a {CDG} grows in mass by cannibalizing its neighbors, its magnitude gap is expected to increase with time. For our sample, we have calculated two gaps: i) the difference in magnitude between the {CDG} and second-rank member, $\Delta m_{12}$, and ii) the difference in magnitude between the second and third-rank members, $\Delta m_{23}$. 

Note that when a {CDG} differs from the original BCG of the cluster (which is the case for 19\% of the clusters in our sample, see \S~\ref{sec:FRGs}), the identification of the second-rank galaxy varies with the type of cluster: for both the Fornax-like clusters and clusters with a fossil candidate in their outskirts, the second-rank galaxy is the second-rank of the main structure, while in Coma-like cluster the second-rank is the initial BCG, brighter than the {CDG}; this produces a negative $\Delta m_{12}$. The various gaps are also reported in Table~\ref{T_Assembl}.

\section{Properties and assembly histories of clusters}
\label{sec:Res}

\subsection{Global cluster properties}
\label{sec:Glob}

\begin{figure}[ht] 
 \centering
  \includegraphics[width=\columnwidth]{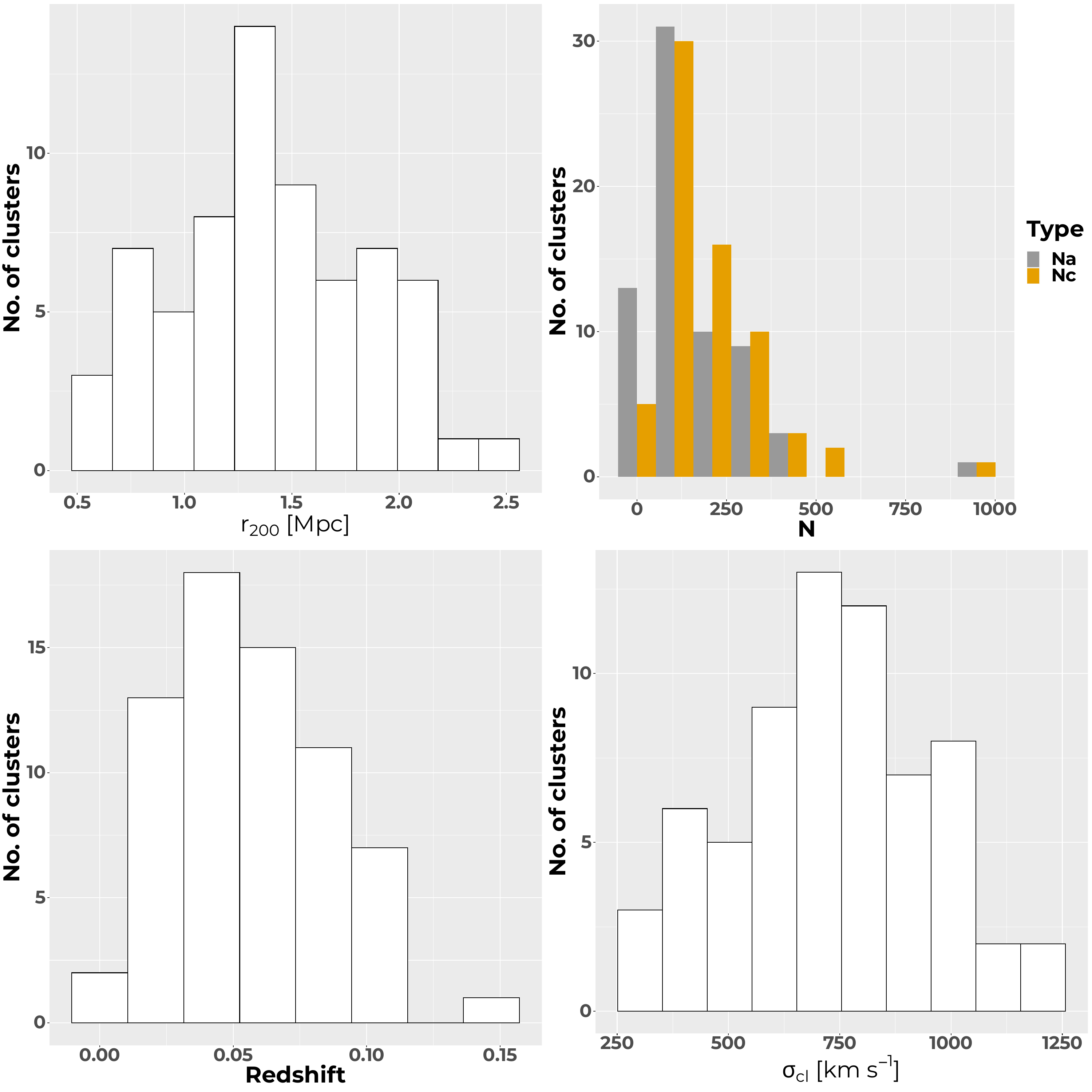}
 \caption{Global properties of the clusters as defined in Table~\ref{T_Clusters}: 
 Upper left, estimated $r_{200}$ radii; 
 upper right, number of galaxies within the caustics ($N_c$, orange) and within the virial radius ($N_a$, gray);
 lower left, redshifts; and 
 lower right, velocity dispersions  ($\sigma_{cl}$).}
 \label{F_Global1}
\end{figure}

Using the optical data related to galaxy membership, we show in Fig.~\ref{F_Global1} the histograms characterizing the ``global'' properties of the clusters identified in Table~\ref{T_Clusters}. 
In the upper left panel, we show the distribution of $r_{200}$, which is commonly used as a reference radius. The median of 1.35 $h_{70}^{-1}$ Mpc, corresponding to only 63\% of the R$_{\mathrm{A}}$, implies the galaxy concentrations in our sample of clusters are relatively high. In the upper right panel, the two distributions for the number of galaxies within the caustic, $N_c$, and galaxies within virial aperture radius, $N_a$, confirm this trend, the medians amounting to 150 and 114 galaxies, respectively. In the lower left panel, the distribution of redshifts is found to be positively skewed, 
since the mode appears before the median at $z = 0.054$. This shows that most of our clusters are nearby, and thus, assuming they formed in the distant past, had had enough time to virialize. Finally, in the lower right panel, the distribution for the velocity dispersion within $1.3 \times r_{200}$ has a median value of 723~km~s$^{-1}$, which is typical for Abell clusters.

\begin{figure}[ht] 
 \centering
\includegraphics[width=\columnwidth]{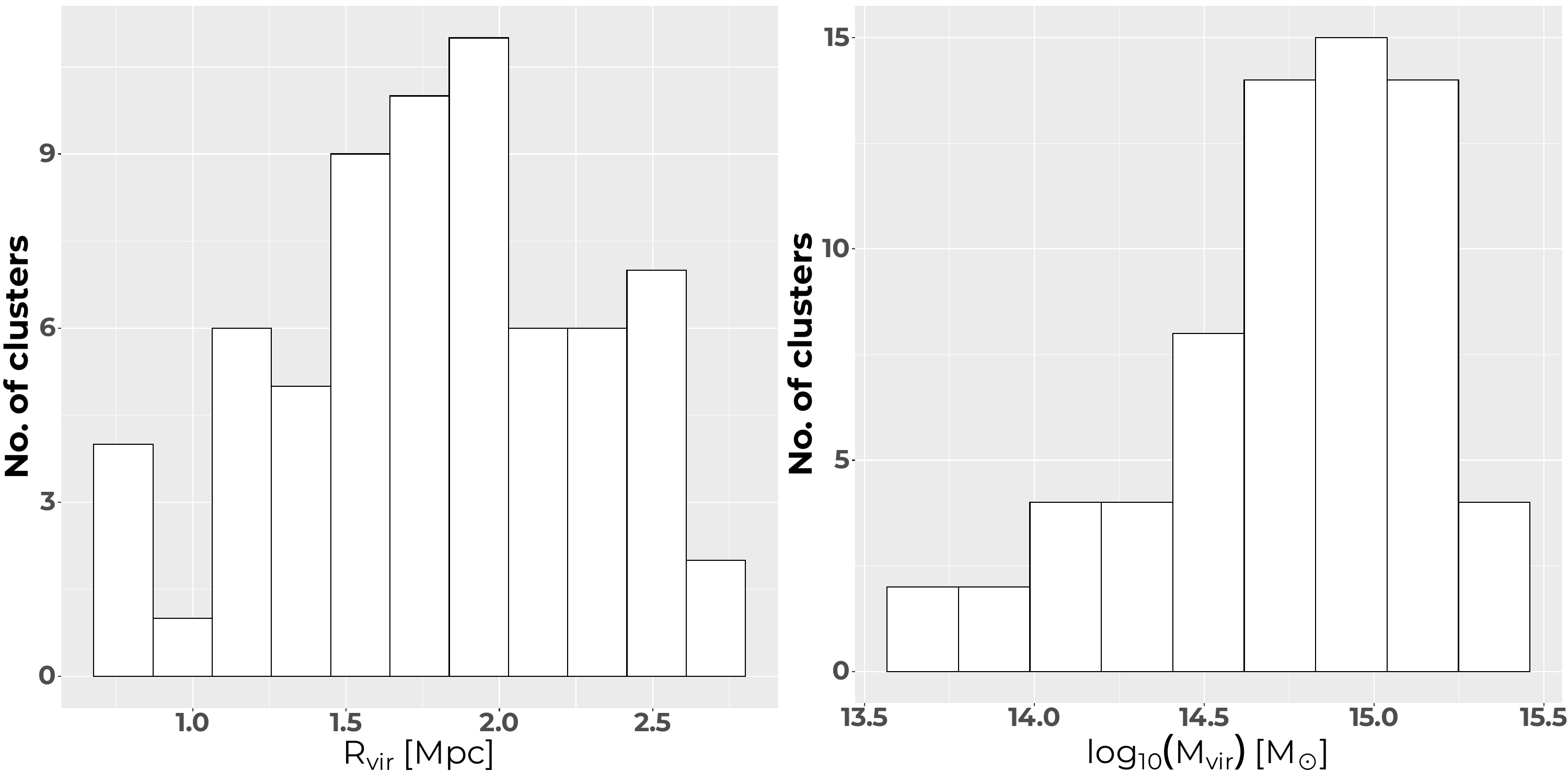}
 \caption{Distributions of virial radius (left panel) and virial mass (right panel).}
 \label{F_Global2}
\end{figure}

Based on the above distributions, we conclude that our sample is composed mostly of nearby, relatively rich clusters, 
where the concentrations of galaxy and velocity dispersion are remarkably high, justifying the assumption that those are systems that had had sufficient time to evolve internally and should then be expected to be close to virialization. Consistent with this assumption, the distribution of the virial masses in the right panel of Fig.~\ref{F_Global2} is found to be significantly negatively skewed, 
with a median 6.4 $\times 10^{14}$ $h_{70}^{-1}$ $M_{\odot}$. However, the distribution for the virial radii, in the left panel, does not follow this trend, spannig a relatively large range, with a median value of 1.82 $h_{70}^{-1}$ Mpc, corresponding to 0.85 times the R$_{\mathrm{A}}$. This peculiarity may suggest that the clusters either have different assembly states or even different assembly histories.

\begin{figure}[ht]
  \centering
  \includegraphics[width=\columnwidth]{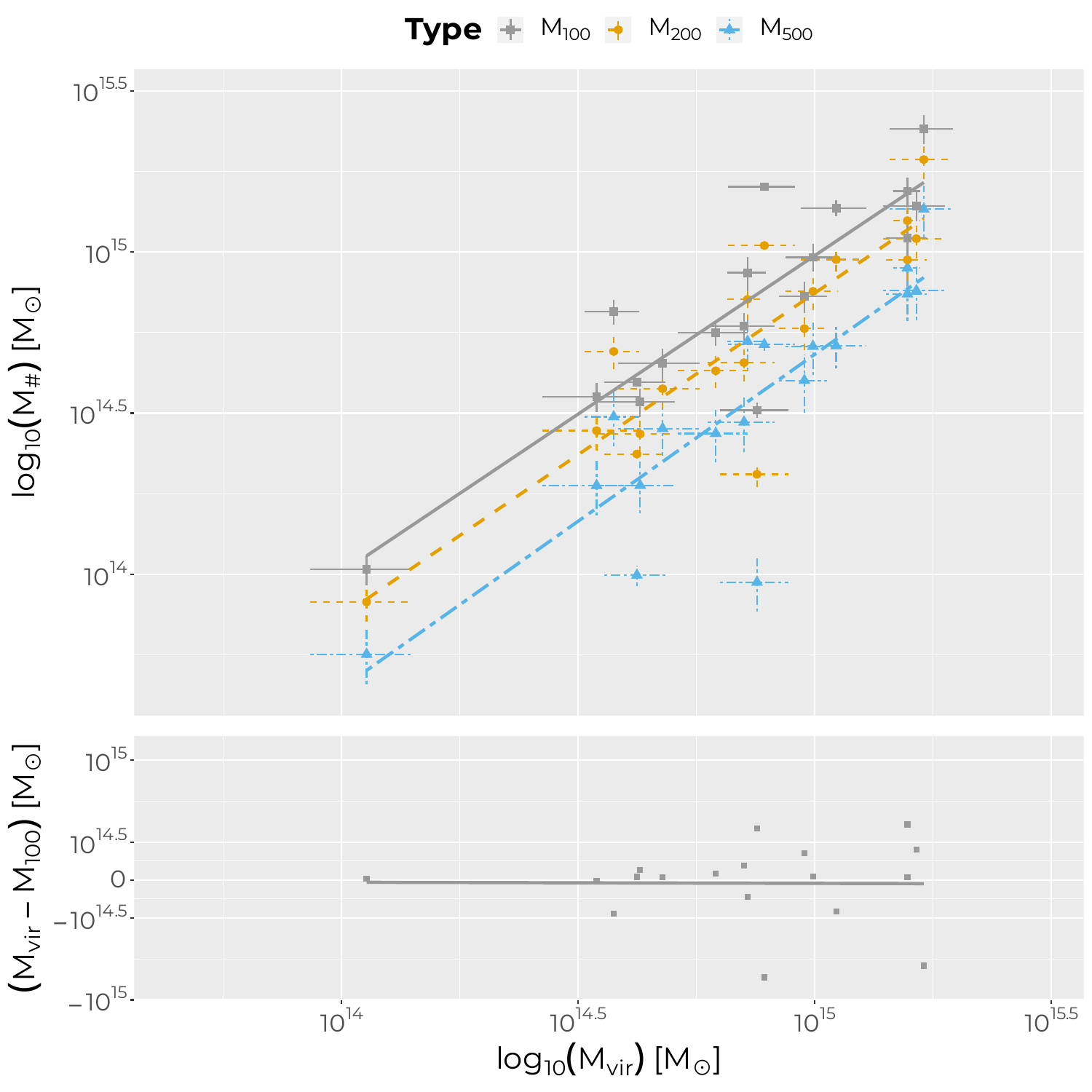}
  \caption{Comparison of our mass estimates with the virial masses in the GalWeight galaxy cluster catalog.  Blue triangles, $M_{500}$, orange circles, $M_{200}$ and gray squares $M_{100}$. The three lines are linear fits to points with the same colors. In the lower panel, the residual corresponds to the comparison of our virial mass with $M_{100}$.}
  \label{F_Mass}
\end{figure}

Comparing our virial masses with those of the literature was not straightforward, since published estimates of these are rare. One suitable source is the GalWeight cluster catalog \citep[GalWCat19;][]{Abd20}, where the masses of 1\,800 clusters were determined within three different projected radii: $r_{100}$, $r_{200}$ and $r_{500}$. 
{There are 18 clusters in this catalog that are also in our sample, and} 
we compare, in Fig.~\ref{F_Mass}, their three different masses, $M_{100}$, $M_{200}$ and $M_{500}$ with our $M_{\mathrm{vir}}$ estimate. The three linear fits we obtained are relatively good, with comparable correlation coefficients R $\sim$ 0.83, 0.87 and 0.89, respectively. However, since our virial masses were estimated using a proxy for $r_{100}$ (cf. \S~\ref{sec:Param}), the comparison that counts for us is with $M_{100}$. In the lower panel, the fit shows our virial mass estimate to be in good agreement with those of GalWCat19 for $r_{100}$, the residuals being due probably to the different ways the membership of galaxies in each cluster was determined, and the mean redshifts of galaxies in our data compared to only SDSS redshifts in theirs.

To disentangle the assembly history of these clusters, we discuss in the following sections the implications of various dynamical parameters and classifications obtained for three different internal regions within the clusters. 
The three regions are the following: 
i) the \emph{outer} region, from $R_{\mathrm{vir}}$ down to $r_{500}$, as traced by the optical data on galaxy membership; 
ii) the \emph{inner} region, inside $r_{500}$, as traced by X-ray emission; 
iii) and the innermost \emph{core}, characterized by gas cooling and {CDG} properties.
Typical radii for these regions are $R_{\mathrm{vir}} \sim 1.8 \, h_{70}^{-1}$ Mpc, $r_{500} \sim 0.9 \, h_{70}^{-1}$ Mpc and core radius ($r_{core}$) about 0.3 $h_{70}^{-1}$ Mpc (note that more specific values will be defined for the various radii as our analysis progresses).

\subsection{Outer region}
\label{sec:Out}
 
The assembly state of the outer region is characterized by the presence or absence of substructures (cf. \S~\ref{sec:Substr}). According to the DS-test, 
the 43 clusters with P$_{\mathrm{sub}} \ge$ 90\% in col.~18 of Table~\ref{T_Clusters} could be considered to have substructures. This represents 64\% of our cluster sample. However, considering all the results for the different tests, substructures appear to be secure in only 39 of these clusters and probable, with P$_{\mathrm{sub}} \lesssim $ 90\%, in 8 further clusters. In total, the number of clusters with substructures in our sample could thus be as high as 70\%. Consequently, since either fraction is relatively high, it seems safe to conclude that substructures are extremely common in nearby clusters. 

Taken at face value, the presence of numerous clusters with substructures suggests that many of these systems did not yet reach equilibrium, and what we observe, consequently, are different phases of a still ongoing process. 
To better characterize these different phases, it seems therefore important to first classify all the substructures in terms of their dynamical significance. This implies determining the gravitational impact that a substructure, when present, has on the whole cluster. 

As a first approximation, this gravitational impact can be estimated by comparing the relative richness, $N_s/N_c$, formed by the ratio of the number of galaxies in each substructure, $N_s$, to the total number of galaxies within the caustics, $N_c$. Using the numbers in Table~\ref{T_ApB} of Appendix~\ref{sec:ap-B}, we distinguish three levels of dynamical significance:

\begin{description}
\item[Main (\textsl{m}):] High relative richness, $N_s/N_c \ge 0.50$. This level characterizes the dynamically dominant substructures in any cluster. In the case of multi-modal clusters (for example, A2804), the sum of the membership ratios of the main modes is indeed larger than 0.50. In Table~\ref{T_ApB}, the substructures with this level of significance are identified by appending the suffix \textbf{m} to their ID.  

\item[Highly significant (\textsl{hs}):] Intermediate relative richness, $0.05\leq N_s/N_c < 0.50$. This level characterises substructures that are sufficiently massive to affect the dynamics of their host clusters. In Table~\ref{T_ApB} the suffixes (\textbf{n, s, e, w, c}, or a combination thereof) are added to the ID of the substructures indicating its location relative to the main structure (North, South, East, West or central, respectively). 
                          
\item[Low-significance (\textsl{ls}):] Low relative richness, $N_s/N_c < 0.05$. These are low-mass clumps of galaxies, attached to a more massive host cluster, that do not affect its dynamics. They are not listed in Table~\ref{T_ApB}. 
\end{description}

{It is worth to note that the \textsl{m} substructure in our sample with the lowest value of $N_s/N_c$ is A1736Am (0.581), while the \textsl{hs} substructure with the highest $N_s/N_c$ is A3027Acw (0.235). 
Thus, the cut in $N_s/N_c = 0.5$ seems to be a good discriminator for this separation.}
The numbers of substructures with relative richness levels \textsl{m}, \textsl{hs} and \textsl{ls} are indicated in col.~19 of Table~\ref{T_Clusters} using three numbers (\textsl{m}, \textsl{hs}, \textsl{ls}). 
For example, while A2798B and A2801 only have one main structure each, $N_{\mathrm{sub}} = (1,0,0)$, A2804, a bimodal cluster, has two, $N_{\mathrm{sub}} = (2,0,0)$. 
A more complex cluster is A0085A, which has one main structure, two~\textsl{hs} and three~\textsl{ls} substructures, $N_{\mathrm{sub}} = (1,2,3)$. A still more complex cluster is A2151, a trimodal cluster marked as $N_{\mathrm{sub}} = (3,2,2)$. 

In col.~20 of Table~\ref{T_Clusters} an extra parameter appears, $\cal{A}$, which is used to qualify the ``assembly state'' of a cluster based on its level of substructuring. 
This classification was inspired by the morphological classifications of ICM X-ray emission proposed in \citet{BeT95} and \citet{JeF99}.
We distinguish five assembly states: high-mass, Unimodal clusters (U); Low-mass unimodal (L); Multi-modal (M); Primary (P) with only  \textsl{ls} substructures attached to the \textsl{m} mode; and finally Substructured (S), formed by the \textsl{m} mode and at least one  \textsl{hs} substructure. 

The way we distinguish between U and L clusters depends on the mass criterion $3.5 \times 10^{14} M_{\odot}$: U is more massive and L less massive or equal to this mass. 
{In fact, the regular (unimodal) clusters may be either the ``beginning'' or the ``end'' of a merging process. They are the beginning if the poor clusters have had time to arrive close to relaxation, while in a relatively isolated environment. As the end, they are the final result of the virialization process of rich clusters. In fact there is no theoretical criterion justifying this distinction, and we chose pragmatically a threshold: the clusters for which we could see relatively relaxed X-ray isophotes were assumed to be close to virialization, while, in the abscence of X-ray emission (implying a less dense or colder ICM, undetectable with ROSAT sensitivity, for example), we assumed the other state.}

To be classified as M, a cluster must be formed by two or more \textsl{m} modes, with comparable richness. 
Consequently, in M clusters the {CDG} of the cluster is ill-defined, since there are {different SDGs competing for this position} (one for each mode, at least). 
For practicality, we choose as the {CDG} the {SDG of} the most central mode (usually the richest in galaxies and/or brightest in X-ray). 
This convention allows us to define a central position for the cluster and serves as reference for the magnitude gaps. 
Of the ten M-type clusters identified in Table~\ref{T_Clusters}, three, namely A0754, A2147 and A2152, show multiplicity only in the optical, while four, A1367, A2151, A2197 and A2804, show multiplicity in both optical and X-rays, and three others, A3356, A3395 and A3716, only in X-rays.

Finally, S and P clusters have both only one main structure and some substructures: in an S cluster there are \textsl{hs} substructures and in a P cluster they are all \textsl{ls}. 

Adopting the above definitions, we count 21\% U, 13\% P, 42\% S, 15\% M and 9\% L clusters. In terms of masses, Table~\ref{T_AvPar} shows that U and P clusters are more massive than S and M clusters, while L clusters are the least massive of all. 
Thus, poor L clumps could represent the building blocks of future 'massive' clusters. 
Also, the distribution in radii presented in Table~\ref{T_AvPar} reveals that the ``size'' of a cluster and, most specifically, its virial radius increases with its mass.

\begin{table*}[!t] 
 \begin{center}
 \tablecols{7}
 \setlength{\tabcolsep}{1mm}
 \setlength{\tabnotewidth}{1\columnwidth}
 \caption{Typical parameters in different assembly class clusters}
 \label{T_AvPar}
   \begin{tabular}{lrccccc}
\toprule
 \multicolumn{1}{c}{Cluster} &
 \multicolumn{1}{c}{N} &
 \multicolumn{1}{c}{$M_{\mathrm{vir}}$} & 
 \multicolumn{1}{c}{$R_{\mathrm{vir}}$} &
 \multicolumn{1}{c}{$r_{200}$} &
 \multicolumn{1}{c}{$r_{500}$} &
 \multicolumn{1}{c}{$r_{core}$} \\
\multicolumn{1}{c}{class}&
\multicolumn{1}{c}{}&
\multicolumn{1}{c}{mean(median)} &
\multicolumn{1}{c}{mean(median)} &
\multicolumn{1}{c}{mean(median)} &
\multicolumn{1}{c}{mean(median)} &
\multicolumn{1}{c}{mean(median)}   \\
\multicolumn{1}{c}{}&
\multicolumn{1}{c}{}&
\multicolumn{1}{c}{[$10^{14} M_{\odot}$]} &
\multicolumn{1}{c}{[Mpc]} &
\multicolumn{1}{c}{[Mpc]} &
\multicolumn{1}{c}{[Mpc]} &
\multicolumn{1}{c}{[Mpc]}  \\
%
\midrule
U & 14 & 9.2(7.7)  & 1.99(1.93)& 1.60(1.52)& 0.94(1.05)& 0.32(0.31) \\
P &  9 & 9.6(11.1) & 2.02(2.16)& 1.61(1.62)& 1.04(1.00)& 0.32(0.34) \\
S & 28 & 8.1(6.3)  & 1.81(1.82)& 1.38(1.39)& 0.79(0.92)& 0.29(0.29) \\
M & 10 & 6.2(6.4)  & 1.72(1.82)& 1.32(1.45)& 0.56(0.63)& 0.27(0.29) \\
L &  6 & 1.9(2.1)  & 1.18(1.25)& 0.82(0.95)& -   (-)   & 0.19(0.20)  \\
\bottomrule
   \end{tabular}
 \end{center}
\end{table*}

Although 70\% of the clusters (M, S and P) show some evidence of substructuring, considering the significance in terms of relative richness and mass, only 57\% (M and S) are expected to be dynamically affected by their substructures. 
This implies that at least 57\% of the clusters in our sample have not yet reached virialization. 
This may be compared to previous numbers reported for local cluster samples, \textit{e.g.} \citet[][]{Lop18}, who found that substructuring ranges between 37--75\%, in 
40 SZ-detected clusters, and between 32--65\%, in 
62 X-ray clusters \citep[both samples taken from][]{2017AndradeSantos}.
As a whole, for 31 clusters in common with these authors, our results agree for 80\% of them.
 
How does this classification of substructures fit the model of cluster formation? 
According to the hierarchical model, clusters form mainly by the mergers of groups of galaxies. 
Within this paradigm U-type clusters would be examples of systems that merged in the distant past and their virialization process would thus be well advanced. 
P-type clusters would also have formed in the past and represent cases that, being massive, have recently attracted small groups in their environment without an important change in their relaxation state. 
This reinforces the idea that the cluster formation process is continuous. 
Consequently, clusters with significant substructures (M and S) would be examples of relatively more recent mergers \citep[which occurred in the last 1--2~Gyr;][]{2018Lisker,2020Benavides,Hag23}. 
Their differences are explained by the importance of the merger: in S-type clusters a massive clump is accreting smaller mass groups (minor mergers), while in M-type clusters the masses of the merging entities are comparable (major mergers). 
Since average masses of M-type clusters are smaller, they could represent the previous step of the formation of the massive main clumps of S-type clusters.
By comparing the sum of the merging masses (main + substructures) with the total mass of the cluster, in S clusters the merging masses are 7\% less massive than the cluster mass (median 10\%), compared to 23\% (median 30\%) in M clusters. 
Obviously, M clusters must be relatively less relaxed than the S clusters.

The best examples of poor clusters might be the six L clusters. 
Indeed, the relatively low masses and small numbers of member galaxies in these systems make them comparable to groups. 
This might also explain why these poor clusters are not observed in X-ray: simply because they do not have sufficiently deep potential wells for infalling gas to heat up and emit detectable amounts of X-rays. 
This is the case of candidate AXU clusters like A0634 and A4012, for which confirmation should be obtained using eROSITA \citep{Pre21}. 
The four remaining L clusters also look like they could be either infalling or satellite groups of more massive clusters (see Table~\ref{T_Pairs}): these are the cases of S0334 related to A3104, S0336 related to A3112B, and possibly A3095 related to A3094A and S0906 related to A3716 (for which binding could not be established). 
Other infalling groups, composed by two clumps either and residing well inside the caustics of their \textbf{main} cluster are: A2870, related to \textbf{A2877}, and A4049, related to \textbf{A4038}. 
These cases would also be excellent candidates to search for evidence, in their galaxies' properties, of pre-processing. 
 

\subsection{Inner region}
\label{sec:Inn}

The region inside $r_{500}$ is where the gas accumulates and gets very hot, emitting in X-rays. In our sample, only eight clusters are undetected in X-rays (Table~\ref{T_Sample}). 
However, quantitative information on ICM evolution is scarce and restricted to two parameters: dynamical status of the cluster ICM (71\% of X-ray detected clusters) and core cooling status (81\%). 
In Table~\ref{T_Assembl} the values for these parameters in clusters with different $\cal{A}$ classes are reported in cols.\ 2 and 4, respectively. 
In col.~2, the symbol {\checkmark} indicates a relaxed status while an {$\ast$} indicates a disturbed ICM. 
The main sources for this information were \citet{Par15} and \citet{Lag19}.
In general, conclusions about the dynamical state of the inner region are confirmed by the different sources, except in four cases, the clusters A2029, A2052, A2063A and A2244, for which evidence of disturbance is still debated.
In col.~3 of Table~\ref{T_Assembl} we added information about the detected presence of radio halos and/or radio relics, mostly from \citet{vWe19}, \citet{Kno22} and \citet{Bot22}.
For the 4 cases above, these data indicate disturbance for A2029, A2063A and A2244, and no information concerning diffuse radio emission for A2052. 
In general, the presence of radio halos and/or relics coincide with the disturbed status of the ICM X-ray data (except for two substructured clusters, A0133A and A1656).

Although the cooling state is related to the core of the clusters, we will discuss it here, together with the remaining X-ray information.
In col.~4 of Table~\ref{T_Assembl}, the core cooling status is codified as follows \citep[see, for example,][]{Kaf19}: 
strong-cool-cores (SCC), which have cooling times $t_{cool} < 1$ Gyr and usually show a temperature drop towards the center and small core radii \citep[$<$ 100 kpc; \textit{e.g.},][]{Ota06}; 
non-cool-cores (NCC), with $t_{cool} > 7.7$ Gyr \citep[see also][]{Hud10}, with flat central temperature profiles and large core radii; 
and weak-cool-cores (WCC), with intermediate characteristics (although sometimes also classified as cool-cores). 
In some clusters the index 's' is also added when ``sloshing'' (the presence of spiral-shaped central cold fronts) is observed. 
In terms of assembly state, clusters with SCC or WCC are expected to be closer to relaxation, that is, they had enough time to settle and radiatively cool. 
NCC clusters, on the other hand, are likely to be younger, or disturbed due to recent mergers, or even re-heated by AGN feedback. Finally, evidence of sloshing in these cores might suggest more recent accretion.
 
How do these classifications in X-ray fit our general interpretation based on the assembly classes, U, P, S and M? 
The best agreement is for the M class: 9 out of 9 clusters in Table~\ref{T_Assembl} are considered disturbed according to the X-ray emission (col.~2), while 5 are NCC and 1 is WCC (col.~4) of a total of 6 with this information. 
Although the situation is more complex for the S and P clusters, the agreement is also relatively good. For the S-type clusters, 11 out of 17 are considered disturbed based on the X-ray distribution, while 11 are NCC, 4 are WCC and only 7 are SCC. 
However, considering the evidence of sloshing in five of the SCC, as much as 20 out of 22 S-type clusters could be considered to have a non-relaxed ICM. 
Note that the diagnostics based on X-ray distribution and core temperature differ in five cases, the ambiguity increasing for the WCC and SCC. 
This ambiguity appears clearer in the P-type clusters: although 7 out of 7 have non-relaxed ICM, 5 out of 9 are WCC and 4 are SCC with sloshing. Considering their particular assembly state  histories --these are old and massive clusters accreting smaller mass systems-- some level of ambiguity in the core cooling status might naturally be expected.
The U class, however, is definitely the most surprising. 
Although we expect all of these clusters to be close to relaxation, based on the X-ray distribution, only 2 out of 9 seem to be, 5 are suggested to be disturbed and 2 are ambiguous. 
The core cooling states draw a similar complex picture: 4 are NCC (usually also disturbed), 5 are WCC (one with sloshing), and 2 are SCC (also one with sloshing). 
This is relatively unexpected, since most of our U-type clusters lying at low redshifts should have had time to reach relaxation through interactions. 

However, the fact that very few clusters are classified as U, combined with the ``unusual'' characteristics of their inner regions compared to their outer region (absence of substructures), suggest that the process of virialization, even in the {most} evolved systems, does not depend solely on time but also on complex processes involved in their assembly history. 
Merging can happen anytime in the history of a cluster --possibly taking it out of a previous equilibrium situation. Also, merging of major and minor subclusters have different consequences, 
{and the same applies to} 
different ICM properties.
Thus, different regions sampled by optical and X-ray observation may show distinct moments of this assembly history.
In fact,  
this could explain the frequent disagreement between optical and X-ray results concerning the dynamical state of galaxy clusters.

\subsection{Core region} 
\label{sec:Core}

In this innermost region, the most 
relevant feature is 
the {CDG}, and possibly other dominant galaxies. Consistent with its definition, the {CDG} position in a cluster is expected to indicate its dynamical center. This is also the expectation for the X-ray emission peak (or centroid), although the two components, galaxies and gas, may be subject to different levels of disturbance with respect to the global potential well, dominated by dark matter. This is why the dynamical status of the {CDG}, with respect to the gas distribution and to the radial velocity distribution of galaxies, is an important information to compare with the assembly status discussed above.  

\begin{figure}[ht] 
 \centering
  \includegraphics[width=0.5\textwidth]{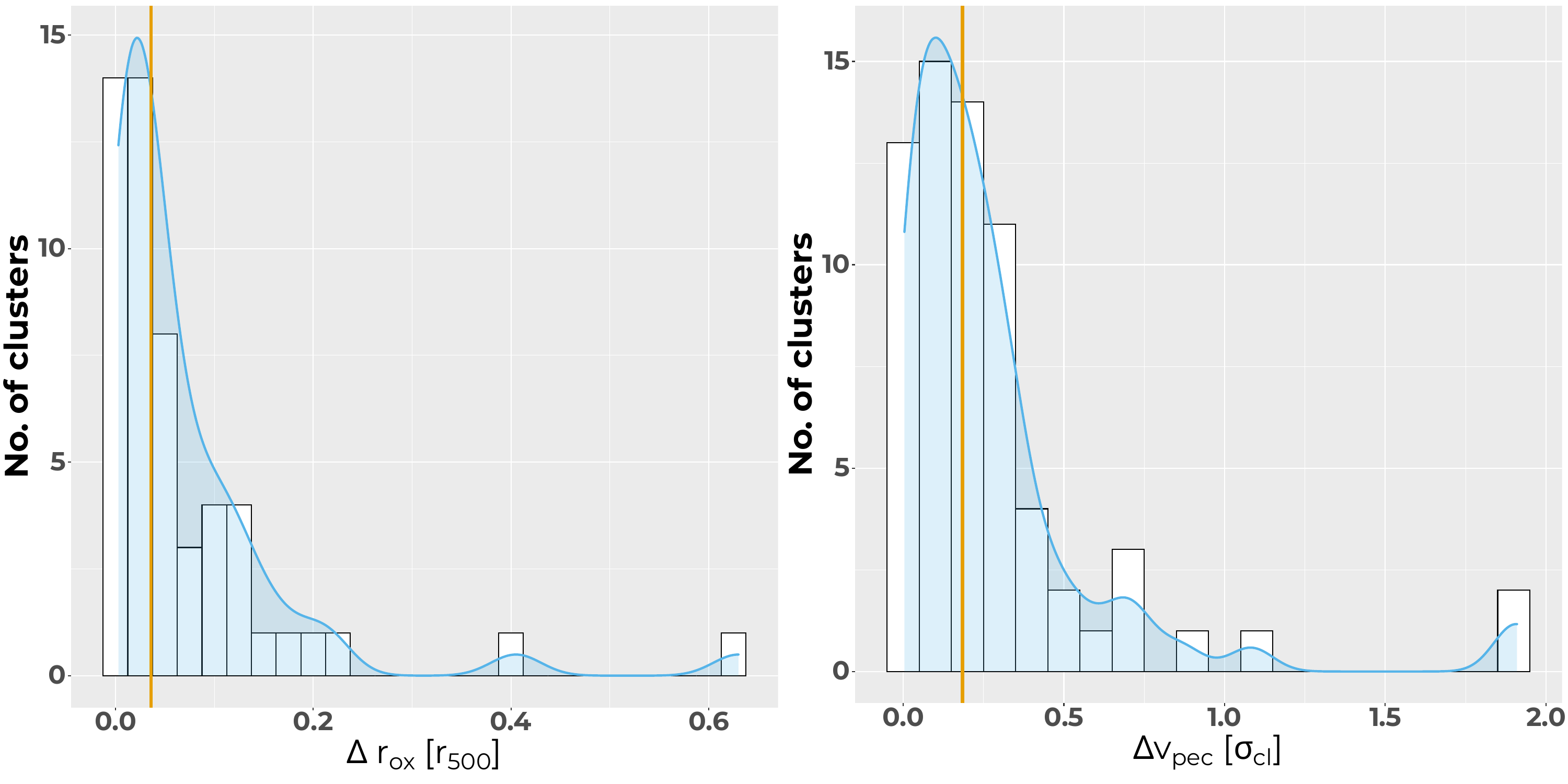}
 \caption{{Left: Distribution of the relative projected positional offset of {CDGs} with respect to the X-ray peak ($ \Delta r_{\mathrm{ox}}$).  Right: distribution of the relative peculiar velocity of {CDGs} with respect to the cluster systemic velocity ($\Delta v_{\mathrm{pec}}$)}. 
The vertical yellow lines in the two panels are the thresholds separating relaxed form non-relaxed clusters.
}
 \label{F_FRG1}
\end{figure}

In the left panel of Fig.~\ref{F_FRG1}, 
the distribution of $\Delta r_{\mathrm{ox}}$ for the total sample is shown, with an assumed upper threshold for relaxed {CDGs}, $\Delta r_{\mathrm{ox}} = 0.03$, marked as a yellow vertical line. 
We see that 55\% of the clusters (29 out of 53) are well above the threshold. The median for $\Delta r_{\mathrm{ox}}$ is 0.04 or about  $r_{\mathrm{ox}} = 40 h_{70}^{-1}$ kpc. This suggests that more than half of the clusters in our sample have a non-relaxed {CDGs}.  

In the right panel of Fig.~\ref{F_FRG1}, we trace the distribution for $\Delta v_{\mathrm{pec}}$. By definition of the {CDG}, in a dynamically relaxed system one would expect $\Delta v_{\mathrm{pec}}$ to tend to zero. However, assuming a typical upper threshold $\Delta v_{\mathrm{pec}} =$ 0.21, the percentage of clusters that have higher values is as high as 46\% (31 out of 67). With a median $\Delta v_{\mathrm{pec}} =$ 0.185 $\sigma_{\mathrm{cl}}$, corresponding to a median $v_{\mathrm{pec}}$ of $\pm$ 147 km s$^{-1}$, once again it is clear that a high number of clusters cannot be assumed to {have relaxed CDGs}. 

\onecolumn
\begin{ThreePartTable}
  \begin{TableNotes}\footnotesize
\item[a] Codes for X-ray (inner region) and Offsets (core) are: 
             [\checkmark] Relaxed; 
             [$\ast$] Disturbed; 
             [$-$] No Data.
             {References for ICM dynamical states are: \citet{Sch01,Rin01,Par15,Vik09,Ich19,Lag19,Tiw21}}.
\item[b] Codes for diffuse radio emission are:
             mini Radio-Halo [mH], Radio-Halo [H], Radio-Relic (shock) [R] and both halo and relic [HR].
\item[c] Codes for cooling status of the core are:
             Strong cool-core [S], Weak-cool-core [W], Non-cool-core [N]; 
             [s] indicates cold gas sloshing (cold gas front) is detected.
             {References for core cooling status are: \citet{Whi01,Fin01,Che07,Sat10,LeS14,Lov15,Kaf19}}.
\item[d] Codes are: 
             [u] unique {CDG} (choice of {CDG} is obvious, $\Delta m_{12} > 0.5$);
             [b] binary central dominant BCGs ({CDG} is the most central or brightest), may also be a ``Coma-like'' system (BCG is brighter than {CDG}); 
             [db] {CDG} is dumbell type;
             [far] binary dominant BCGs, but 2$^{nd}$-rank is far out of core radius;
             [m] multiple central dominant BCGs (3 or more BCGs inside 0.5 magnitud range);
             [n] weakly dominant {CDG} (giant elliptical);
             [fos] fossil group candidate BCG;
             [BSp] there is a bright spiral among the BCGs.
  \end{TableNotes}
\begin{center}
\begin{scriptsize}
 \setlength{\tabcolsep}{1mm}
\begin{longtable}{lrrlr rrrrrr r} 
 \caption{Assembly state of the clusters.}
 \label{T_Assembl} \\
\toprule
  \multicolumn{1}{l} {ID$_{\mathrm{cl}}$} &
  \multicolumn{1}{c} {Inner\tnote{a}} &
  \multicolumn{1}{l} {Radio\tnote{b}} &
  \multicolumn{1}{c} {CC\tnote{c}} & 
  \multicolumn{1}{c} {$\Delta r_{\mathrm{ox}}$} &
  \multicolumn{1}{c} {$\Delta v_{\mathrm{pec}}$} &
  \multicolumn{1}{c} {Offsets\tnote{a}} &
  \multicolumn{1}{r} {$\Delta m_{12}$} &
  \multicolumn{1}{r} {$\Delta r_{12}$} &
  \multicolumn{1}{r} {$\Delta m_{23}$} &
  \multicolumn{1}{r} {$\Delta r_{13}$} &
  \multicolumn{1}{r} {Comments\tnote{d}} \\
%
(1) & (2) & (3) & (4) & (5) & (6) & (7) & (8) & (9) & (10) & (11) & (12) \\
%
\midrule
\endfirsthead
\toprule
  \multicolumn{1}{l} {ID$_{\mathrm{cl}}$} &
  \multicolumn{1}{c} {Inner\tnote{a}} &
  \multicolumn{1}{l} {Radio\tnote{b}} &
  \multicolumn{1}{c} {CC\tnote{c}} &
  \multicolumn{1}{c} {$\Delta r_{\mathrm{ox}}$} &
  \multicolumn{1}{c} {$\Delta v_{\mathrm{pec}}$} &
  \multicolumn{1}{c} {Offsets\tnote{a}} &
  \multicolumn{1}{r} {$\Delta m_{12}$} &
  \multicolumn{1}{r} {$\Delta r_{12}$} &
  \multicolumn{1}{r} {$\Delta m_{23}$} &
  \multicolumn{1}{r} {$\Delta r_{13}$} &
  \multicolumn{1}{r} {Comments\tnote{d}} \\
(1) & (2) & (3) & (4) & (5) & (6) & (7) & (8) & (9) & (10) & (11) & (12) \\
%
\midrule
\endhead
\hline \multicolumn{12}{|r|}{{Continued on next page}} \\ \hline
\endfoot
\insertTableNotes
\endlastfoot 
&&&&& U &&&&&& \\
\midrule
  A0122     & $-$              & $-$ & $-$ & 0.062    & 0.342 & $\ast$     &  0.000 & 0.01 & 1.618  & 0.41 & b(db) \\   
  A0399     & $\ast$           & R   & N   & 0.099    & 0.331 & $\ast$     &  0.916 & 0.37 & 0.433  & 0.65 & u \\
  A0401     & $\ast$           & R   & N   & 0.015    & 0.214 & $\ast$     &  1.056 & 1.08 & 0.133  & 0.74 & u \\
  A1650     & \checkmark       & $-$ & W   & 0.025    & 0.100 & \checkmark &  0.512 & 1.06 & 0.082  & 1.07 & u \\
  A1795     & \checkmark       & mH? & S   & 0.011    & 0.095 & \checkmark &  0.636 & 0.91 & 0.015  & 0.96 & u \\
  A2029    & \checkmark/$\ast$ & mH  & S,s & 0.100    & 0.301 & $\ast$     &  1.954 & 0.22 & 0.161  & 0.93 & u \\
  A2065     & $\ast$           & H?  & W   & 0.045    & 0.044 & $\ast$     &  0.000 & 0.02 & 0.054  & 0.09 & m(db),For \\
  A2244    & $\ast$/\checkmark & H   & W   & 0.013    & 0.184 & \checkmark &  1.025 & 1.11 & 0.186  & 0.92 & u \\
  A2670     & $-$              & $-$ & W   & 0.035    & 0.342 & $\ast$     &  1.131 & 0.71 & 0.183  & 0.93 & u \\
  A2798B    & $-$              & R   & $-$ & 0.129    & 0.244 & $\ast$     &  0.000 & 0.34 & 0.172  & 0.35 & m(db) \\
  A2801     & $-$              & $-$ & N   & \nodata  & 0.025 & $\ast$     &  1.125 & 0.39 & 0.080  & 0.60 & u \\
  A3094A    & $-$              & $-$ & $-$ & 0.215    & 0.019 & $\ast$     &  0.855 & 0.23 & 0.546  & 0.88 & u \\
  A3391     & $\ast$           & $-$ & N   & 0.027    & 0.481 & $\ast$     &  0.000 & 0.02 & 1.502  & 0.87 & b(db) \\ 
  A3562     & $\ast$           & H   & W,s & 0.046    & 0.211 & $\ast$     &  0.330 & 1.07 & 0.795  & 0.45 & u \\
\midrule \midrule
&&&&& P &&&&&& \\
\midrule
  A0426A    & $\ast$           & mH  & S,s & 0.003    & 0.029 & \checkmark &  0.436 & 0.11 & 0.407  & 0.07 & b \\
  A1060     & $-$              & $-$ & W/S & 0.007    & 0.166 & \checkmark &  0.230 & 0.03 & 0.005  & 0.46 & b,BSp \\ 
  A1644     & $\ast$           & $-$ & S,s & 0.040    & 0.123 & $\ast$     &  1.186 & 0.67 & 0.295  & 0.67 & u \\
  A1651     & $\ast$           & $-$ & W   & 0.023    & 0.177 & \checkmark &  1.128 & 0.58 & 0.097  & 0.35 & u \\
  A2063A   & $\ast$/\checkmark & R   & W   & 0.018    & 0.040 & \checkmark &  0.870 & 1.03 & 0.014  & 1.44 & u \\
  A2142     & $\ast$           & mH  & W,s & 0.030    & 0.241 & $\ast$     &  0.295 & 0.17 & -0.075 & 0.61 & b \\
  A2199     & $\ast$           & $-$ & S,s & 0.006    & 0.256 & $\ast$     &  1.271 & 1.14 & 0.179  & 0.55 & u \\
  A3526A    & $-$              & $-$ & S,s & 0.005    & 0.078 & \checkmark &  1.223 & 0.74 & 0.015  & 0.82 & u \\
  A3558     & $\ast$           & H   & W,s & 0.023    & 0.402 & $\ast$     &  0.801 & 0.49 & 0.273  & 0.15 & u \\
\midrule \midrule
&&&&& L &&&&&& \\
\midrule
  A0634     & $-$              & $-$ & $-$ & \nodata  & 0.019 & \checkmark &  0.332 & 0.29 & 0.214  & 0.75 & n \\
  A3095     & $-$              & $-$ & $-$ & \nodata  & 0.697 & $\ast$     &  0.405 & 0.39 & 0.428  & 0.37 & n \\
  A4012A    & $-$              & $-$ & $-$ & \nodata  & 0.013 & \checkmark &  1.947 & 0.93 & 0.018  & 0.27 & u \\
  S0334     & $-$              & $-$ & $-$ & \nodata  & 0.066 & \checkmark &  0.359 & 0.04 & 0.587  & 0.10 & n,For,BSp \\
  S0336     & $-$              & $-$ & $-$ & \nodata  & 0.581 & $\ast$     &  0.514 & 0.38 & 0.352  & 0.23 & n,Sp \\
  S0906     & $-$              & $-$ & $-$ & \nodata  & 1.082 & $\ast$     &  0.732 & 0.20 & 0.367  & 0.47 & u,BSp \\
\midrule \midrule
&&&&& S &&&&&& \\
\midrule
  A0085A    & $\ast$           & R   & S   & 0.007    & 0.031 & \checkmark &  1.532 & 0.54 & 0.059  & 0.49 & u \\
  A0118     & $-$              & $-$ & $-$ & \nodata  & 0.313 & $\ast$     & -0.228 & 0.12 & 0.467  & 0.49 & b,For \\
  A0119     & $\ast$           & $-$ & N   & 0.133    & 0.024 & $\ast$     &  0.532 & 0.13 & 0.194  & 0.75 & u \\
  A0133A    & \checkmark       & R   & S,s & 0.036    & 0.255 & $\ast$     &  1.506 & 0.17 & 0.036  & 1.23 & u,For \\
  A0400     & $\ast$           & $-$ & N   & 0.061    & 0.450 & $\ast$     &  0.000 & 0.01 & 1.175  & 0.10 & b(db) \\ 
  A0496     & $\ast$           & $-$ & S,s & 0.009    & 0.114 & \checkmark &  1.126 & 0.85 & 0.087  & 0.90 & u \\
  A0539     & $-$              & $-$ & N   & 0.008    & 0.520 & $\ast$     &  0.000 & 0.01 & 0.404  & 0.28 & m(db) \\ 
  A0576     & $\ast$           & $-$ & W   & 0.102    & 0.094 & $\ast$     &  0.000 & 0.02 & 1.088  & 0.14 & b(db) \\
  A1656     & \checkmark       & HR  & N   & 0.051    & 0.157 & $\ast$     & -0.534 & 0.20 & 1.117  & 0.70 & b,BSp \\
  A1736A    & $-$              & $-$ & $-$ & \nodata  & 0.017 & \checkmark &  0.142 & 0.73 & 0.798  & 1.16 & b(far) \\
  A1736B    & $\ast$           & $-$ & N   & 0.630    & 0.117 & $\ast$     &  0.916 & 0.98 & 0.369  & 0.48 & u \\
  A2040B    & $-$              & $-$ & $-$ & \nodata  & 0.283 & $\ast$     &  0.727 & 0.51 & 0.080  & 0.40 & u,For \\ 
  A2052    & $\ast$/\checkmark & $-$ & S,s & 0.010    & 0.125 & \checkmark &  0.928 & 0.92 & 0.072  & 0.22 & u \\
  A2204A    & \checkmark       & mH  & S,s & 0.037    & 0.024 & $\ast$     &  0.000 & 0.57 & 0.646  & 0.37 & m(db) \\ 
  A2255     & $\ast$           & HR  & N   & 0.172    & 1.904 & $\ast$     &  0.122 & 0.08 & 0.006  & 0.60 & m \\
  A2256     & $\ast$           & HR  & N   & 0.115    & 0.163 & $\ast$     &  0.107 & 0.18 & 0.000  & 0.07 & m(db) \\
  A2634     & \checkmark       & $-$ & W   & 0.069    & 0.204 & $\ast$     &  0.102 & 0.79 & 0.914  & 0.55 & u \\
  A2811B    & $-$              & H   & N   & 0.005    & 0.130 & \checkmark &  0.828 & 1.00 & 0.049  & 1.09 & u \\
  A2877-70  & $-$              & $-$ & W   & 0.044    & 0.100 & \checkmark &  1.231 & 0.59 & 0.125  & 0.37 & u \\
  A3027A    & $-$              & $-$ & $-$ & 0.158    & 0.061 & $\ast$     & -0.146 & 0.57 & 0.977  & 0.70 & b(far),For \\ 
  A3104     & $-$              & $-$ & $-$ & 0.040    & 0.195 & $\ast$     &  0.927 & 0.48 & 0.049  & 0.28 & u \\
  A3112B    & \checkmark       & $-$ & S   & 0.012    & 0.125 & \checkmark &  1.070 & 0.91 & 0.646  & 0.74 & u \\
  A3158     & $\ast$           & $-$ & N   & 0.018    & 0.361 & $\ast$     &  0.268 & 0.08 & 0.755  & 0.18 & b \\
  A3526B    & $-$              & $-$ & $-$ & \nodata  & 0.133 & \checkmark &  0.450 & 0.73 & 1.099  & 0.74 & u \\
  A3530     & $-$              & $-$ & N   & 0.083    & 0.185 & $\ast$     &  1.002 & 0.06 & 0.798  & 0.89 & b \\
  A3532     & \checkmark       & $-$ & N   & 0.096    & 0.867 & $\ast$     &  0.000 & 0.07 & 0.952  & 0.49 & b(db) \\ 
  A4038A-49 & $\ast$           & R   & W   & 0.016    & 0.259 & $\ast$     &  0.141 & 0.14 & 0.443  & 0.13 & b,For \\ 
  S0373     & $-$              & $-$ & S,s & 0.004    & 0.022 & \checkmark &  0.159 & 0.43 & 0.306  & 0.06 & u,For,BSp \\ 
\midrule \midrule
&&&&& M &&&&&& \\
\midrule
  A0754     & $\ast$           & HR  & N   & 0.209    & 0.223 & $\ast$     &  0.933 & 0.42 & 0.139  & 0.41 & u \\
  A1367     & $\ast$           & R   & N   & 0.405    & 0.301 & $\ast$     &  0.511 & 0.66 & 0.030  & 0.86 & b(far) \\
  A2147     & $\ast$           & $-$ & N   & 0.128    & 0.303 & $\ast$     &  0.416 & 0.18 & 0.238  & 0.55 & b,fos \\ 
  A2151     & $\ast$           & $-$ & $-$ & 0.005    & 1.910 & $\ast$     & -0.252 & 0.07 & 0.121  & 0.37 & m,For \\ 
  A2152A    & $-$              & $-$ & $-$ & 0.073    & 0.004 & $\ast$     &  0.408 & 0.76 & 0.541  & 0.80 & n \\
  A2197     & $\ast$           & $-$ & W   & 0.033    & 0.672 & $\ast$     &  0.823 & 1.07 & 0.409  & 0.95 & u,For \\ 
  A2804     & $\ast$           & $-$ & N   & \nodata  & 0.214 & $\ast$     &  0.051 & 0.50 & 0.165  & 0.65 & b(far) \\
  A3395     & $\ast$           & R   & N   & 0.012    & 0.391 & $\ast$     &  0.228 & 0.59 & 0.151  & 0.51 & b(far) \\
  A3556     & $\ast$           & $-$ & $-$ & \nodata  & 0.054 & $\ast$     &  0.339 & 0.75 & 0.315  & 0.23 & b(far) \\
  A3716     & $\ast$           & $-$ & $-$ & \nodata  & 0.738 & $\ast$     &  0.088 & 0.30 & 0.019  & 0.64 & m \\
\bottomrule
\end{longtable}
\end{scriptsize}
\end{center}
\end{ThreePartTable}
\twocolumn

Note that, compared with the literature \citep[\textit{e.g.},][]{Coz09, Lau14, Lop18}, our median value for $\Delta v_{\mathrm{pec}}$ is relatively low. This is not due to a difference in sample but rather to a difference in the identification of the {CDG}. For example, in the cluster A2197, \citet{Lau14} assumed NGC\,6173 is the BCG, instead of NGC\,6160, which we identified as the real {CDG}. Since NGC\,6173 turned out to be the {SDG} of a substructure of A2197 (A2197me in Appendix~\ref{sec:ap-B}), its $v_{\mathrm{pec}}$ is naturally estimated to be higher than for NGC\,6160. This emphasizes that a thorough analysis of the substructures in clusters is necessary to better determine the assembly state of the clusters. However, despite our careful analysis, the upgraded cluster peculiar velocities and velocity dispersion, we must still conclude that a significantly high number of nearby clusters do not have a relaxed core. 

\begin{figure}[h] 
\setlength{\unitlength}{1cm}
\includegraphics[width=0.5\textwidth]{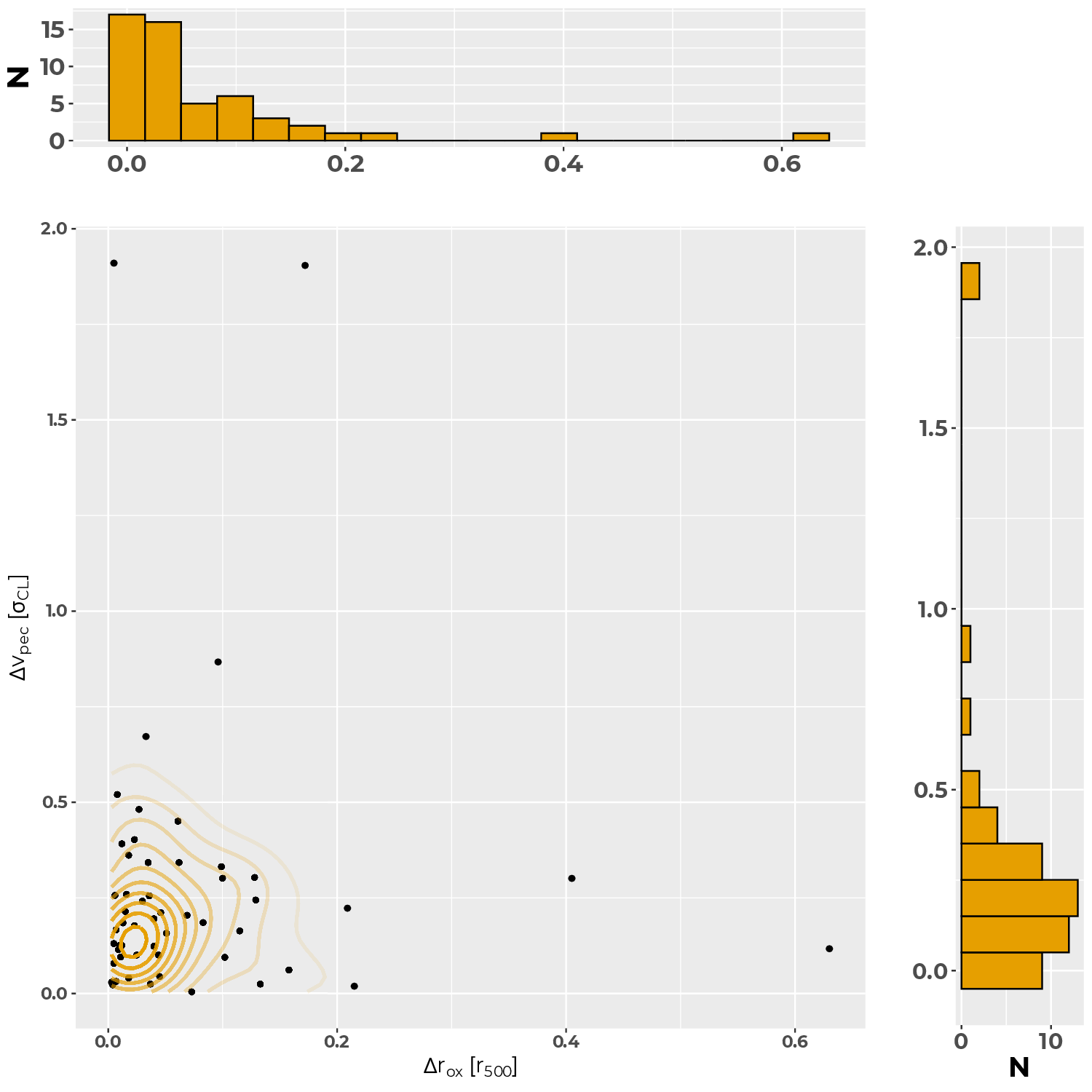}
 \caption{Distribution of offsets, parameterized.}
 \label{F_Offs2}
\end{figure}

In fact, considering the clusters individually or in any of the assembly state class, we found no correlation between $\Delta r_{\mathrm{ox}}$ and $\Delta v_{\mathrm{pec}}$, as can be seen in Fig.~\ref{F_Offs2} (compare, also, col.~21 with col.~22 of Table~\ref{T_Clusters}), a fact already noted in the literature \citep[\textit{e.g.},][]{Lau14,DeP21}. 
This advocates against the use of only one of these parameters as the proxy for the shift from the bottom of the cluster potential well, as proposed by, \textit{e.g.}, \citet{Lop18}. In the present work we consider both together as indicators of the displacement of the {CDG} with respect to the bottom of the cluster potential well. Thus, we find that 70\% of the clusters in our sample present significant disturbance in their mere core.
The parameters $\Delta r_{\mathrm{ox}}$ and $\Delta v_{\mathrm{pec}}$ are reported respectively in cols.~5 and 6 of Table~\ref{T_Assembl}.
In col.~7, both offsets are used to classify the 
state of the {CDG}, adopting the same code as for col.~2, that is, the mark $\ast$ is assigned when any of them indicates dynamical disturbance.
 

\begin{figure}
 \centering
\setlength{\unitlength}{1.cm}
\begin{minipage}[t]{5.9cm}
  \begin{picture}(5.5,5.5)
\includegraphics[width=\textwidth]{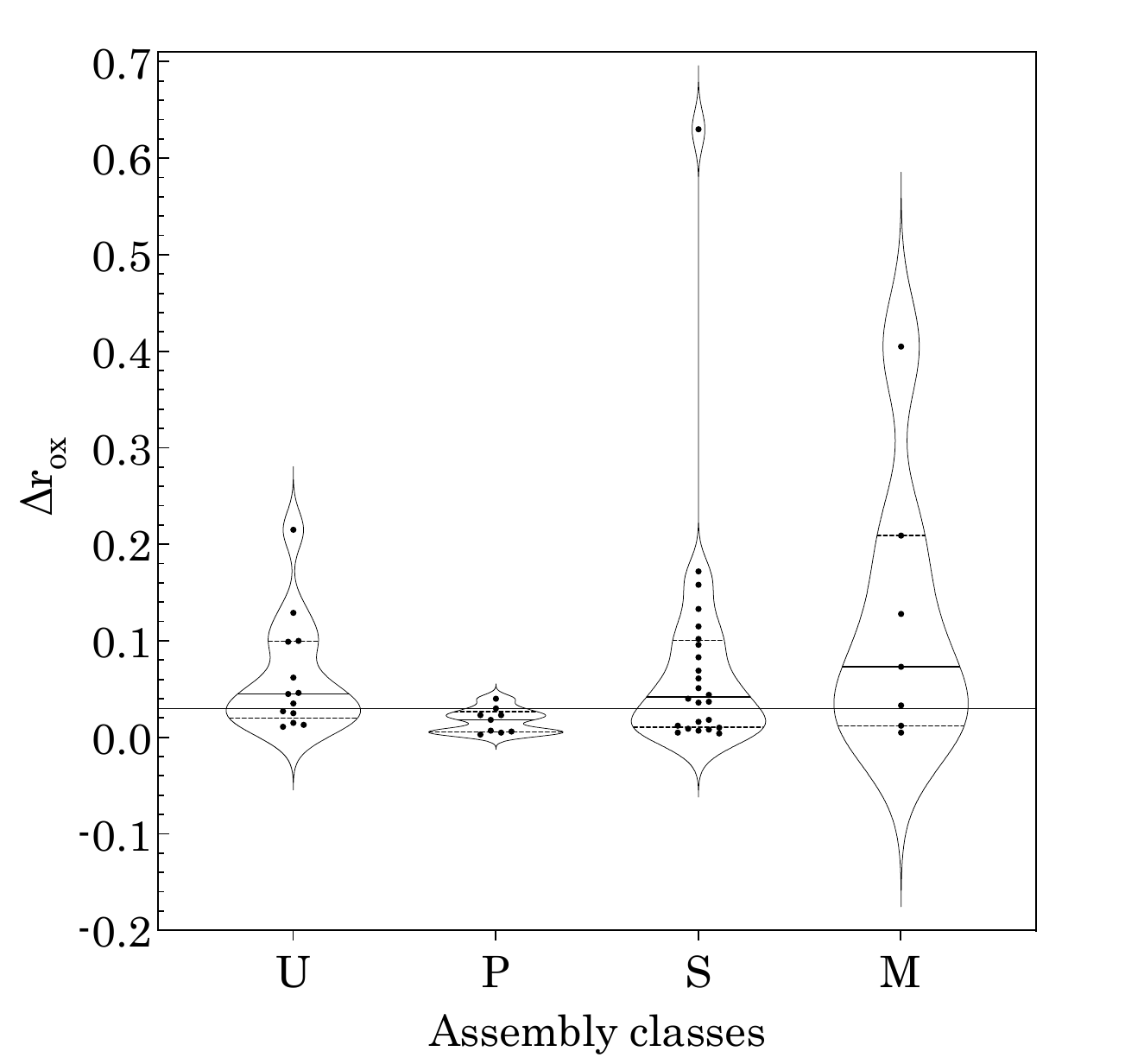}
  \end{picture}\par
 \end{minipage}
\hfill
\begin{minipage}[t]{5.9cm}
  \begin{picture}(5.5,5.5)
\includegraphics[width=\textwidth]{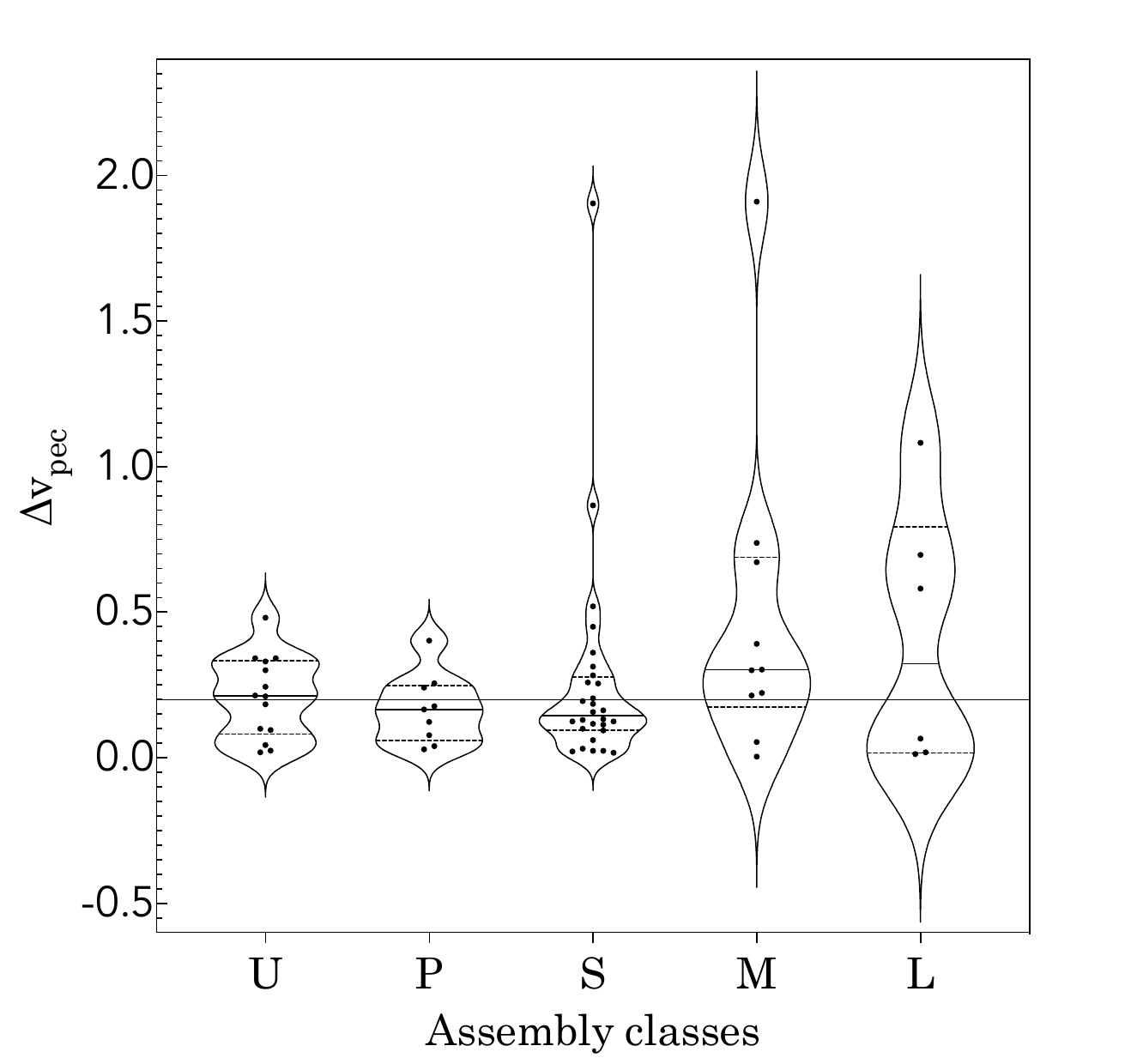}
  \end{picture}\par
 \end{minipage}
\hfill
 \begin{minipage}[t]{5.9cm}
  \begin{picture}(5.5,5.5)
\includegraphics[width=\textwidth]{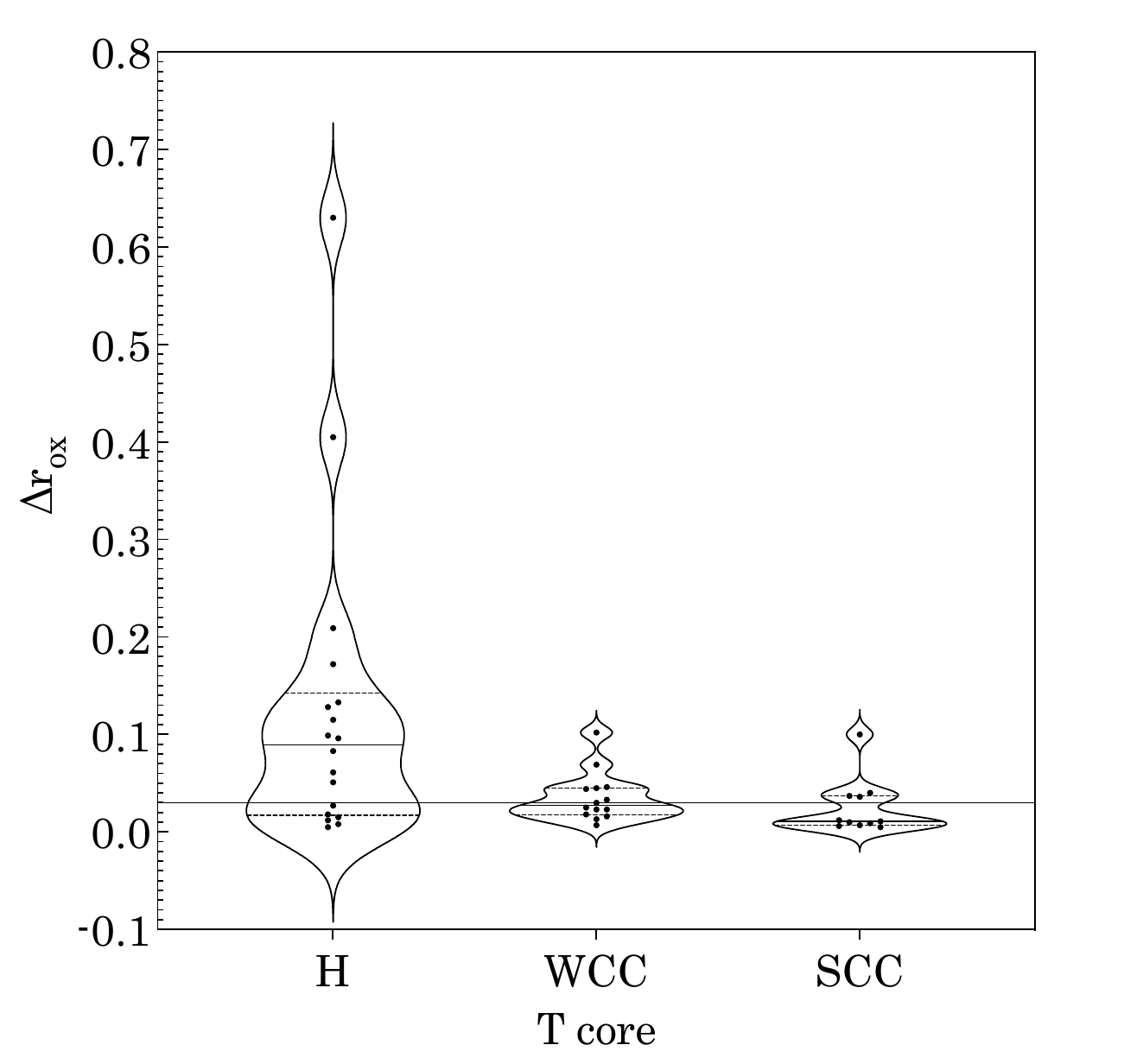}
  \end{picture}\par
 \end{minipage}
\hfill 
 \begin{minipage}[t]{5.9cm}
  \begin{picture}(5.5,5.5)
\includegraphics[width=\textwidth]{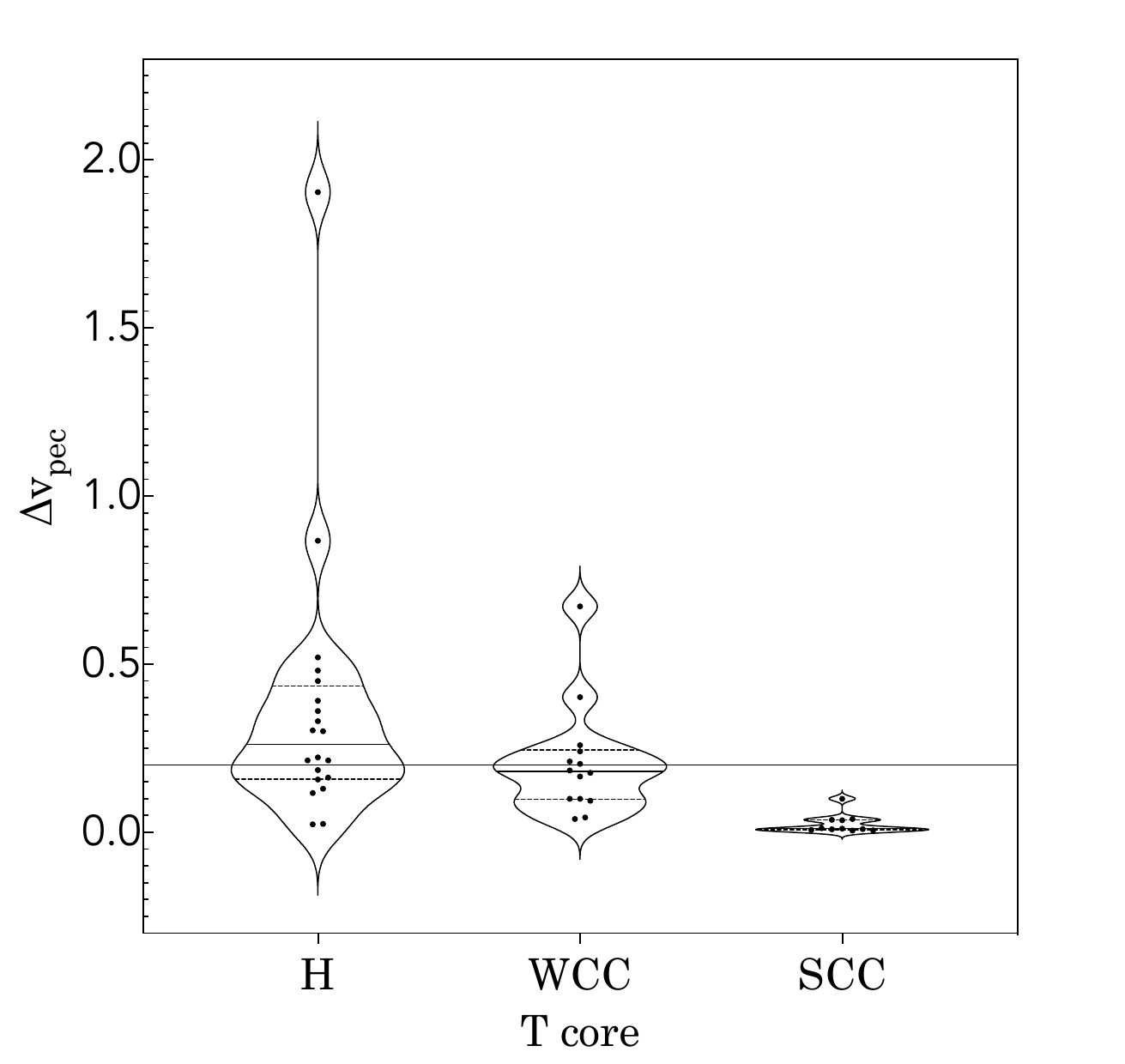}
  \end{picture}\par
 \end{minipage}
 \caption{Distributions of {CDG} parameters in different assembly classes (upper panel) and different core cooling status (lower panel): 
 Left panels, $\Delta r_{ox}$, 
 right panels, $\Delta v_{pec}$.
 In each graphic, the threshold for relaxation associated to the parameter is shown as a horizontal line.}
 \label{F_Violin1}
\end{figure}

To obtain a more comprehensive view of the impact of substructures, we 
show in Fig.~\ref{F_Violin1} violin plots for $\Delta r_{\mathrm{ox}}$ and $\Delta v_{\mathrm{pec}}$, for each  class of assembly state. 
In the upper left panel ($\Delta r_{\mathrm{ox}}$), the only trend visible is for the {CDGs} in the P class to lie below the threshold. 
This is confirmed for the pairs (U,P) and (P,S), a difference in the distribution being found using a Mann-Withney test at 95\% CL, with $P = 0.008$ and $P = 0.025$, respectively. 
However, a Kruskal-Wallis test performed comparing the whole classes (with Dunn's post-tests) found no statictically significant differences. 
Similarly, a Kruskal-Wallis test for $\Delta v_{\mathrm{pec}}$, in the upper right panel, is also negative, with $P = 0.512$. 
Thus, we see no evidence for $\Delta r_{\mathrm{ox}}$ and $\Delta v_{\mathrm{pec}}$ to be related to the classes of assembly state.

In the lower panels of Fig.~\ref{F_Violin1}, we compare the distribution of $\Delta r_{\mathrm{ox}}$ (left) and $\Delta v_{\mathrm{pec}}$ (right) separating the clusters based on the core cooling status. 
Performing a Kruskal-Wallis test for $\Delta r_{\mathrm{ox}}$, we find a statistically significant difference between the NCC and SCC (with $P = 0.011$) but not between NCC and WCC or WCC and SCC. 
However, for $\Delta v_{\mathrm{pec}}$ the difference is much more significant ($P \ll 0.0001$) between both NCC and SCC and NCC and SCC (but no difference between NCC and WCC as before). 
We find a 60\% probability for WCC and NCC clusters to be associated with clusters that have both high $\Delta r_{\mathrm{ox}}$ and $\Delta v_{\mathrm{pec}}$, the trends being more obvious in S and M clusters than in U and P clusters. 
Considering that the latter two classes represent more massive clusters than the two former ones (cf. Table~\ref{T_AvPar}), the U and P clusters, consequently, are possibly slightly more relaxed than the S and M clusters. 
This is consistent with the complex assembly history of the clusters suggested by the assembly state classes. 
This correlation has already been pointed out in the literature: \citet{Zha11}, for instance, found that the {CDG}--X-ray offset is related to the central cooling time of the clusters, for the HIFLUGCS X-ray flux limited galaxy cluster sample, suggesting that the system must be close to relaxed to have its cooling flow enhanced or a CC formed.

In general, the fact that the probability of association between the parameters related to the galaxies and gas in the core are not higher than 60\% suggests that these two components most probably follow different paths towards equilibrium, the virialization time-scale, most specifically, being much smaller for the gas than for galaxies.

\subsection{Results on the co-evolution of {CDGs} and clusters}
\label{sec:CEvln}

The remaining columns in Table \ref{T_Assembl} are dedicated to report the evolutionary parameters for the {CDGs}: the magnitude gaps $\Delta m_{12}$ (col. 8) and $\Delta m_{23}$ (col. 10); the projected separation (clustercentric distance) of the 2$^{nd}$-rank, $\Delta r_{12}$, and 3$^{rd}$-rank, $\Delta r_{13}$ (cols. 9 and 11, respectively); and additional comments in column 12.

\begin{figure}[h] 
 \centering
  \includegraphics[width=0.5\textwidth]{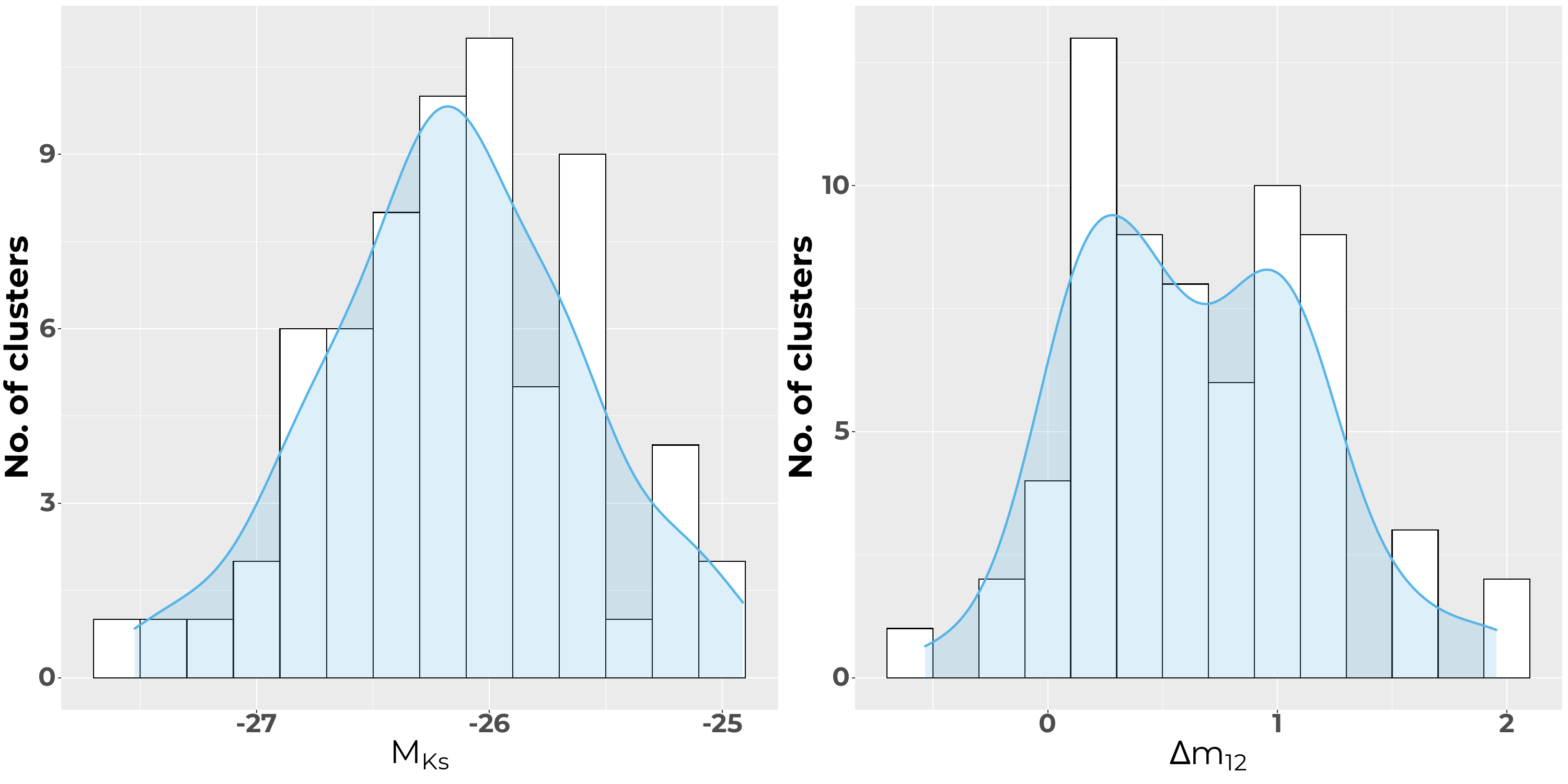}
 \caption{Distributions of {CDGs} 
 {properties}: Left panel, CDG M$_\mathrm{Ks}$, right panel, $\Delta m_{12}$.
}
 \label{F_FRG2}
\end{figure}

Using M$_\mathrm{Ks}$ as a proxy for the stellar mass of the {CDG} \citep[\textit{e.g.},][]{SGH83}, we traced in the left panel of Fig.~\ref{F_FRG2} its distribution for all the {CDGs} in our sample.  
A relatively good Gaussian fit suggests some level of similarity in the evolution of these {CDGs}.
However, the distribution of the magnitude gaps, $\Delta m_{12}$, in the right panel of Fig.~\ref{F_FRG2}, is clearly bimodal, indicating different assembly histories for the {CDGs} themselves.

\begin{figure}[h]
 \centering
 \includegraphics[width=84mm]{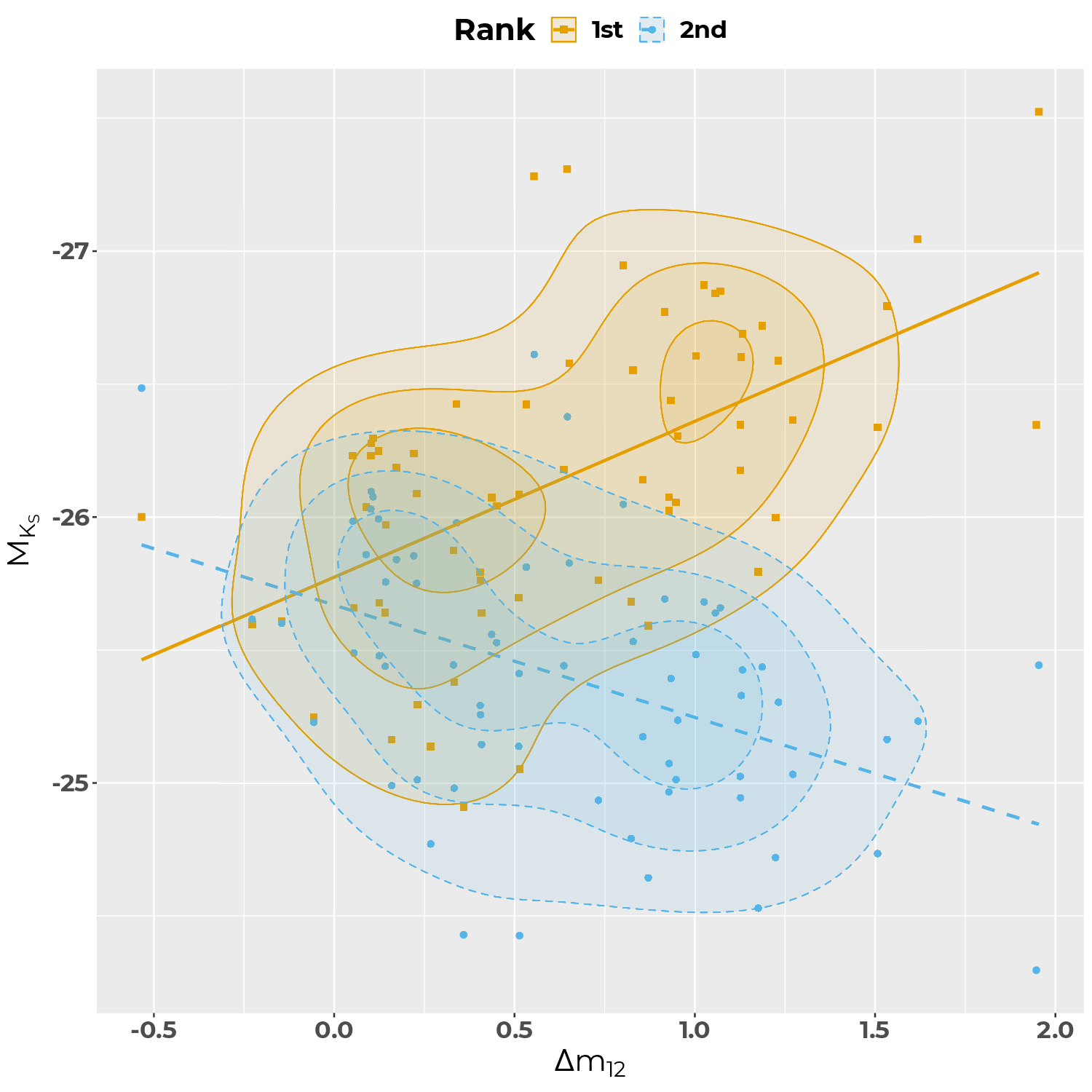}
 \caption{Distributions of absolute magnitudes for {CDGs} and 2$^{nd}$-rank {galaxies} as function of the magnitude gap.}
 \label{F_GMabs}
\end{figure}

This can also be appreciated in Fig.~\ref{F_GMabs}, which shows the distribution of absolute magnitudes for the {CDGs} and {their respective} second-rank {galaxies} as a function of the magnitude gaps $\Delta m_{12}$. As the gap increases, the luminosity of the {CDG} (its mass) grows almost linearly, while the luminosity of the second-rank {galaxy} slowly declines. 
Note that, due to our thorough analysis of substructures and definition of {CDGs}, it is not surprising that the change in mass we find is much faster than what was seen before \citep[e.g.,][]{Smi10}.
Although systems showing large magnitude gaps are consistent with a model where the {CDG} co-evolve with its cluster, the bimodality clearly suggests more complex assembly histories for the {CDGs}.

\begin{figure}
\centering
\setlength{\unitlength}{1.cm}
\begin{minipage}[t]{5.9cm}
  \begin{picture}(5.5,5.5)
\includegraphics[width=\textwidth]{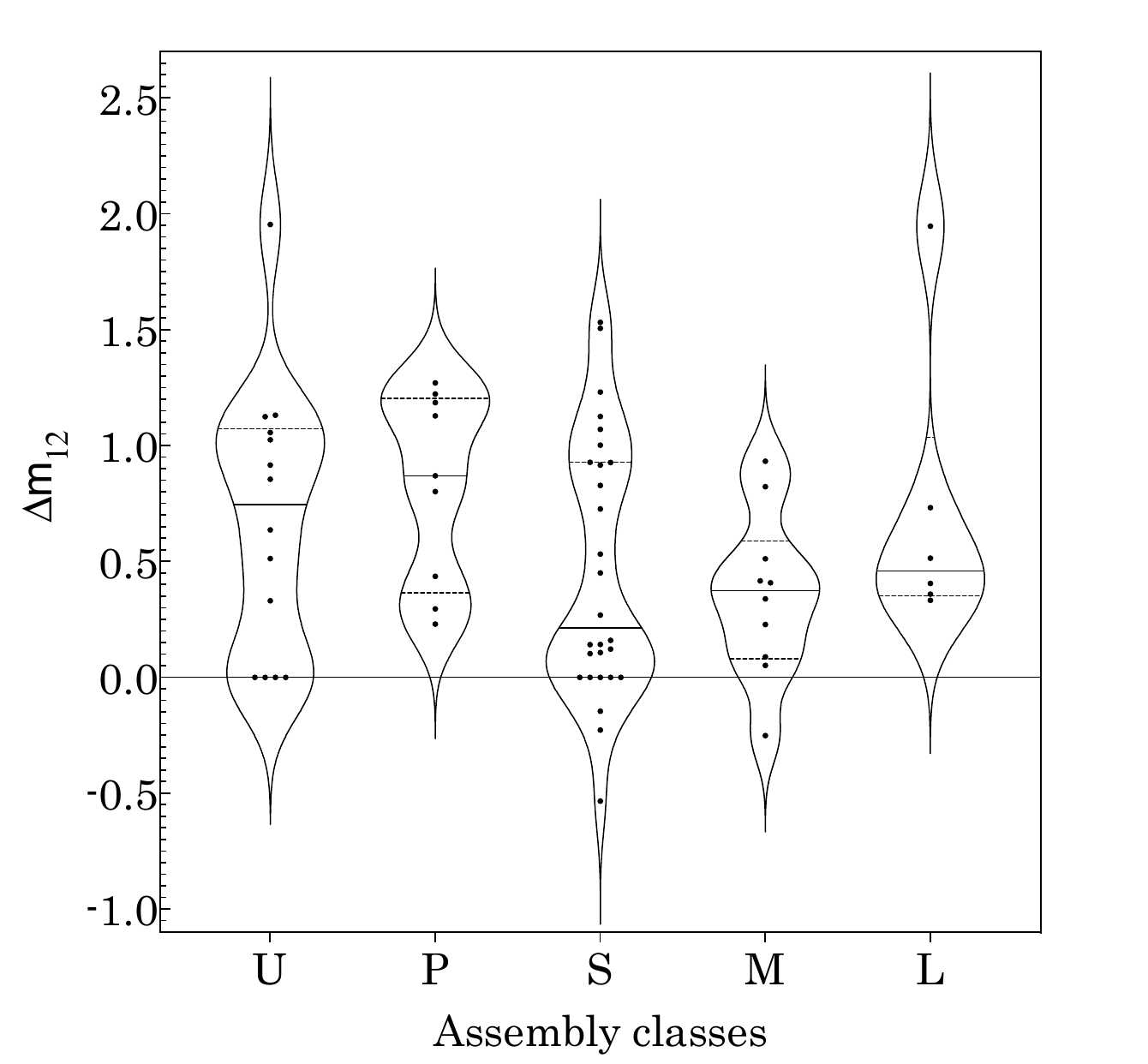}
  \end{picture}\par
 \end{minipage}
\hfill
\begin{minipage}[t]{5.9cm}
  \begin{picture}(5.5,5.5)
\includegraphics[width=\textwidth]{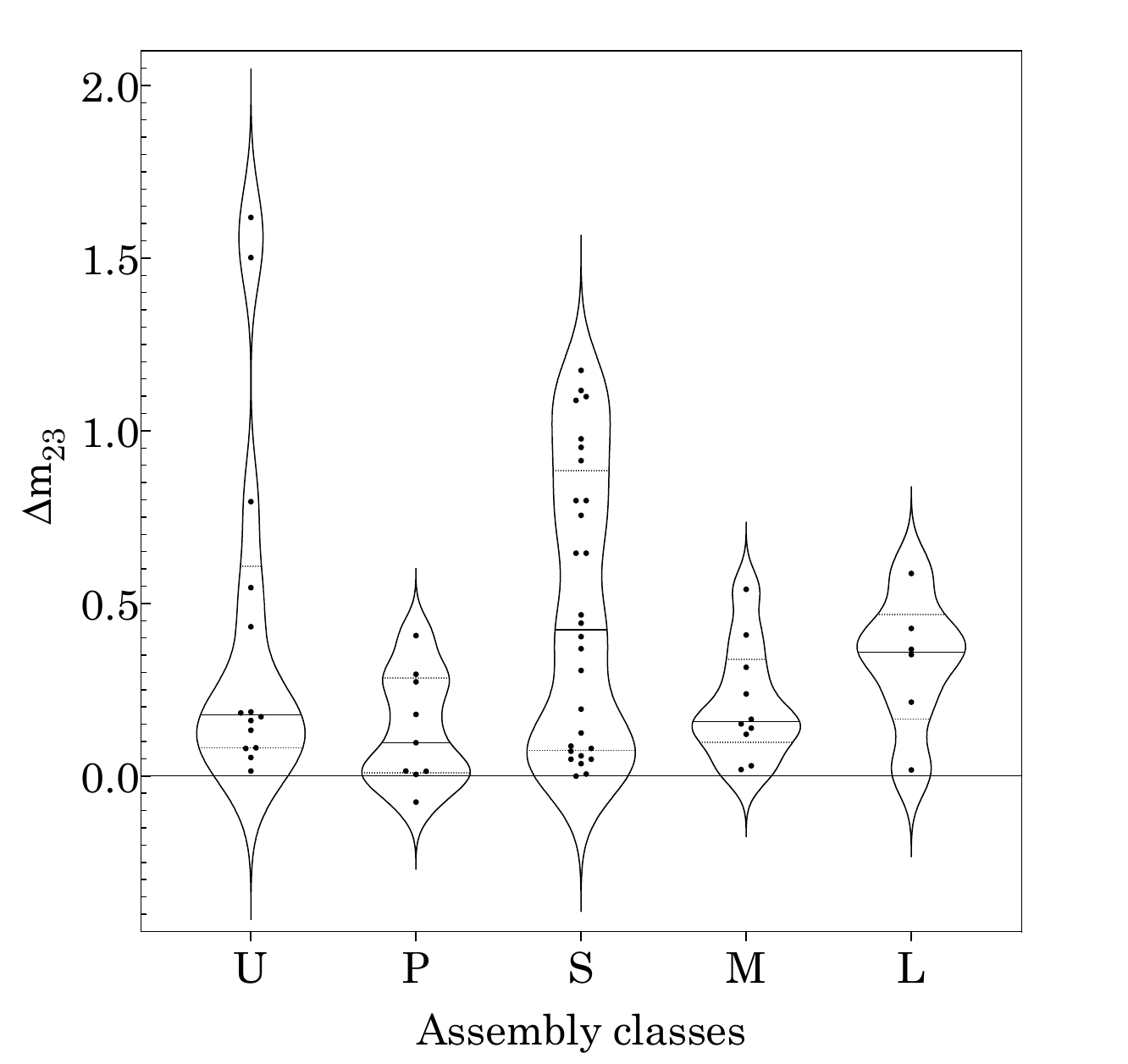}
  \end{picture}\par
 \end{minipage}
\hfill
 \begin{minipage}[t]{5.9cm}
  \begin{picture}(5.5,5.5)
\includegraphics[width=\textwidth]{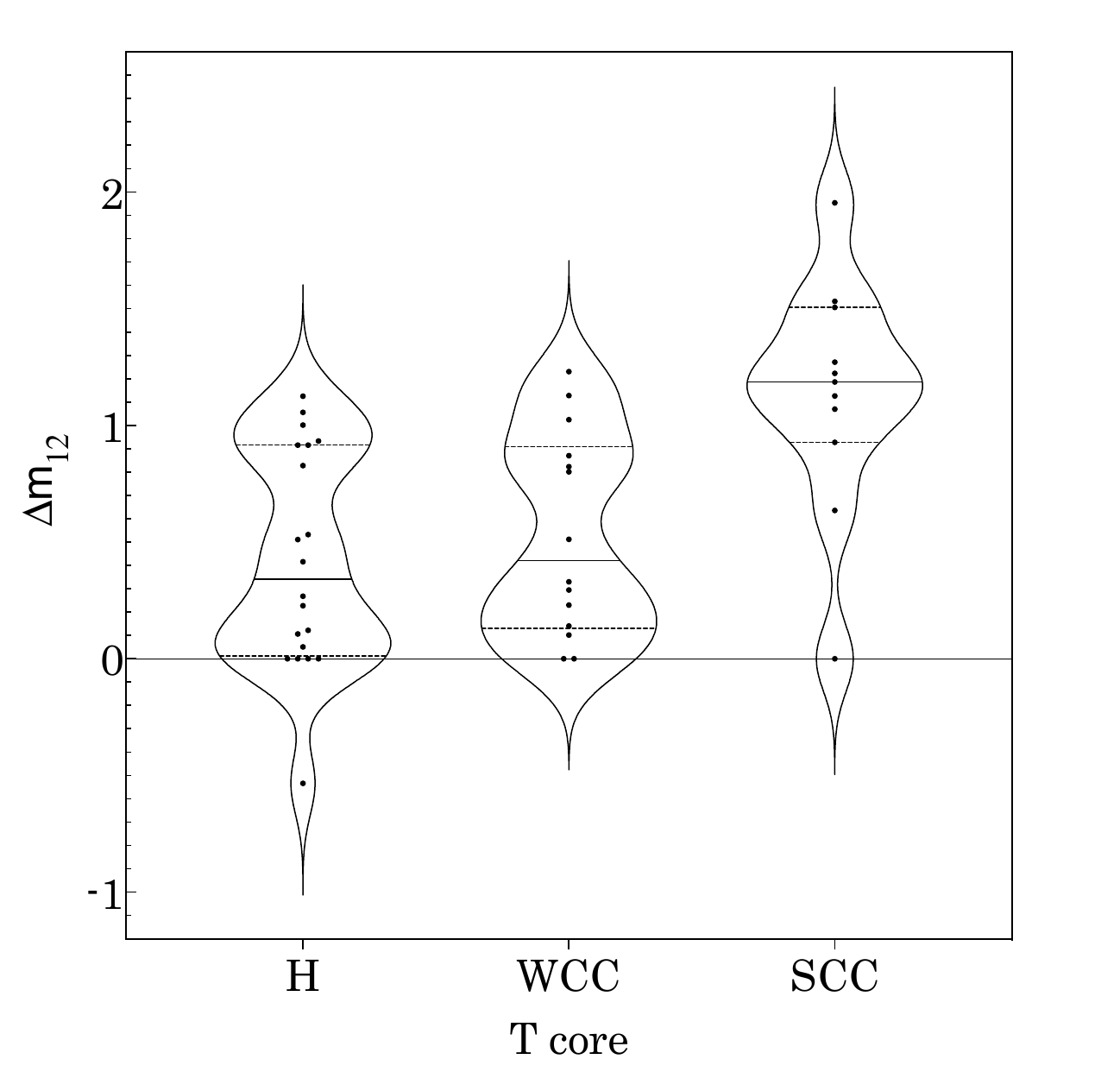}
  \end{picture}\par
 \end{minipage}
\hfill 
 \begin{minipage}[t]{5.9cm}
  \begin{picture}(5.5,5.5)
\includegraphics[width=\textwidth]{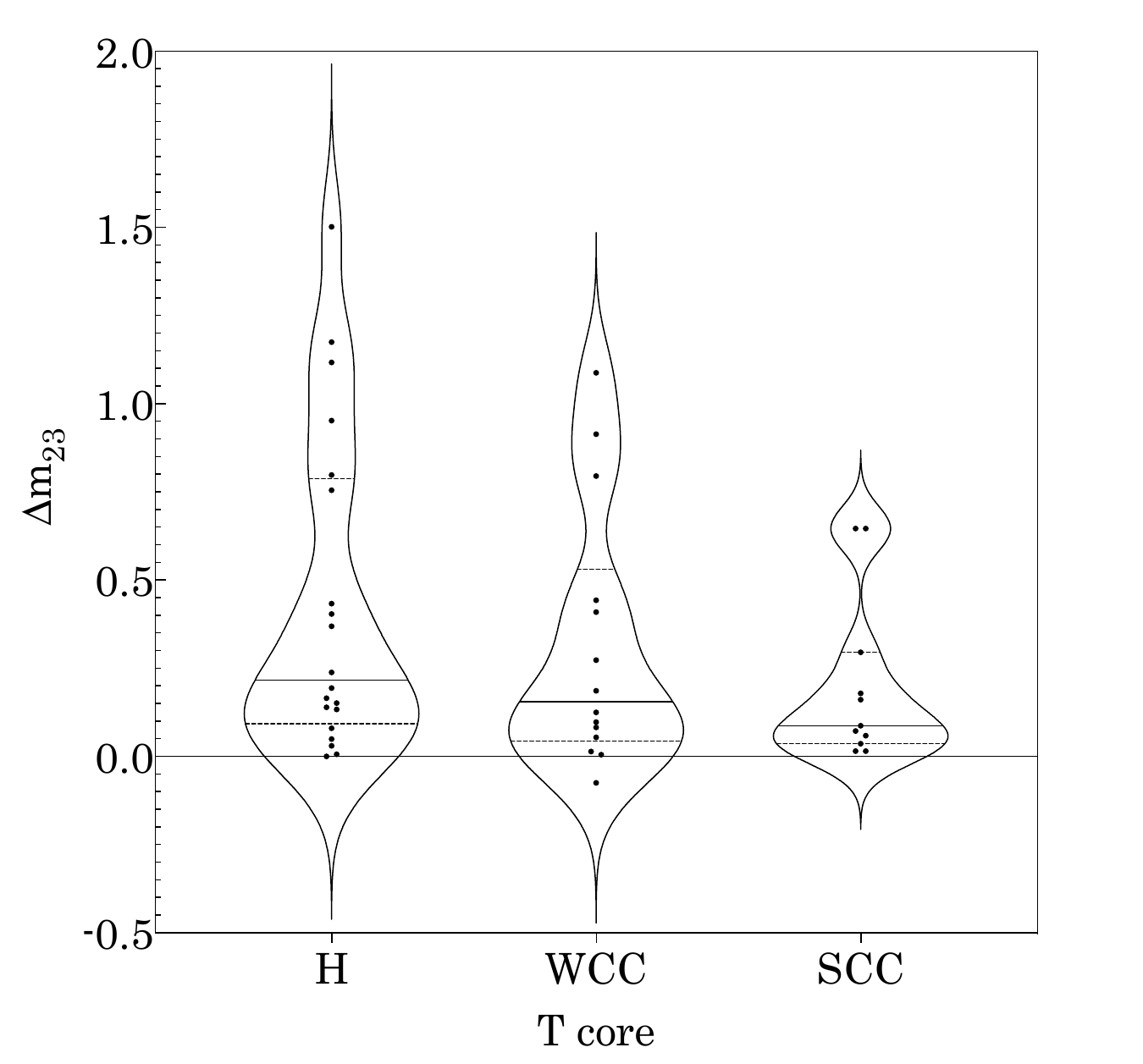}
  \end{picture}\par
 \end{minipage}
 \caption{Comparison of magnitude gaps in different assembly classes (upper panels and cooling states (lower panels).}
 \label{FRG_Violin2}
\end{figure}

To shed more light on the co-evolution of the {CDGs} and their clusters, we compare how the two magnitude gaps, $\Delta m_{12}$ and $\Delta m_{23}$, vary in the different assembly classes. 
This is done in the upper panels of Fig.~\ref{FRG_Violin2}. 
Although there is an apparent trend for U and P cluster to have higher magnitude gaps between the {CDG} and second-rank {galaxy} than for the S and M clusters, a Kruskal-Wallis test found no statistically significant difference ($P = 0.190$). 
Similarly, a Kruskal-Wallis test found no statistically significant difference ($P = 0.125$) for $\Delta m_{23}$.

In the lower panels of Fig.~\ref{FRG_Violin2} we compare the two magnitude gaps for the different core cooling states. 
This time the Kruskal-Wallis test clearly identified a statistically significant difference for $\Delta m_{12}$, with $P = 0.002$, between the pairs (NCC, SCC) and (WCC, SCC), but not for $\Delta m_{23}$. 
This is consistent with what we found before for $\Delta r_{\mathrm{ox}}$ and $\Delta v_{\mathrm{pec}}$. 
Consequenlty, despite the clear evidence of co-evolution of the clusters and their {CDGs}, especially in the core, the fact that the assembly state classes contain a mixture of core cooling states seems to confirm the complex assembly history of clusters in general.

\section{Summary and conclusions}
\label{sec:Con}

In this study we traced the assembly histories for a sample of 67 relatively rich (median $N_c = 150$ spectroscopic members) and nearby ($z < 0.15$) galaxy clusters, by classifying their level of substructuring in their outer regions (mostly beyond $r_{500}$) and estimating the dynamical impact of such subclumps on the host clusters. We also identified and characterized the dynamical properties of the 
{CDGs} of the clusters and compared them to {the} ICM equilibrium state, from X-ray  literature data, mapped in the inner part and innermost core regions. 

On the accompanying webpage\footnote{\ \url{www.astro.ugto.mx/recursos/HP_SCls/Top70.html}}, we offer the complete set of figures describing all clusters presented in this article: distribution of galaxies in each complex, system and significant substructure, projected number densities (like in Fig.~\ref{F_Isod}), X-ray contour images (like in Fig.~\ref{F_Xray}), {CDG} optical images, together with further information completing the data presented in the various tables included in the present article.

The following is a brief summary of our findings and conclusions:

\begin{itemize}
 \item In 19\% of the cluster in our sample, the classical BCG (directly identified from photometry) is not the {CDG} (gravitationally dominant galaxy). Among the discrepant cases we distinguish most specifically three different groups: binary {central dominant galaxies} with the second brightest as {the CDG} (Coma-like clusters), a BCG that is {the SDG} of a substructure (Fornax-like clusters), and clusters with a peripheric fossil candidate, where the BCG appears relatively isolated in the outskirt of the cluster.

 \item Using robust methods to determine cluster membership allowed us to more thoroughly determine the global dynamical parameters of the clusters: radial velocity of the system, velocity dispersion of galaxy members, virial mass and radius.

 \item Using different algorithms to detect substructures and estimate their gravitational impact on their host clusters, our analysis allowed us to determine that, although as many as 70\% of nearby clusters show evidence of substructures, those dynamically significant only appear in 57\% of the clusters. 

 \item Based on the significance level of the impact of the substructure, we defined five classes of assembly states: high-mass, Unimodal (U); Low-mass unimodal (L); Multi-modal (M); Primary (P), with low-mass substructures attached to a main structure; and, finally, Substructured (S), formed by a main structure and high-mass substructures. We count 21\% U, 13\% P, 42\% S, 15\% M and 9\% L {clusters}. In terms of masses, U and P clusters are more massive than S and M clusters, while L clusters are less massive, explaining why they are not detected in X-ray.  
\end{itemize}

Our classification of clusters in terms of substructures seems consistent with a hierarchical model of formation, where clusters form by the mergers of groups of galaxies: 

\begin{itemize}
 \item U clusters are examples of massive systems that merged in the distant past and, consequently, their virialization process is well advanced. 
 
 \item P clusters also formed in the past, and, because they are massive, they still accrete small groups from their environment. 
 
 \item M and S clusters, which have significant substructures, are examples of relatively more recent mergers: in S clusters massive clumps are accreting smaller mass groups {(minor mergers)}, while in M clusters the masses of the merging entities are comparable {(major mergers)}. 
 
 \item L clusters are the best examples of poor clusters in our sample: their masses and richness are comparable to those of massive groups, and, like the latter, are usually poor in gas. Their environment suggests some of them are either infalling or satellites of more massive clusters. 
\end{itemize}

{The classes can be interpreted as a ``sequence'' of different global assembly states possible for the clusters: they begin as a poor cluster (L) or a pile-up of small systems (M), then grow and pass to a state where a main structure starts to dominate (S), then become massive, although still accreting small groups (P), and finally become massive and regular/relaxed (U). 
Note that this is a snapshot of the assembly state, which can evolve in time: a U cluster can still accrete (becoming a P or S) or merge (becoming a M), 
for example.
However, although there is a dispersion in masses, this dispersion is not that high. This is because the era of cluster evolution is relatively recent, and they did not have time to pass the process of a major merger much more than once or twice. Capturing smaller clumps (minor mergers), on the other hand, may have been frequent, but with smaller impact in their global masses. Also, the availability of new systems to be captured is decreasing with time 
because of the accelerated expansion of the Universe.}

Our comparison of the properties of {CDGs} ($\Delta r_{\mathrm{ox}}$, $\Delta v_{\mathrm{pec}}$, 
$\Delta m_{12}$ and $\Delta m_{23}$) in the clusters with different assembly state classes and with the characteristics of the ICM in the inner region (different core cooling status) allows us to obtain a more precise view about the assembly process of the clusters. 

\begin{itemize}
 \item Considering the clusters individually or in any of the assembly states, we found no correlation between $\Delta r_{\mathrm{ox}}$ ({CDG}--X-ray offset) and $\Delta v_{\mathrm{pec}}$ ({CDG} peculiar velocity). We suggest the use of both together to characterize the dynamical state of the {CDGs}.
 
 \item We found a 60\% probability for WCC and NCC clusters to be associated with clusters that have both high $\Delta r_{\mathrm{ox}}$ and $\Delta v_{\mathrm{pec}}$, the trends being more obvious in more massive clusters. Considering the difference in masses, this suggests U and P clusters are more relaxed than S and M clusters. 
 
 \item Comparing how the two magnitude gaps, $\Delta m_{12}$ and $\Delta m_{23}$, vary in the different assembly states, we found only an apparent trend for U and P cluster to have higher magnitude gaps between the {CDG} and second-rank galaxy than the S and M clusters, while no trend is visible for $\Delta m_{23}$. 
 
 \item However, we also found a significant difference for $\Delta m_{12}$, the gap being smaller in NCC and WCC than in SCC, while no difference is detected for $\Delta m_{23}$.  
\end{itemize}

We conclude that, despite of clear evidence of co-evolution of the clusters and their {CDGs}, especially considering the gas in the core, the fact that the assembly state classes contain a mixture of core cooling states seems to confirm a complex assembly history of clusters. 
In general, the two baryonic components of clusters, galaxies and gas, probably follow different paths towards equilibrium, the relaxation time-scale, most specifically, being much smaller for the gas than for the galaxies. This difference implies that, even in the {most} evolved systems, the virialization and evolution of the {CDG} are complex processes that do not depend solely on time but also on the frequency and impact of mergers, cooling and heating of the ICM by shocks and feedback, cannibalism, pre-processing, among others. 
How much of this evolution is due to pre-processing in the initial groups, however, is still an open question.  


\section*{Acknowledgments}

We acknowledge financial support from \textit{Universidad
de Guanajuato} (DAIP), \textit{Convocatoria Institucional de Investigaci\'on Cient\'ifica}, projects 087/2010, 219/2013, 205/2019 and 138/2022. Y.V. acknowledges financial support from the 
postdoctoral-fellowship ``\textit{2do. a\~no de estancia posdoctoral en M\'exico}'' from CONACyT and UNAM-DGAPA-PAPIIT IN111620 grant.

This research has made use of ``Aladin sky atlas'' developed at CDS, 
Strasbourg Observatory, France, the NASA/IPAC Extragalactic Database (NED) which is operated by the Jet Propulsion Laboratory, California Institute of Technology, under contract with the National Aeronautics and Space Administration, the SuperCOSMOS Science Archive, prepared and hosted by the Wide Field Astronomy Unit, Institute for Astronomy, University of Edinburgh, which is funded by the UK Science and Technology Facilities Council and the Two Micron All Sky Survey, which is a joint project of the University of Massachusetts, the Infrared Processing and Analysis Center/California Institute of Technology, funded by the National Aeronautics and Space Administration and the National Science Foundation and the Digitized Sky Surveys, produced at the Space Telescope Science Institute under U.S. Government grant NAG W-2166. 

\begin{appendices}

\section{New temperature measurements}
\label{sec:ap-A}
We estimated the temperature for seven galaxy clusters of our sample using data from both 
XMM-Newton\footnote{\ \url{http://nxsa.esac.esa.int/nxsa-web/}}
and Chandra\footnote{\ \url{https://chandra.harvard.edu/}}
(Table \ref{T_ApA1}), 
four of which are previously unreported in the literature.
This section briefly discusses the procedure for spectral fitting 
and computing temperatures.

\begin{table*}[!t] 
 \begin{center}
 \tablecols{7}
 \begin{small}
 \setlength{\tabcolsep}{1mm}
 \setlength{\tabnotewidth}{1\columnwidth}
 \caption{X-ray data for the targeted galaxy clusters.}
 \label{T_ApA1}
   \begin{tabular}{lcccccr}
\toprule
 \multicolumn{1}{l}{Cluster} &
 \multicolumn{1}{c}{$z_{spec}$} &
 \multicolumn{1}{c}{X-ray peak} &
 \multicolumn{1}{c}{Telescope} &
 \multicolumn{1}{c}{Observation} &
 \multicolumn{1}{c}{Date of Observation} &
 \multicolumn{1}{c}{Exposure time} \\
 \multicolumn{1}{c}{ID} &
 \multicolumn{1}{c}{} &
 \multicolumn{1}{c}{$\alpha_{J2000},\delta_{J2000}$} &
 \multicolumn{1}{c}{} &
 \multicolumn{1}{c}{ID} &
 \multicolumn{1}{c}{} &
 \multicolumn{1}{c}{(ks)} \\
 \multicolumn{1}{c}{(1)} & 
 \multicolumn{1}{c}{(2)} & 
 \multicolumn{1}{c}{(3)} & 
 \multicolumn{1}{c}{(4)} & 
 \multicolumn{1}{c}{(5)} & 
 \multicolumn{1}{c}{(6)} & 
 \multicolumn{1}{c}{(7)} \\
\midrule
A0122 & 0.113 & 00 57 24.7, -26 16 50 & XMM     & 0504160101   & 2007 Dec 3     & 56.92 \\
A2811 & 0.108 & 00 42 08.7, -28 32 09 & XMM     & 0404520101   & 2006 Nov 28    & 25.91 \\
A2870 & 0.024 & 01 07 43.2, -46 54 59 & XMM     & 0205470301   & 2004 May 15    & 11.91 \\
A0399 & 0.072 & 02 57 56.4,  13 00 59 & Chandra & 3230         & 2002 Nov 18    & 49.28 \\
A3094 & 0.068 & 03 11 25.0, -26 53 59 & Chandra & 5799         & 2005 Nov 28    & 40.15 \\
A3716 & 0.046 & 20 51 16.7, -52 41 43 & Chandra & 15133, 15583 & 2012 Dec 24/20 & 14.77, 16.08 \\
A4038 & 0.028 & 23 47 43.2, -28 08 29 & Chandra & 4992         & 2004 Jun 28    & 33.97 \\
\bottomrule
\tabnotetext{1}{X-ray data obtained from XMM-Newton (http://nxsa.esac.esa.int/nxsa-web/)
and Chandra (https://chandra.harvard.edu/). 
Columns: (1) Cluster ID; 
         (2) spectroscopic redshift coming from NED (http://ned.ipac.caltech.edu); 
         (3) Optical RA, Dec (J2000.0); 
         (4) Telescope;
         (5) Observation ID;
         (6) Date of Observation; and
         (7) Exposure Time.}
   \end{tabular}
 \end{small}
 \end{center}
\end{table*}                                

The data from XMM-Newton were reduced using the 
\textsl{XMM-Newton Science Analysis Software} (SAS), version 14.0.0.
The raw data, downloaded in the form of observation data files (ODF), 
were processed in the following steps: 
i) Generation of calibrated event lists for the EPIC (MOS1, MOS2, 
and PN) cameras using the latest calibration data;
this step was done using the SAS packages \texttt{cifbuild}, 
\texttt{odfingest}, \texttt{epchain}, and \texttt{emchain}. 
ii) Creation of the background light curves to identify 
time intervals with poor quality data, and filtering of the EPIC 
event lists to exclude periods of high background flaring and bad 
events.
iii) Creation of a sky image of the filtered data set; these steps 
were performed using SAS packages \texttt{evselect}, 
\texttt{tabgtigen}, and \texttt{xmm\_select}. 

Finally, we extracted the spectra from the source and background 
using the task \texttt{especget} from SAS. 
This task produced two sets of files called the response matrix 
files (redistribution matrix file, RMF, and ancillary response 
file, ARF). 
Similarly, the Chandra data were obtained from the Chandra Data 
Archive (CDA)\footnote{\ \url{https://cxc.harvard.edu/cda/}} 
and operated by the Chandra Interactive Analysis of Observations 
\textsf{CIAO}, version 4.6.1, with calibration database version 4.6.1. 
In addition, the CIAO tool \texttt{chandra\_repro} was applied 
to perform initial processing and obtain a new event file. 
These files are used for spectral fittings.

\subsection{Spectral model fitting:}
We extracted the spectra within a fixed radius of 0.5 $h_{70}^{-1}$ Mpc 
and excluded the point sources from this region. 
In some cases, the object was not centered on the observed 
field, such that we had to reduce the size of the extraction
circle (see Table \ref{T_ApA2}).
The spectra from both, XMM-Newton and Chandra, were fitted 
using \textsl{XSPEC} spectral fitting software, version 12.5.1
\citep{Arn96}.
The photon counts of each cluster spectrum were grouped into bins 
with at least one count per bin.
The spectral model was fitted to the data using the 
Ftools task \texttt{grppha}.
The Galactic HI column (nH) was derived from the HI map from the 
Leiden/Argentine/Bonn (LAB) survey \citep{Kal05}.
This parameter was fixed while fitting the X-ray spectrum. 
The redshift of the spectral model was fixed to 
the cluster spectroscopic redshift coming from the NED 
database\footnote{\ \url{http://ned.ipac.caltech.edu}}.
Finally, we employed a fitting model to multiply a TBABS 
absorption model \citep{Wil00}
and a single-temperature 
optically thin thermal plasma component 
\citep[the MEKAL code in XSPEC terminology,][]{Mew86}
to model the X-ray emission from ICM plasma.

\begin{table}[!h] 
 \begin{center}
 \tablecols{4}
 \setlength{\tabcolsep}{1mm}
 \setlength{\tabnotewidth}{1\columnwidth}
 \caption{Estimated X-ray temperatures for the targeted galaxy clusters.}
 \label{T_ApA2}
   \begin{tabular}{lccc}
\toprule
 \multicolumn{1}{c}{Cluster} &
 \multicolumn{1}{c}{Projected radius} &
 \multicolumn{1}{c}{Physical radius} &
 \multicolumn{1}{c}{$k$T$_\mathrm{X}$} \\
 \multicolumn{1}{c}{} &
 \multicolumn{1}{c}{(armin)} &
 \multicolumn{1}{c}{(Mpc)} &
 \multicolumn{1}{c}{(keV)} \\
 \multicolumn{1}{c}{(1)} & 
 \multicolumn{1}{c}{(2)} & 
 \multicolumn{1}{c}{(3)} & 
 \multicolumn{1}{c}{(4)} \\
\midrule
A0122  & 4.0 & 0.5  & 3.70 $\pm$ 0.07 \\
A2811  & 4.2 & 0.5  & 5.04 $\pm$ 0.05 \\
A2870  & 8.6 & 0.25 & 1.07 $\pm$ 0.07 \\
A0399  & 6.0 & 0.5  & 6.49 $\pm$ 0.27 \\
A3094  & 6.4 & 0.5  & 3.15 $\pm$ 0.48 \\
A3716N & 4.6 & 0.25 & 2.19 $\pm$ 0.26 \\
A3716S & 4.6 & 0.25 & 3.65 $\pm$ 0.27 \\
A4038  & 7.4 & 0.25 & 3.15 $\pm$ 0.05 \\
\bottomrule
\tabnotetext{1}{Columns: (1) Cluster ID; 
                      (2) radius used to extract spectra, 
                      (3) relative physical radius covered in the plan of sky, 
                      (4) estimated temperature from Model TBabs*mekal.}
   \end{tabular}
 \end{center}
\end{table}

\section{List of \textsl{m} and \textsl{hs} substructures for P, S and M clusters.}
\label{sec:ap-B}

\clearpage
\onecolumn
\begin{center}
\begin{scriptsize}
 \setlength{\tabcolsep}{1mm}
\begin{longtable}{lrrlrrrrrrrrrr} 
 \caption{Substructures detected in the clusters of our sample. SDG: Substructure Dominant Galaxy}
 \label{T_ApB} \\
\toprule
  \multicolumn{1}{l}{ID$_{\mathrm{sub}}$} &
  \multicolumn{1}{l}{RA$_{\mathrm{SDG}}$} &
  \multicolumn{1}{l}{Dec$_{\mathrm{SDG}}$} &
  \multicolumn{1}{l}{SDG} &
  \multicolumn{1}{r}{$N_{s}$} &
  \multicolumn{1}{r}{$v_{s}$} &
  \multicolumn{1}{r}{$\sigma_{s}$} &
  \multicolumn{1}{r}{$r_{200}$} &
  \multicolumn{1}{r}{$N_{\mathrm{a}}$} &
  \multicolumn{1}{r}{$v_{\mathrm{sub}}$} &
  \multicolumn{1}{r}{$\sigma_{\mathrm{sub}}$} &
  \multicolumn{1}{r}{$R_{\mathrm{p}}$} &
  \multicolumn{1}{r}{$R_{\mathrm{vir}}$} &
  \multicolumn{1}{r}{$M_{\mathrm{vir}}$} \\
 \multicolumn{1}{l}{(1)} & 
 \multicolumn{1}{l}{(2)} & 
 \multicolumn{1}{l}{(3)} & 
 \multicolumn{1}{l}{(4)} & 
 \multicolumn{1}{c}{(5)} &  
 \multicolumn{1}{c}{(6)} & 
 \multicolumn{1}{c}{(7)} &  
 \multicolumn{1}{c}{(8)} & 
 \multicolumn{1}{c}{(9)} & 
 \multicolumn{1}{c}{(10)} &
 \multicolumn{1}{c}{(11)} & 
 \multicolumn{1}{c}{(12)} & 
 \multicolumn{1}{c}{(13)} & 
 \multicolumn{1}{c}{(14)} \\ 
\midrule
\endfirsthead
\toprule
  \multicolumn{1}{l}{ID$_{\mathrm{sub}}$} &
  \multicolumn{1}{l}{RA$_{\mathrm{SDG}}$} &
  \multicolumn{1}{l}{Dec$_{\mathrm{SDG}}$} &
  \multicolumn{1}{l}{SDG} &
  \multicolumn{1}{r}{$N_{s}$} &
  \multicolumn{1}{r}{$v_{s}$} &
  \multicolumn{1}{r}{$\sigma_{s}$} &
  \multicolumn{1}{r}{$r_{200}$} &
  \multicolumn{1}{r}{$N_{\mathrm{a}}$} &
  \multicolumn{1}{r}{$v_{\mathrm{sub}}$} &
  \multicolumn{1}{r}{$\sigma_{\mathrm{sub}}$} &
  \multicolumn{1}{r}{$R_{\mathrm{p}}$} &
  \multicolumn{1}{r}{$R_{\mathrm{vir}}$} &
  \multicolumn{1}{r}{$M_{\mathrm{vir}}$} \\
 \multicolumn{1}{l}{(1)} & 
 \multicolumn{1}{l}{(2)} & 
 \multicolumn{1}{l}{(3)} & 
 \multicolumn{1}{l}{(4)} & 
 \multicolumn{1}{c}{(5)} &  
 \multicolumn{1}{c}{(6)} & 
 \multicolumn{1}{c}{(7)} &  
 \multicolumn{1}{c}{(8)} & 
 \multicolumn{1}{c}{(9)} & 
 \multicolumn{1}{c}{(10)} &
 \multicolumn{1}{c}{(11)} & 
 \multicolumn{1}{c}{(12)} & 
 \multicolumn{1}{c}{(13)} & 
 \multicolumn{1}{c}{(14)} \\ 
\midrule
\endhead
\hline \multicolumn{14}{|r|}{{Continued on next page}} \\ \hline
\endfoot
\endlastfoot
  A2804\textbf{m}w  &   9.907535 & $-$28.906199 & 6dF J0039377-285422       & 48  & 33790 & 323  & 0.631 & 27  & 33836 & 404  & 0.842 & 1.036 &  1.256 \\
  A2804\textbf{m}e  &  10.008486 & $-$28.902040 & GALEX J004002.00-285407.9 & 32  & 32688 & 380  & 0.743 & 6   & 32268 & 314  & 0.540 & 0.757 &  0.487 \\
  A085Acw           &  10.376354 &  $-$9.262768 & KAZ 364                   & 22  & 14352 & 473  & 0.954 & 20  & 14414 & 482  & 0.529 & 1.019 &  1.121 \\
  A085A\textbf{m}   &  10.460515 &  $-$9.303040 & MCG -02-02-086            & 272 & 16745 & 799  & 1.603 & 218 & 16738 & 852  & 1.813 & 2.242 & 12.000 \\
  A085Ase           &  10.792272 &  $-$9.861486 & GIN 011                   & 25  & 15083 & 272  & 0.547 & 15  & 15132 & 271  & 0.515 & 0.687 &  0.345 \\
  A2811B\textbf{m}  &  10.537175 & $-$28.535772 & 2MASX J00420892-2832087   & 123 & 32345 & 891  & 1.745 & 107 & 32340 & 947  & 1.766 & 2.345 & 14.500 \\
  A0118\textbf{m}   &  13.743478 & $-$26.375153 & 2MASX J00545843-2622305   & 51  & 34586 & 702  & 1.369 & 46  & 34470 & 728  & 1.666 & 1.925 &  8.056 \\
  A0118c            &  13.874743 & $-$26.396147 & SARS 013.26663-26.66683   & 14  & 33958 & 338  & 0.660 & 13  & 33996 & 368  & 0.329 & 0.712 &  0.407 \\
  A0118e            &  13.958530 & $-$26.365671 & SARS 013.35062-26.63606   & 15  & 34171 & 584  & 1.139 & 15  & 34171 & 584  & 0.381 & 1.017 &  1.189 \\
  A0119\textbf{m}   &  14.067088 &  $-$1.255492 & UGC 00579                 & 277 & 13276 & 876  & 1.768 & 266 & 13281 & 893  & 1.302 & 2.080 &  9.488 \\
  A0119n            &  14.258582 &  $-$0.875172 & UGC 00588                 & 25  & 13402 & 450  & 0.909 & 23  & 13417 & 456  & 0.626 & 1.040 &  1.187 \\
  A0119ne           &  14.365250 &  $-$0.471357 & GIN 021                   & 23  & 13197 & 422  & 0.852 & 17  & 13188 & 461  & 0.651 & 1.062 &  1.263 \\
  A0133Anw          &  15.008555 & $-$21.488472 & ESO 541-G007              & 13  & 16585 & 474  & 0.951 & 13  & 16585 & 474  & 1.113 & 1.289 &  2.283 \\
  A0133A\textbf{m}  &  15.674046 & $-$21.882154 & ESO 541-G013              & 109 & 16762 & 723  & 1.453 & 87  & 16833 & 774  & 1.340 & 1.901 &  7.333 \\
  A0133Ane          &  15.999613 & $-$21.372469 & ESO 541-G016              & 10  & 17550 & 154  & 0.308 & 2   & 17436 & 205  & 0.698 & 0.630 &  0.267 \\
  A2870w            &  16.226784 & $-$46.999523 & 2MASX J01045442-4659582   & 27  & 6396  & 329  & 0.672 & 24  & 6418  & 336  & 0.787 & 0.923 &  0.812 \\
  A2870e            &  16.927452 & $-$46.907574 & IC 1625                   & 23  & 6919  & 235  & 0.480 & 17  & 6964  & 239  & 0.455 & 0.611 &  0.237 \\
  A2877\textbf{m}   &  17.481663 & $-$45.931217 & IC 1633                   & 124 & 7231  & 647  & 1.318 & 99  & 7169  & 676  & 0.856 & 1.511 &  3.574 \\
  A3027Acw          &  37.482681 & $-$33.177345 & GALEX J022955.81-331036.6 & 24  & 23769 & 406  & 0.806 & 21  & 23821 & 423  & 1.054 & 1.164 &  1.721 \\
  A3027A\textbf{m}  &  37.706005 & $-$33.103752 & 2dFGRS S518Z162           & 73  & 23283 & 697  & 1.384 & 63  & 23318 & 731  & 1.363 & 1.828 &  6.653 \\
  A0400\textbf{m}   &  44.423164 &     6.026997 & NGC 1128                  & 51  & 7011  & 336  & 0.684 & 42  & 7027  & 347  & 0.561 & 0.841 &  0.616 \\
  A0400ne           &  44.587616 &     6.095203 & CGCG 415-046              & 10  & 6769  & 73   & 0.149 & 3   & 6768  & 94   & 0.228 & 0.260 &  0.018 \\
  A3104\textbf{m}   &  48.590549 & $-$45.420238 & LCRS B031238.4-453620     & 38  & 21784 & 446  & 0.887 & 28  & 21736 & 511  & 0.785 & 1.201 &  1.874 \\
  A3104e            &  49.033882 & $-$45.391937 & 2MFGC 02678               & 10  & 21609 & 271  & 0.541 & 9   & 21589 & 278  & 0.698 & 0.769 &  0.493 \\
  A3104se           &  49.237873 & $-$45.540817 & LCRS B031514.7-454323     & 5   & 22160 & 125  & 0.248 & 2   & 22128 & 53   & 0.152 & 0.152 &  0.004 \\
  A3112Bn           &  49.472656 & $-$44.041534 & 2MASX J03175343-4402295   & 7   & 23838 & 303  & 0.601 & 7   & 23838 & 303  & 0.718 & 0.819 &  0.601 \\
  A3112B\textbf{m}  &  49.490250 & $-$44.238213 & ESO 248-G006              & 90  & 22549 & 596  & 1.185 & 56  & 22526 & 595  & 1.637 & 1.695 &  5.283 \\
  A0426A\textbf{m}  &  49.950980 &    41.511680 & NGC 1275                  & 297 & 5289  & 1023 & 2.092 & 296 & 5295  & 1024 & 1.322 & 2.309 & 12.700 \\
  S0373sw           &  50.674120 & $-$37.208200 & NGC 1316                  & 20  & 1705  & 200  & 0.410 & 16  & 1728  & 230  & 0.238 & 0.484 &  0.115 \\
  S0373n            &  52.081950 & $-$31.068180 & NGC 1340                  & 23  & 1283  & 249  & 0.513 & 16  & 1248  & 229  & 0.624 & 0.664 &  0.298 \\
  S0373\textbf{m}   &  54.621180 & $-$35.450740 & NGC 1399                  & 178 & 1454  & 343  & 0.705 & 98  & 1458  & 390  & 0.307 & 0.748 &  0.427 \\
  A3158nw           &  55.549149 & $-$53.390938 & 2MASX J03421179-5323273   & 16  & 18631 & 207  & 0.414 & 6   & 18711 & 104  & 0.541 & 0.367 &  0.053 \\
  A3158\textbf{m}   &  55.720634 & $-$53.631302 & ESO 156-G008 NED01        & 190 & 17373 & 1077 & 2.161 & 190 & 17373 & 1077 & 1.193 & 2.279 & 12.600 \\
  A3158cs           &  55.873669 & $-$53.692135 & 2MASX J03432968-5341316   & 20  & 18594 & 287  & 0.575 & 15  & 18603 & 334  & 0.616 & 0.836 &  0.628 \\
  A0496nw           &  67.818871 & $-$12.455068 & IC 0377                   & 36  & 9376  & 452  & 0.917 & 33  & 9384  & 468  & 0.943 & 1.219 &  1.890 \\
  A0496\textbf{m}   &  68.407669 & $-$13.261956 & MCG -02-12-039            & 315 & 9957  & 682  & 1.383 & 272 & 9933  & 715  & 1.330 & 1.812 &  6.210 \\
  A0539\textbf{m}   &  79.155548 &     6.440917 & UGC 03274 NED05           & 110 & 8649  & 637  & 1.295 & 100 & 8645  & 674  & 0.982 & 1.576 &  4.073 \\
  A0539se           &  79.819481 &     5.757124 & 2MASX J05191667+0545256   & 15  & 8735  & 200  & 0.407 & 9   & 8777  & 243  & 0.646 & 0.694 &  0.348 \\
  A0539e            &  80.000107 &     6.680067 & CGCG 421-028              & 7   & 8902  & 118  & 0.240 & 2   & 8809  & 69   & 0.428 & 0.261 &  0.018 \\
  A3395nw           &  96.518044 & $-$54.029495 & LEDA 423636               & 35  & 14534 & 614  & 1.238 & 34  & 14514 & 613  & 0.670 & 1.295 &  2.299 \\
  A3395\textbf{m}   &  96.901047 & $-$54.449364 & ESO 161-G008              & 166 & 14980 & 740  & 1.491 & 161 & 14995 & 746  & 1.032 & 1.703 &  5.241 \\
  A3395se           &  97.606720 & $-$54.762615 & ESO 161-IG012 NED01       & 13  & 14587 & 398  & 0.801 & 13  & 14587 & 398  & 0.544 & 0.905 &  0.786 \\
  A0576sw           & 108.784760 &    55.419525 & CGCG 261-039              & 16  & 11328 & 305  & 0.618 & 5   & 11213 & 108  & 1.023 & 0.471 &  0.110 \\
  A0576\textbf{m}   & 110.375999 &    55.761581 & CGCG 261-056 NED02        & 191 & 11359 & 861  & 1.743 & 183 & 11351 & 878  & 1.563 & 2.189 & 11.000 \\
  A0576ne           & 110.941490 &    56.581875 & 2MFGC 05892               & 13  & 11483 & 421  & 0.851 & 9   & 11474 & 231  & 1.051 & 0.787 &  0.512 \\
  A0754nw           & 136.941254 &  $-$9.392439 & 2MASX J09074590-0923327   & 24  & 16494 & 303  & 0.609 & 15  & 16487 & 295  & 0.615 & 0.770 &  0.488 \\
  A0754s            & 137.009079 &  $-$9.993835 & 2MASX J09080217-0959378   & 30  & 15827 & 541  & 1.088 & 30  & 15827 & 541  & 0.595 & 1.144 &  1.592 \\
  A0754\textbf{m}w  & 137.134949 &  $-$9.629739 & 2MASX J09083238-0937470   & 193 & 16168 & 820  & 1.647 & 173 & 16182 & 880  & 0.981 & 1.867 &  6.931 \\
  A0754\textbf{m}e  & 137.330139 &  $-$9.699759 & 2MASX J09091923-0941591   & 118 & 16438 & 776  & 1.559 & 116 & 16450 & 782  & 0.881 & 1.665 &  4.922 \\
  A1060\textbf{m}   & 159.177963 & $-$27.528584 & NGC 3311                  & 343 & 3698  & 676  & 1.385 & 323 & 3701  & 694  & 0.884 & 1.560 &  3.886 \\
  A1367\textbf{m}nw & 176.009048 &    19.949820 & NGC 3842                  & 152 & 6546  & 512  & 1.045 & 122 & 6532  & 556  & 0.818 & 1.308 &  2.313 \\
  A1367\textbf{m}se & 176.270782 &    19.606382 & NGC 3862                  & 117 & 6330  & 599  & 1.222 & 109 & 6309  & 608  & 0.882 & 1.424 &  2.979 \\
  A3526A\textbf{m}  & 192.203918 & $-$41.311665 & NGC 4696                  & 210 & 3061  & 510  & 1.045 & 124 & 2991  & 569  & 0.817 & 1.332 &  2.414 \\
  A3526B\textbf{m}  & 192.516449 & $-$41.382072 & NGC 4709                  & 90  & 4563  & 276  & 0.566 & 47  & 4634  & 317  & 0.501 & 0.764 &  0.459 \\
  A3530\textbf{m}   & 193.900009 & $-$30.347490 & ESO 443-G011              & 101 & 16087 & 611  & 1.227 & 88  & 16105 & 633  & 1.195 & 1.601 &  4.367 \\
  A3530s1           & 193.929596 & $-$30.718328 & 2MASX J12554310-3043059   & 9   & 15480 & 246  & 0.495 & 7   & 15509 & 256  & 0.511 & 0.660 &  0.306 \\
  A3532s2           & 193.980011 & $-$30.721376 & 2MASX J12555520-3043169   & 7   & 16945 & 198  & 0.398 & 5   & 16878 & 163  & 0.574 & 0.506 &  0.139 \\
  A3532n            & 194.253647 & $-$29.951527 & 2MASX J12570087-2957054   & 16  & 17021 & 279  & 0.561 & 9   & 17062 & 342  & 0.717 & 0.894 &  0.765 \\
  A3532\textbf{m}   & 194.341339 & $-$30.363482 & PKS 1254-30               & 80  & 16671 & 427  & 0.857 & 57  & 16700 & 443  & 0.908 & 1.151 &  1.628 \\
  A1644\textbf{m}   & 194.298248 & $-$17.409575 & 2MASX J12571157-1724344   & 288 & 14077 & 1011 & 2.039 & 283 & 14088 & 1018 & 1.473 & 2.363 & 13.900 \\
  A1651\textbf{m}   & 194.843826 &  $-$4.196117 & 2MASX J12592251-0411460   & 177 & 25463 & 862  & 1.707 & 160 & 25465 & 877  & 1.808 & 2.262 & 12.700 \\
  A1656sw           & 194.351242 &    27.497778 & NGC 4839                  & 54  & 7412  & 404  & 0.822 & 54  & 7412  & 404  & 0.540 & 0.918 &  0.803 \\
  A1656\textbf{m}   & 194.898788 &    27.959389 & NGC 4874                  & 828 & 6921  & 1046 & 2.132 & 813 & 6927  & 1039 & 1.611 & 2.487 & 15.900 \\
  A3526Be           & 196.608917 & $-$40.414490 & ESO 323-G077              & 5   & 4615  & 101  & 0.208 & 4   & 4656  & 113  & 0.255 & 0.308 &  0.030 \\
  A3556\textbf{m}   & 201.027893 & $-$31.669956 & ESO 444-G025              & 102 & 14424 & 504  & 1.016 & 90  & 14436 & 520  & 1.047 & 1.346 &  2.586 \\
  A1736Anw          & 201.544220 & $-$26.826834 & 6dF J1326106-264937       & 14  & 9974  & 158  & 0.321 & 6   & 9976  & 148  & 0.208 & 0.342 &  0.042 \\
  A1736A\textbf{m}  & 201.683777 & $-$27.439398 & ESO 509-G008              & 43  & 10500 & 271  & 0.549 & 18  & 10481 & 311  & 0.517 & 0.759 &  0.457 \\
  A1736Acn          & 201.898788 & $-$27.042744 & ESO 509-G016              & 8   & 10977 & 168  & 0.340 & 7   & 11059 & 186  & 0.449 & 0.512 &  0.141 \\
  A1736Ase          & 202.062210 & $-$27.976803 & MCG -05-32-027            & 9   & 9862  & 140  & 0.283 & 2   & 9850  & 23   & 0.172 & 0.091 &  0.001 \\
  A1736Bcn          & 201.703033 & $-$27.143835 & ESO 509-G009              & 28  & 14064 & 568  & 1.145 & 28  & 14064 & 568  & 0.505 & 1.120 &  1.488 \\
  A1736B\textbf{m}  & 201.866852 & $-$27.324682 & IC 4252                   & 107 & 13600 & 866  & 1.748 & 97  & 13572 & 878  & 1.372 & 2.092 &  9.668 \\
  A3558\textbf{m}   & 201.987015 & $-$31.495474 & ESO 444-G046              & 525 & 14408 & 953  & 1.920 & 448 & 14419 & 961  & 1.850 & 2.452 & 15.600 \\
  A2040Bsw          & 227.880768 &     7.251906 & CGCG 049-033              & 14  & 13399 & 164  & 0.331 & 8   & 13330 & 211  & 0.649 & 0.631 &  0.265 \\
  A2040B\textbf{m}  & 228.197815 &     7.434258 & UGC 09767                 & 136 & 13491 & 600  & 1.210 & 97  & 13584 & 663  & 1.309 & 1.708 &  5.263 \\
  A2052nw           & 228.094116 &     7.727029 & CGCG 049-041              & 18  & 9881  & 253  & 0.514 & 8   & 9934  & 309  & 1.054 & 0.958 &  0.918 \\
  A2052\textbf{m}   & 229.185364 &     7.021667 & UGC 09799                 & 158 & 10355 & 588  & 1.192 & 120 & 10416 & 648  & 1.115 & 1.599 &  4.276 \\
  A2063A\textbf{m}  & 230.772095 &     8.609181 & CGCG 077-097              & 189 & 10312 & 672  & 1.364 & 145 & 10347 & 758  & 1.196 & 1.818 &  6.284 \\
  A2142\textbf{m}   & 239.583450 &    27.233349 & 2MASX J15582002+2714000   & 182 & 27031 & 827  & 1.634 & 155 & 27046 & 830  & 1.745 & 2.152 & 11.000 \\
  A2147\textbf{m}s  & 240.570862 &    15.974513 & UGC 10143                 & 185 & 10706 & 854  & 1.732 & 178 & 10706 & 867  & 1.181 & 1.979 &  8.111 \\
  A2147\textbf{m}n  & 240.582687 &    16.346182 & UGC 10144                 & 210 & 11039 & 950  & 1.925 & 200 & 11030 & 964  & 1.479 & 2.288 & 12.600 \\
  A2147s            & 240.981934 &    14.902552 & IC 1168                   & 27  & 10564 & 467  & 0.946 & 15  & 10686 & 557  & 0.651 & 1.208 &  1.846 \\
  A2147se           & 241.606033 &    15.685868 & UGC 10201                 & 31  & 11740 & 489  & 0.989 & 31  & 11740 & 489  & 0.755 & 1.162 &  1.648 \\
  A2151sw           & 240.883575 &    17.198523 & NGC 6034                  & 32  & 10351 & 438  & 0.889 & 30  & 10348 & 434  & 0.552 & 0.968 &  0.950 \\
  A2151\textbf{m}w  & 241.148987 &    17.721445 & NGC 6041                  & 63  & 11006 & 871  & 1.764 & 63  & 11006 & 871  & 1.083 & 1.927 &  7.498 \\
  A2151\textbf{m}c  & 241.287537 &    17.729971 & NGC 6047                  & 92  & 10436 & 652  & 1.323 & 89  & 10434 & 663  & 0.645 & 1.353 &  2.589 \\
  A2151\textbf{m}n  & 241.566620 &    18.249800 & NGC 6061                  & 77  & 11186 & 341  & 0.691 & 57  & 11229 & 359  & 0.801 & 0.965 &  0.942 \\
  A2151e            & 241.663986 &    17.761154 & IC 1194                   & 22  & 11632 & 452  & 0.915 & 22  & 11632 & 452  & 0.371 & 0.870 &  0.692 \\
  A2152\textbf{m}nw & 241.360168 &    16.442734 & 2MASX J16052644+1626338   & 56  & 13504 & 478  & 0.964 & 39  & 13480 & 466  & 0.857 & 1.172 &  1.698 \\
  A2152\textbf{m}se & 241.371750 &    16.435793 & UGC 10187 NED2            & 60  & 13126 & 255  & 0.514 & 19  & 13203 & 296  & 0.586 & 0.763 &  0.469 \\
  A2197\textbf{m}w  & 246.293030 &    40.892746 & NGC 6146                  & 67  & 8900  & 343  & 0.697 & 47  & 8870  & 331  & 0.782 & 0.909 &  0.782 \\
  A2197\textbf{m}c  & 246.921143 &    40.926899 & NGC 6160                  & 111 & 9575  & 463  & 0.939 & 64  & 9574  & 541  & 0.958 & 1.348 &  2.558 \\
  A2197\textbf{m}e  & 247.436890 &    40.811710 & NGC 6173                  & 92  & 8786  & 395  & 0.802 & 68  & 8774  & 380  & 0.709 & 0.965 &  0.934 \\
  A2199\textbf{m}   & 247.159485 &    39.551380 & NGC 6166                  & 461 & 9086  & 785  & 1.595 & 441 & 9083  & 795  & 1.244 & 1.903 &  7.175 \\
  A2204Aw           & 247.792801 &     5.530654 & 2MASX J16311027+0531503   & 7   & 45236 & 371  & 0.710 & 7   & 45236 & 371  & 0.571 & 0.849 &  0.718 \\
  A2204An           & 248.111099 &     5.839127 & 2MASX J16322666+0550208   & 8   & 44975 & 264  & 0.507 & 4   & 45025 & 237  & 0.120 & 0.375 &  0.062 \\
  A2204A\textbf{m}  & 248.195404 &     5.575833 & VLSS J1632.7+0534         & 77  & 45378 & 856  & 1.639 & 42  & 45406 & 1062 & 1.985 & 2.595 & 20.500 \\
  A2256cf           & 255.294220 &    78.726463 & 2MASX J17011061+7843352   & 27  & 17181 & 719  & 1.444 & 27  & 17181 & 720  & 0.599 & 1.384 &  2.832 \\
  A2256cb           & 255.700409 &    78.740837 & 2MASX J17024809+7844270   & 16  & 19704 & 250  & 0.500 & 12  & 19718 & 209  & 0.453 & 0.552 &  0.182 \\
  A2256\textbf{m}   & 256.113525 &    78.640564 & UGC 10726                 & 231 & 17530 & 1168 & 2.341 & 231 & 17530 & 1167 & 1.427 & 2.552 & 17.800 \\
  A2255sw           & 257.713409 &    63.853771 & 2MASX J17105121+6351135   & 12  & 24487 & 383  & 0.760 & 9   & 24556 & 442  & 0.567 & 0.974 &  1.011 \\
  A2255\textbf{m}   & 258.119812 &    64.060699 & ZwCl 1710.4+6401A         & 155 & 24063 & 1072 & 2.128 & 154 & 24050 & 1069 & 1.564 & 2.465 & 16.300 \\
  A2255e            & 258.788116 &    64.048248 & 2MASX J17150914+6402536   & 14  & 24244 & 287  & 0.569 & 6   & 24078 & 206  & 0.599 & 0.597 &  0.232 \\
  A3716\textbf{m}   & 312.987152 & $-$52.629829 & ESO 187-G026              & 140 & 13509 & 744  & 1.501 & 123 & 13508 & 783  & 1.246 & 1.877 &  6.986 \\
  A2634sw           & 354.218323 &    26.509964 & UGC 12708                 & 13  & 9490  & 272  & 0.552 & 11  & 9434  & 268  & 0.646 & 0.741 &  0.425 \\
  A2634\textbf{m}   & 354.622437 &    27.031303 & NGC 7720                  & 172 & 9243  & 716  & 1.454 & 160 & 9235  & 736  & 1.217 & 1.796 &  6.031 \\
  A4038A\textbf{m}  & 356.937683 & $-$28.140705 & IC 5358                   & 196 & 8872  & 725  & 1.474 & 166 & 8910  & 773  & 1.056 & 1.769 &  5.765 \\
  A4049s            & 357.903015 & $-$28.365068 & IC 5362                   & 23  & 8307  & 270  & 0.549 & 23  & 8307  & 270  & 0.639 & 0.742 &  0.424 \\
  A4049n            & 357.976715 & $-$27.929789 & MCG -05-56-025            & 18  & 8809  & 88   & 0.180 & 6   & 8837  & 67   & 0.159 & 0.184 &  0.007 \\
\bottomrule
\end{longtable}
\end{scriptsize}
\end{center}
\clearpage
\twocolumn


\end{appendices}


\appendix


\label{lastpage}

\end{document}